\documentclass[traditabstract]{aa} 
\usepackage{txfonts}
\usepackage{graphicx}
\usepackage{txfonts}
\def\la{\raise.5ex\hbox{$<$}\kern-.8em\lower 1mm\hbox{$\sim$}}
\def\ma{\raise.5ex\hbox{$>$}\kern-.8em\lower 1mm\hbox{$\sim$}}

\def\msol{M$_{\odot}$ }

\def\kms{$\rm km\, s^{-1}$}
\def\cm3{$\rm cm^{-3}$}
\def\Ts{$\rm T_{*}$~}
\def\Vs{$\rm V_{s}$}
\def\n0{$\rm n_{0}$}
\def\B0{$\rm B_{0}$}
\def\ne{$\rm n_{e}$~}
\def\Ne{$\rm N_{e}$~}
\def\Te{$\rm T_{e}$~}

\def\erg{$\rm erg\, cm^{-2}\, s^{-1}$}

\def\L12{L$_{12\mu m}$~}
\def\F12{F$_{12\mu m}$~}
\def\agr{a$_{gr}$}
\def\Hb{H${\beta}$}
\def\Ha{H${\alpha}$}
\def\Hg{H${\gamma}$}

\def\Ly{Ly$\alpha$~}

\def\Tef{T$_{eff}$}

\def\RO3{R$_{[OIII]}$}

\begin{document}

   \title{
Element abundances  in metal poor local star-forming
galaxies constrained by  the weak spectral lines}


   \author{M. Contini \inst{1,2}
}

   \institute{Dipartimento di Fisica e Astronomia, University of Padova, Vicolo dell'Osservatorio 2. I-35133 Padova, Italy
         \and
             School of Physics and Astronomy, Tel Aviv University, Tel Aviv 69978, Israel\\
}

   \date{Received }


  \abstract{
We have  collected  from different surveys some significant spectroscopic data observed from star-forming
galaxies in the local Universe.
The objects showing a relatively rich spectrum in number of lines  from different elements were selected
in order to constrain the models.
In particular, we  looked at
the  relatively weak lines such as  [OIII]4363,   HeII4686,  HeI4471 and HeI5876.
We have modelled  in detail the spectra by the coupled effect of photoionization  from  the  stars and shocks.
We  have found that  the abundances relative to H of most of the elements are  lower than solar but not as
low as  those evaluated  by the direct strong line and  the \Te methods.
Sulphur which appears  through the
[SII]6717,6731 and [SIII]6312 lines is not depleted,  revealing a   strong contribution from the ISM.
We have added to the sample the optical-UV spectra of local low-metallicity dwarf galaxies
containing the   CIV/\Hb~ and CIII]/\Hb~ line ratios in order to determine with relative precision the C/H
relative abundance.
The  results show  He/H   lower than  solar in some objects
and suggest that the  geometrical thickness of the clouds  constrains the HeII/\Hb~ line ratios.
We explain the low He/H by  mixing of the wind from the star-forming region with ISM clouds.
}


\titlerunning{
	Element abundances  in metal poor local star-forming galaxies}
\authorrunning{M. Contini}

\maketitle

\section{Introduction}

The  transformation of the line  spectra  emitted from galaxies throughout the redshift  is a leading argument 
because it is connected with reionization at a certain epoch (Izotov et al 2020 and references therein). 
In local galaxies, however,  the reionization era is less directly recognized from the element  abundances because of 
physical and chemical phenomena acting on gas and dust, respectively, in their evolution towards the local era.
 Even for high redshift galaxy spectra Rupke et al (2005) claim that 'the prominence of \Ly  makes it a 
tempting target for 
parametrizing outflows. However, radiation transfer effects make it an ambiguous indicator.'

In this paper we deal with the spectra emitted from local star-forming (SF) galaxies.
In the latest years, the number of  lines that  could be observed in each spectrum  from  local galaxies 
substantially increased. 
In fact, the observations  (e.g.    
Izotov et al 1997, 2006, 2018a,b, 2019a,b, 2021,  Guseva et al 2020, P\'{e}rez-Montero et al 2005, Pustilnik et al 2004, 
Berg et al 2012, 2016, etc.)  provide high precision spectra accounting for 
significant lines in the optical-- near-IR range  which are generally weak ([OIII]4363, HeII 4684, HeI 5876, 
[OI]6300,6363, [SIII]6312, etc.) besides the relatively strong 
[NeIII]3869,3969, [OIII]5007,4959, [OII]3727,3729, [NII]6548,6584,  [SII]6717, 6731 doublets  and
the Balmer   \Ha~ and  \Hb~ lines.
Berg et al (2016) presented  high precision spectra accounting  
also for UV lines,  in particular CIV1548,1550 and  CIII]1906, 1909.
The weak lines  that  appear in nearly all  local galaxy  surveys   
 allow a  more  accurate classification of  galaxy types  such as AGN, starburst  and  host galaxies  in general.
For galaxy surveys at high redshift z$>$1  the strongest lines which are provided by the observations are  
necessarly sufficient to obtain the characteristic  physical conditions of the gas although  with some 
uncertainty (see e.g. Contini 2014 and references therein).

The increasing number  of objects observed  in  recent surveys  demands fast classification methods dictated 
directly by the observations
such as e.g. the BPT diagrams (Baldwin, Phillips \& Terlevich 1981, Kauffmann 2003, Kewley 2001) for AGN, 
star-formation and shock dominated objects.  
Most authors use the \Te method that was succesfully  adopted for AGN (e.g. Monteiro \& Dors 2021 and references therein).
Direct methods such as the \Te ones and those
based on the strong lines  and on the electron temperature and density obtained  from the characteristic 
line ratios (e.g. Izotov et al 2020, 2021 and references therein)
yield metallicity results (in terms in particular of the O/H relative abundance) for single objects with a 
relatively high precision. By these methods very low element abundances     
compared to solar  were generally found  in local star-forming   galaxies (SFG) and  HII regions,
while  abundances  less far from   solar  in  objects at similar redshifts were found  by the detailed 
modelling of the spectra  (e.g. Contini 2017 and references therein).  
However, detailed modelling methods are  long-time spending  even  when used to fit a single spectrum 
therefore, they are less adapted to  model  surveys  presenting  a high number of galaxy spectra.

Detailed modelling methods were  used to investigate AGN spectra (e.g. {\sc cloudy}, Ferland et al 2017). 
They can be adapted to distinguish the spectra
emitted from gas affected  by the   flux from different photoionization sources. 
 However,  models based on pure photoionization   cannot always reproduce all the lines
 in a single spectrum. For instance,
in some spectra emitted from SFG   the HeII4686 line  is evident.  In order to reproduce  
the observed  HeII/\Hb~ line ratio the emitting gas should be heated to a  relatively high temperature  
 because the ionization potential  of the He$^{++}$ ion  is relatively high (54.17 eV). 
Therefore, in the present investigation we use the  code {\sc suma} (e.g. Contini 2009)  which shows  the role 
of shocks  coupled  to the  photoionization flux. 
 To reproduce the  spectra emitted  from local SF galaxies, shocks propagating throughout a ionized medium 
with a relatively high velocity (300-500 \kms) were  proposed   by Izotov et al (2012)  in order
 to heat the gas to temperatures high enough to obtain strong HeII4686 lines.  
Shocks propagating throughout a neutral medium, on the other hand,  were  suggested by Allen et al (2008). 

 In this paper we  present the  method adopted  to evaluate the abundances of the leading elements
 in local  SFG
and   we compare   our results  with those obtained by the \Te and strong line methods.
We have selected the spectra more adapted to modelling, i.e. those containing  
at least  \Hb, \Ha,  [OIII]5007, [OII]3727   and the [NII] lines.  We have chosen  to explore in particular    
 local  SFG spectra  characterized by significantly 
 high [OIII]5007/[OII]3727 ($>$ 10) and low  [OII]3727/\Hb~ ($\leq$1), [NII]/\Hb, [SII]/\Hb~ and [OI]/\Hb, as those 
 presented by  Izotov et al (2020), Izotov et al (2019a,b), Izotov et al (2018a,b,), Guseva et al (2020),  
 Pustilnik et al (2004), etc.
The sample of galaxies selected in this paper is presented in Sect. 2.  
Results are  given in Sect. 3  and discussed  in Sect. 4. Concluding remarks  follow in Sect. 5.

\section{Selected galaxies}

 The SFG sample selected for our investigation is  shown in Table 1.
Each galaxy is presented by the corresponding redshift, by the observed \Hb~ line intensity, the observed 
\Ha/\Hb~ and \Hg/\Hb~ line ratios. The corresponding calculated values are  reported from the next sections.
These Balmer lines were chosen because a) from the comparison of calculated  with observed \Hb~  the radius of the
emitting nebulae  can be evaluated, b) from the comparison of the calculated with observed  \Ha/\Hb~ the physical 
conditions in the emitting nebulae are roughly evaluated and c) from the comparison of calculated with 
observed \Hg/\Hb~  line blending can be revealed.

The SFG sample   includes the  galaxies  of  the
Izotov et al (2020, hereafter I20) survey which  covers the  0.02811$\leq$z$\leq$0.06360  redshift range.
The  spectra were classified as   extremely metal-poor types.
The survey contains  eight objects with similar line ratios except for  [OIII]5007+/\Hb~ (the plus indicates
that  the 5007, 4959  doublet is summed) which ranges between 5.49 and 10.35. 
Moreover, the [OII]3727+3729/\Hb~ ratios are as low as  the [OIII]4363/\Hb~ ones. 
Characteristic of these spectra are the   HeI4471 and HeII4686 lines which are often  lacking in  SFG spectra.
We have added to the I20 survey the spectrum of the most metal-poor galaxy J1234+3901 at z=0.133  presented
by Izotov et al (2019b, hereafter I19b) which shows a relatively low [OIII]5007+/\Hb=2.7, for comparison. 
J0811+4730 spectrum presented by Izotov et al (2018a hereafter I18a) at z=0.04444 contains the HeII line.
I18a found an extremely low metallicity (12+log(O/H)=6.98).  J0811+4730 is  also included in our sample (Table 1).
Moreover,  the survey of SFG presented by Izotov et al (2018b, hereafter I18b) at z=0.2993-0.4317 
with relatively high [OIII]5007+/[OII]3727+  (but lower than those included in the I20 survey) 
is also accounted for. I18b observations were aimed to
detect Lyman continuum emission.  
I18b claim that they discovered 'a class of galaxies in the local Universe
which are leaking ionizing radiation and sharing many properties of high-redshift galaxies'. 
I18b have found very low metallicities by the strong line methods.
They  suggest about local SFG galaxies  with low metallicities
 that  a considerable fraction of the galaxy stellar mass was formed during the most recent burst
of star formation. 
Guseva et al (2020, hereafter G20) presented observations of SFG, in particular  J0901+2119 and J1011+1947 already 
included  within the I18b survey but containing more significant lines (e.g. MgII2796).
We included the G20 survey in our sample.

P\'{e}rez-Montero et al (2011) observed  blue compact dwarf (BCD) galaxies
at z between 0.016 and 0.042 with the aim  of investigating galaxies with an  outstandingly high  N/H relative abundance. 
The spectrum of one of these galaxies (HS0837.4717 at z=0.04195) - particularly rich in  significant lines - 
was already observed by Pustilnik et al (2004, hereafter P04)  with the same aim. It  appears in our sample.
Two galaxies, J0314-0108 (z=0.02027) and J1433+1544 (z=0.02741) were selected from the Izotov et al 
(2019a, hereafter I19a) survey which includes in  the spectra the [OII]7327+ doublet, but it lacks some other 
lines such as e.g. 
the He ones. Unfortunately most of the spectra do not show all the significant lines, in particular the  [OII]3727+. 
Therefore only  two galaxies are  added for comparison in  our sample.
 Berg et al (2016) presented the spectra of selected metal poor local dwarf galaxies at z between 0.003 and 0.04
containing some characteristic UV lines (e.g. CIV and CIII]). We have  included this survey in order to calculate 
the C/O and C/N relative abundances.

\begin{table*}
\centering
\caption{Balmer lines}
\begin{tabular}{lcccccccccccccccc} \hline  \hline
\ galaxy     & z      & \Hb(obs)$^1$ & \Hb(calc)$^2$ &\Ha/\Hb(obs) &\Ha/\Hb(calc)& \Hg/\Hb(obs) & \Hg/\Hb(calc)\\ \hline 
\ J0007+0226$^{3}$ & 0.06360& 52.3         &120            &2.77         & 2.84        &0.473         &0.47      \\
\ J0159+0751$^{3}$ &0.06105 & 99.3         &4.             &2.75         & 2.93        &0.475         &0.46      \\
\ J0820+5431$^{3}$ &0.03851 & 20.2         &1.9            &2.32         & 2.9         &0.486         &0.46       \\
\ J0926+4504$^{3}$ &0.04232 & 34.4         &1.4            &2.75         & 2.93        &0.44          &0.46       \\
\ J1032+4919$^{3}$ &0.04420 & 109.6        &2.2            & -           & 3.4         &0.448         &0.45       \\
\ J1205+4551$^{3}$ &0.06540 & 115.0        &5.             &2.75         & 3.27        &0.475         &0.42       \\
\ J1242+4851$^{3}$ &0.06226 & 40.1         &2.             &2.72         & 2.94        &0.497         &0.46         \\
\ J1355+4651$^{3}$ &0.02811 & 51.2         &1.9            &2.73         & 2.9         &0.483         &0.46        \\
\ J1234+3901$^{4}$ &0.13297 & 9.74         &9.             &2.71         & 3.18        & 0.46         &0.45         \\
\ J0901+2119$^{5,6}$&0.2993 &29.1          &36             &2.88         &2.88         & 0.48         &0.46         \\
\ J1011+1947$^{5,6}$&0.3322 &27.0          &4.2            &2.83         &2.88         & 0.445        &0.465       \\
\ J1243+4646$^{5}$ &0.4317  &14.1          &5.1            &2.8          &2.95         & 0.47         &0.462        \\
\ J1248+4259$^{5}$ &0.3629  &35.2          &4.6            &2.79         &2.95         &0.496         &0.461         \\
\ J1256+4509$^{5}$ &0.3530  &11.4          &5.7            &2.80         &2.9          &0.482         &0.464   \\
\ J0925+1403$^6$   &0.3010  &21.61         &30.0           &2.85         &2.9          &0.47           &0.46   \\
\ J1154+2443$^6$   &0.3690  &9.77          &3.7            &2.77         &2.92         &0.45           &0.46   \\
\ J1442+0209$^6$   &0.2937  &19.44         &7.0            &2.82         &2.84         &0.47           &0.47 \\
\ J0811+4730$^{7}$ &0.04444 &12.60         &7.             &3.12         &3.25         & 0.48          &0.45        \\
\ HS0837+4717$^{8}$&0.04195 &313           &90.            &2.76         &2.89         &0.48          &0.46  \\  
\ J0314-0108$^9$   &0.02741 &  26.2        &  9.4          & -           &3.1          & -             &0.455  \\
\ J1433+1544$^9$   &0.02027 &   2.0        &  8.5          & -           &3.1          & -            &0.456  \\ 
\  J082555$^{10}$&0.003& 230.8$^{11}$ &2.6            & 2.76        & 2.96        & 0.476         &0.46   \\
\  J104457$^{10}$&0.013&413.7$^{11}$  &2.35           &2.75         & 3.0         &0.506          &0.46    \\
\  J120122$^{10}$&0.003&114.3$^{11}$  &0.8            &2.78         & 2.96        &0.53           &0.46   \\
\  J124159$^{10}$&0.009&98.3$^{11}$   &31.            &2.79         & 2.97        &0.495          &0.46   \\
\  J122622$^{10}$&0.007&81.31$^{11}$  &10.            &2.79         &3.34         &0.493          &0.46\\
\  J122436$^{10}$&0.040&138.4$^{11}$  &1.7            &2.79         &3.26         &0.458          &0.45 \\
\  J124827$^{10}$&0.030&78.1$^{11}$   &1.7            &2.77         &3.26         &0.47           &0.46 \\  \hline
\end{tabular} 

$^1$ in units of 10$^{-16}$ \erg; $^2$ in 10$^{-4}$ \erg; $^3$ I20 survey; $^4$  I19b for J1234.3901;
$^5$  I18b survey; $^6$ G20 survey ;$^7$  I18a; $^8$  P04); $^9$ I19a survey;  $^{10}$ B16 galaxy sample;
$^{11}$ observed  flux in units of 10$^{-16}$ \erg .

\end{table*}

 \begin{figure}
 \centering
\includegraphics[width=7.2cm]{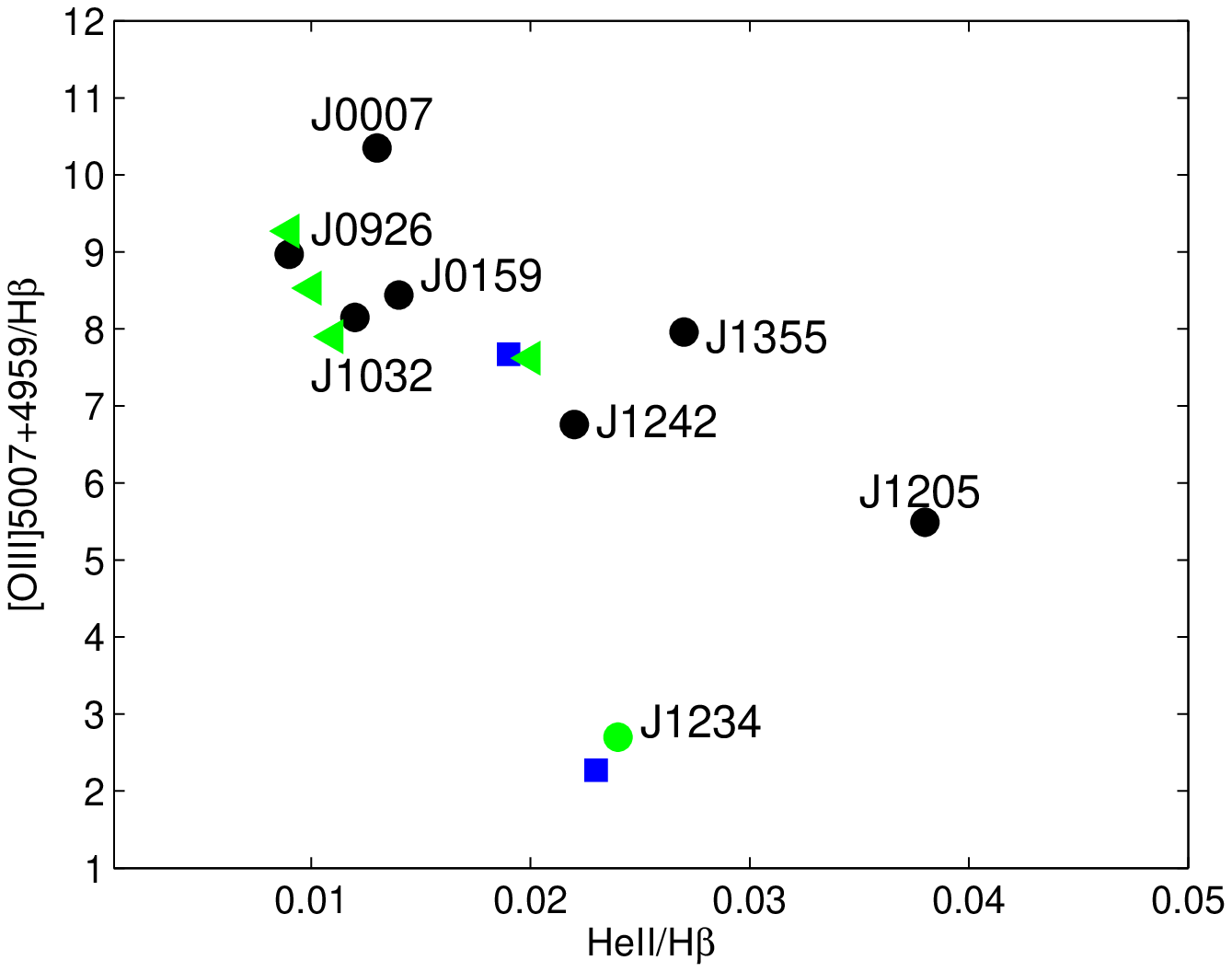}
\includegraphics[width=7.2cm]{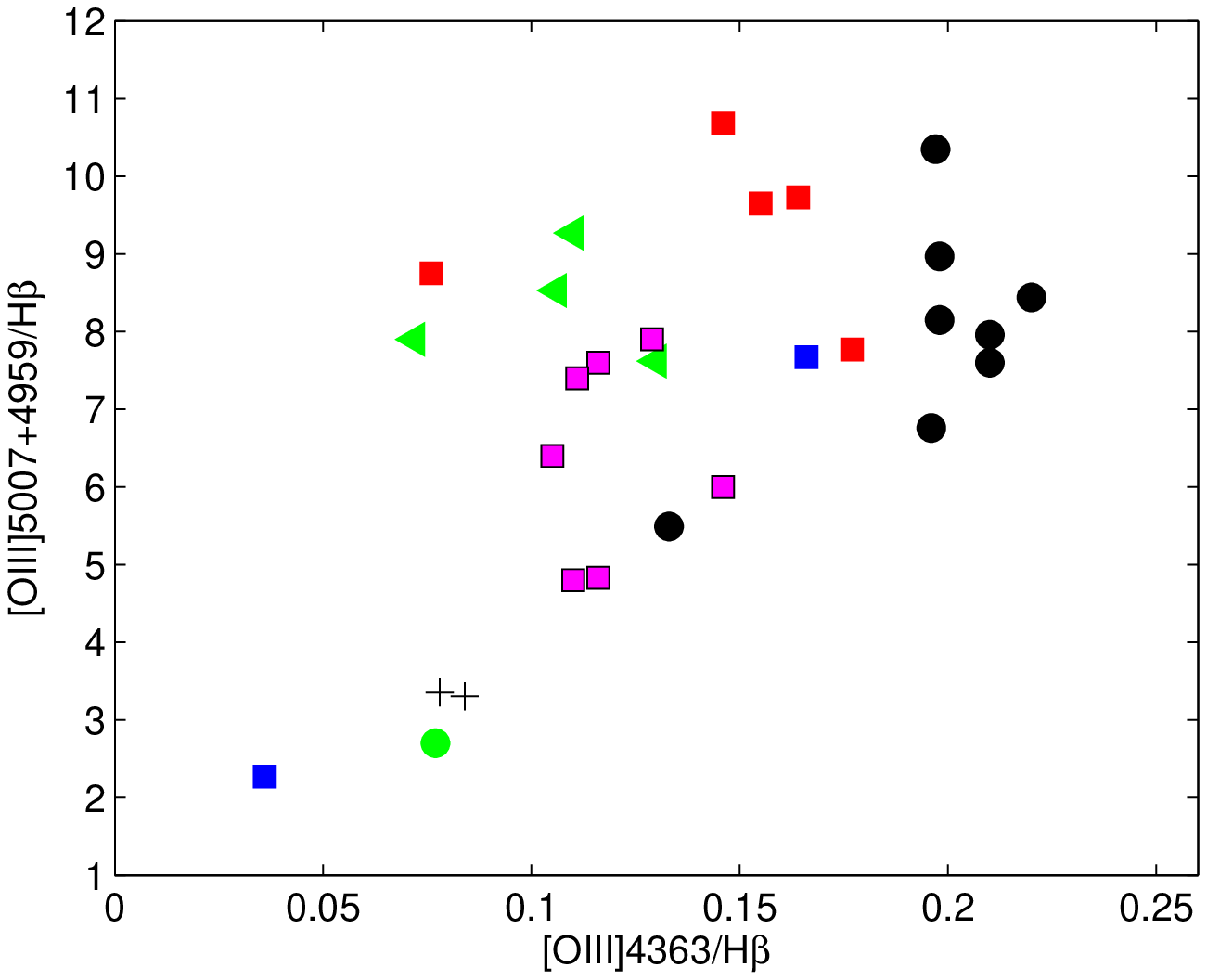}
\includegraphics[width=7.2cm]{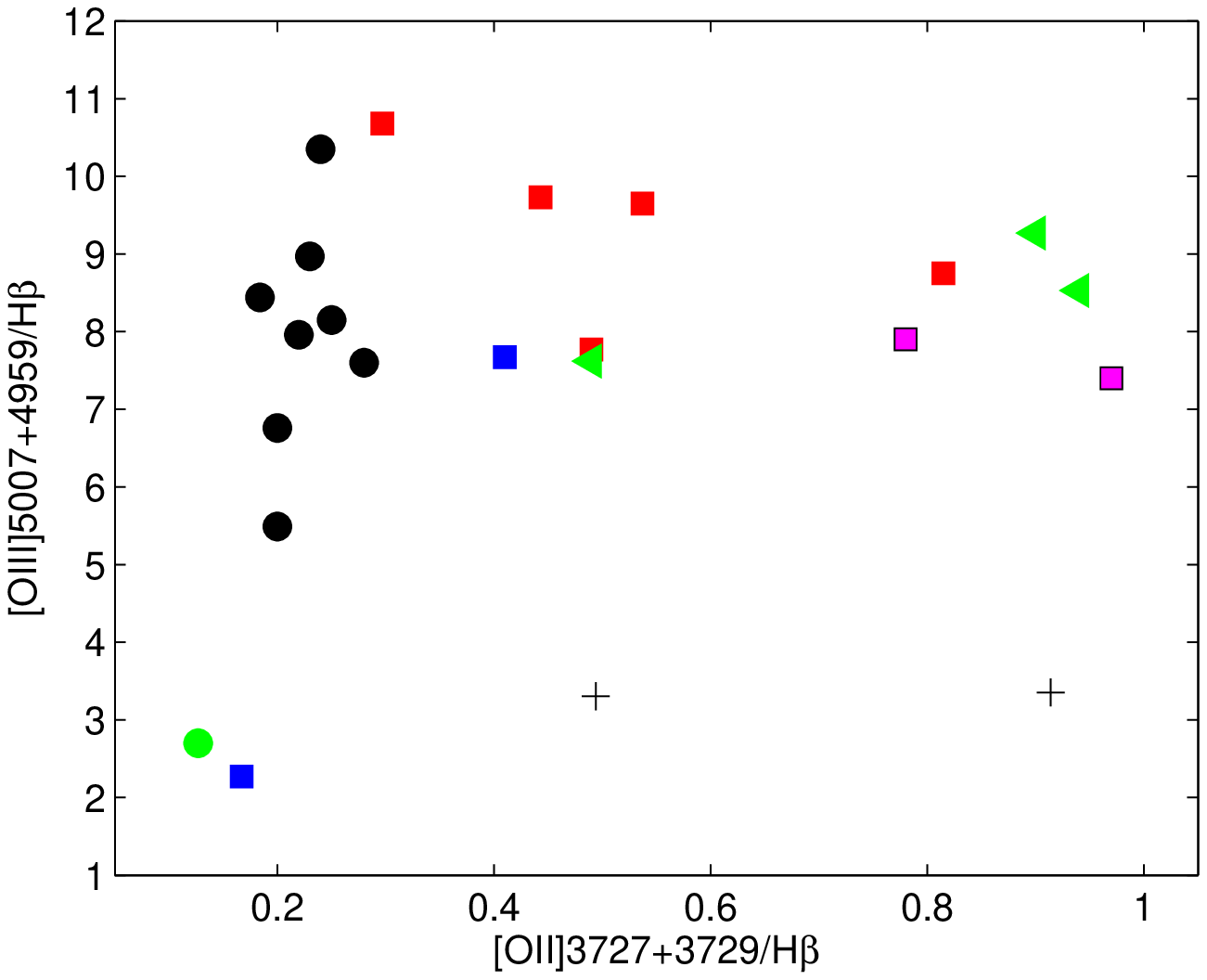}
 \caption{
Black circles: I20 survey galaxies;
green circle: J1234 (I19b);
red squares: I18b survey objects;
blue squares: J0811 (I18a) and HS0837 (P04);
green triangles: G20 survey galaxies;
black plus: I19a survey;  magenta squares: B16
}
\end{figure}

 To have a  first hint about the  nature of the  objects  in our  sample we present
 in Fig. 1 the correlations of some significant observed line ratios. The data  come from Tables 2, 5, 7, 9  and 10.
 In the  top diagram,  [OIII]5007+/\Hb~ versus HeII/\Hb~  indicates that [OIII]5007+/\Hb~
 increases when HeII/\Hb~ decreases.
 Generally, both [OIII]5007+/\Hb~ and HeII/\Hb~ should increase with the temperature of the emitting gas.
 Fig. 1 (top) clearly  shows an opposite trend. This suggests that the emission lines come from a region within the cloud
 where the O$^{3+}$ ion is  relatively strong,  in agreement with  I20 claim.
 J1234 is  dislocated from the general trend for perhaps two reasons. First, it was  inserted in Table 2 because its
 spectrum  shows the
 HeII line but this galaxy is at redshift z=0.133, while the I20 sample is  at z$\leq$ between 0.028 and 0.0654.
 Second, the [OIII]5007+/\Hb~ ratio is the lowest observed one   among the sample  galaxies presented in Table 1.
 In fact,  spectra  with a relatively  high [OIII]/\Hb~ were selected by I20.
 J0007, J1355 and J1205 are slightly shifted towards higher HeII/\Hb.
 This could suggest that these   galaxies  belonging to the I20 survey  correspond to quasi solar 
 He/H ([He/H]$_{\odot}$=0.1). 
 However,  the trend is not only due to the  He/H relative abundance.
The other diagrams show that J0007 and J1205 correspond to different parameters on a large scale.
[OIII]5007+/\Hb~ versus [OIII]4363/\Hb~ (Fig. 1, middle diagram) and  [OIII]5007+/\Hb~ versus [OII]3727+/\Hb~  (bottom)
  confirm that in general the spectra do not depend  on a single specific parameter.
 Interestingly, even considering the relatively  small  number of galaxies selected  in this work,
 the correlations of different line ratios in the middle and bottom diagrams of Fig. 1 roughly indicate that objects 
 within different redshift ranges show  different trends.

\section{Modelling galaxy spectra}

\subsection{BPT diagrams}

\begin{figure*}
\centering
\includegraphics[width=8.4cm]{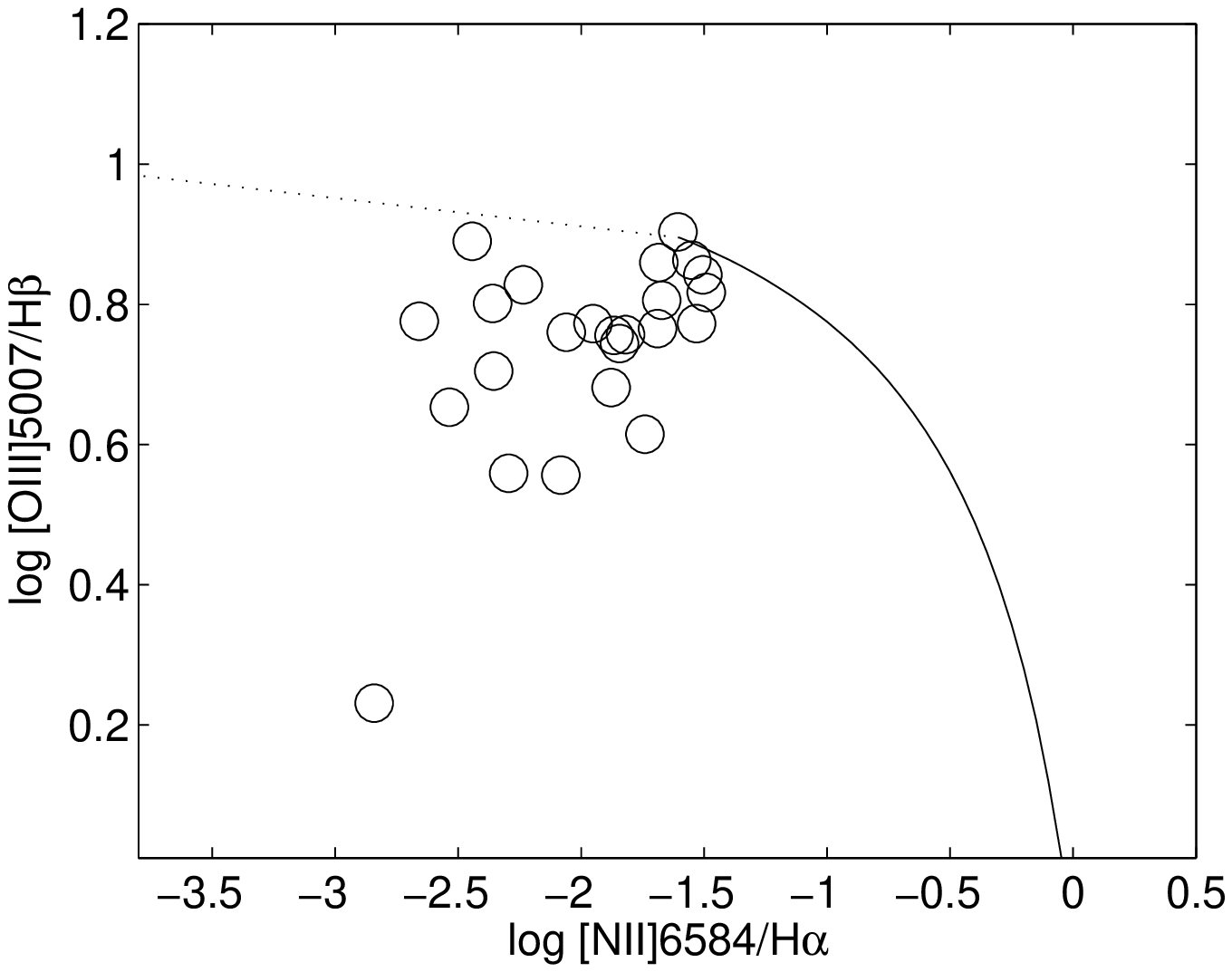}
\includegraphics[width=8.4cm]{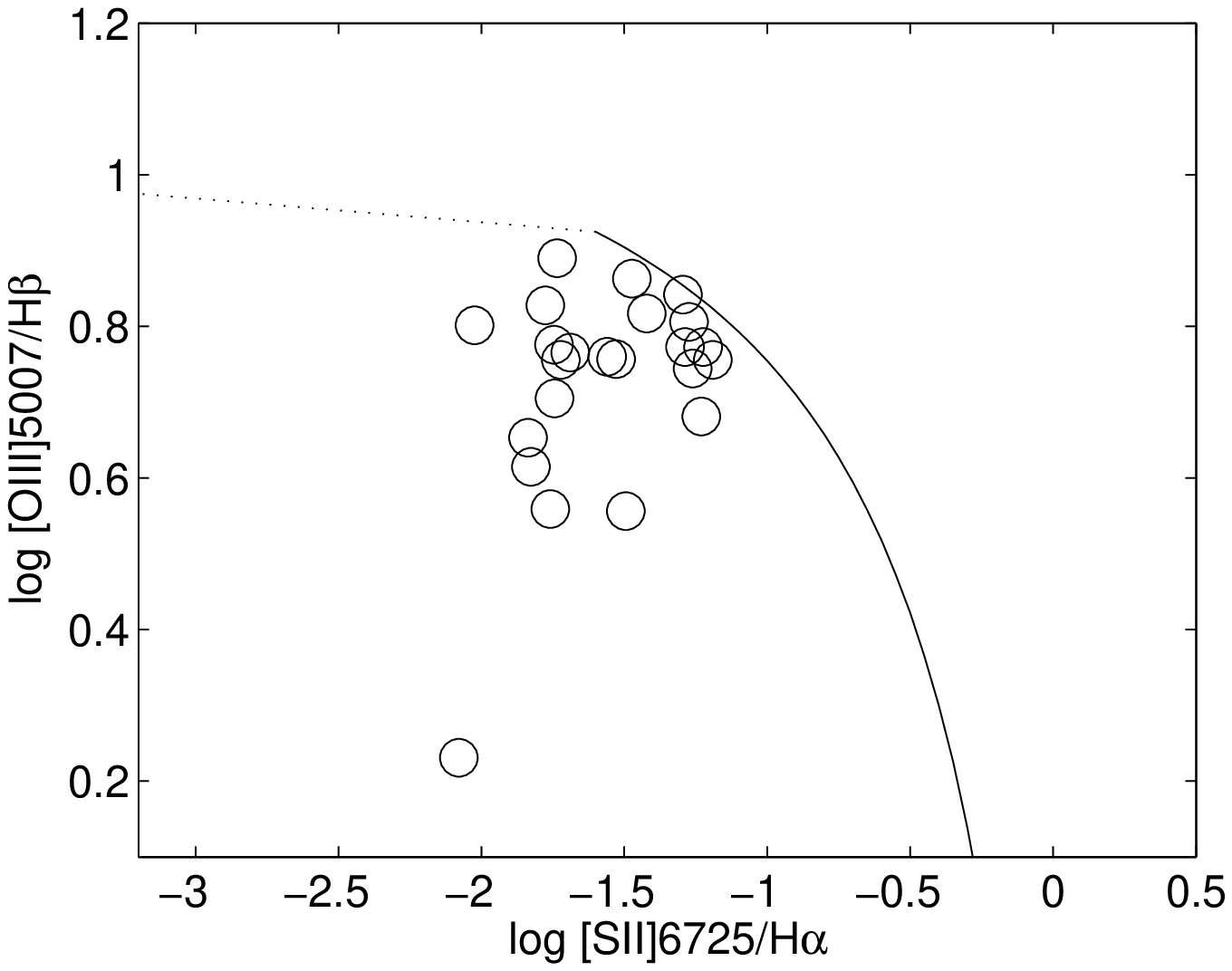}
\caption{BPT diagrams including all the objects (open circles) described in Table 1.
The black solid line is taken by Kauffmann et al (2003). The black dotted line is a rough continuation
of Kauffmann line towards very low [NII]/\Ha~ and [SII]/\Ha~ line ratios.}
\end{figure*}

Our galaxy  sample  (Table 1) is  shown in the   BPT diagrams (Fig. 2). 
The objects are located in the region of very low
[NII]/\Ha~ (left panel) and relatively low [SII]/\Ha~ (right panel)   beyond the  lower limit of the observed
line ratios e.g. for AGN,  starburst  and HII region galaxies (solid line). They may indicate low N/H and S/H relative 
abundances and/or matter-bounded  emitting clouds  such that  a large part of the recombination zone is excluded.
 J0811+4730 corresponds to the lowest [OIII]5007/\Hb~ and to the lowest O/H.

BTP diagrams (Kauffmann 2003,  Kewley et al 2001) 
for the [OIII]5007/\Hb~ and [NII]6583/\Ha~ line ratios are  generally adopted
by the  author community in order  to   identify the galaxy  type in terms of the 
radiation source. 
However, extreme physical conditions and relative abundances far from  solar, in particular  for O/H and N/H, 
may  shift the observed [OIII]/\Hb~ and [NII]/\Ha~ line ratios throughout the BTP diagram
towards sectors which were  assigned to   different galaxy types.
For example, a  low N/H relative abundance may shift the observed [NII]/\Ha~  line ratio emitted 
from a  AGN towards the SB galaxy domain  
although other features of the same object are characteristic of an  AGN.
These  diagrams  provide an approximated but  rapid information of the gas physical conditions  
within the emitting clouds. (The line and continuum  flux are  emitted from the nebula).
The [OIII]/\Hb~ and [NII]/\Ha~ line ratios alone cannot  definitively constrain the models because we deal 
with two line ratios 
referring to different elements. [OIII]/\Hb~ ratios depend  on  the effective temperature of the star, on the
ionization parameter, on  the shock velocity, etc., more  than on the  O/H relative abundance,
whereas  the [NII]/\Ha~ ratios depend also strongly on the N/H relative
abundances.  We cannot determine a priori whether the best fit  to the observed [NII]/\Ha~ line ratio could be  
reached by changing one or more input parameters
representing  the physical conditions (see Appendix A) or by modifying the 
N/H relative abundance. 
  N$^+$ and H$^+$ ions as well as  O$^+$ and H$^+$   are correlated by charge exchange reactions, therefore 
 [NII]/\Hb~ and [OII]/\Hb~ have a similar trend throughout a cloud.
 When  [OIII]/\Hb~ (and [OII]/\Hb)  are well reproduced by modelling  and [NII]/\Hb~ is less fitted by solar N/H 
  the  N/H  relative abundances can be calculated directly from 
([NII]/\Hb)$_{obs}$ = ([NII]/\Hb)$_{calc}$ (N/H)/(N/H)$_{\odot}$,  
where (N/H)$_{\odot}$ is the solar N/H relative abundance.  In fact, N is  not a strong coolant. 

\subsection{Detailed modelling results}

  Our modelling method  consists in trying to  reproduce all the line ratios reported by the observations
 in a single spectrum by a particular model characterised by a  set of physical parameters and
 element  abundances (see Appendix A for a detailed presentation of the model) which are briefly described  
 in the following.
 The shock velocity \Vs, the atomic preshock density \n0 and the preshock
 magnetic field \B0 define the hydrodynamical field.
 The primary radiation from  the stars which is represented by  the
 effective star temperature  \Ts and the ionization parameter $U$
 affects the surrounding gas.   This  region  is  considered
 as a  sequence of plane-parallel slabs.
 The geometrical thickness $D$ of the clouds is also an input parameter.
 The fractional abundances of the ions are calculated resolving the ionization equations
 for each element (H, He, C, N, O, Ne, Mg, Si, S, Ar, Cl, Fe) in each ionization level.
 The calculated line ratios, integrated throughout the cloud geometrical width, are compared with the
 observed ones. The calculation process is repeated
 changing  the input parameters until the observed data are reproduced by the model results,  at maximum
 within 10-20 percent for the strong lines and 50 percent for the weak ones.
 A grid of models is built for each spectrum in order to select the best fit to the data.
 The observed  spectra  which strongly constrain  the models are those showing
MgII2798+,  [OII]3727+, [NeIII]3686,   [OIII]4363, HeI4471,  HeII4686, [OIII]5007+, HeI5876,   [OI]6300+,  [SIII]6312,
 [NII]6548+, [SII]6717, [SII]6731, [ArIII]7136, etc (the + indicated that the doublet is considered)
 and the Balmer lines \Hb, \Ha~ and \Hg.
 The higher the number of lines referring to different elements
 in different ionization levels, the   higher is the number of models considered in each grid.
 Our aim  is to find out  which lines
 are the most  critical ones  trying to  fit  a rich spectrum by a single-cloud model.
When   the required  precision is not reached, a pluri-cloud model is adopted (e.g. Fonseca-Faria et al 2021). 
which  yields  a more detailed picture of the physical parameter and element abundance distribution throughout 
the observed region.

To chose the grid  models we should consider that
in a hydrodynamical regime of gas outflowing from the starburst (e.g. Yu et al 2021) the shock front is
on the external edge of the clouds.
The gas throughout the shock front and downstream is heated to  high temperatures  depending on the shock velocity
and it is compressed, increasing the cooling rate.
 Consequently, through the emitting cloud, the  regions of gas emitting the different lines have different sizes.
 The internal edge of the clouds  - facing the radiation source -
 is heated by  the photoionization flux  from  the stars to temperatures of $\sim$ 2-3 $\times10^4$K which
 correspond to strong lines from intermediate ionization levels.
 Therefore, strong lines from high ionization levels
 can be predicted in a single spectrum  together with lines from  intermediate levels and recombination lines due to
 the rapid temperature  drop downstream.
 Each line intensity is integrated through the cloud following the profile of \Te and \ne.
 Consequently, the spectra can show  unexpected line ratios.
In the following we will  identify the galaxies only by their right ascension  for sake of table formatting. 

 The  observed line ratios  to \Hb~  of the selected galaxies are   presented in Tables 1, 2, 5, 7,  9  and 10
and the calculated  ones which approximatively reproduce the observations are shown
in the line (or column as in Table 10) next  to  the data  for each object  
(models mis1a-mis9a, mis1b-mis9b, miss1-miss6, mG1-mG4, mpa-mpb, misss1-misss2,  and B1m-B7m).
For a few galaxies we present different models for the same observed  spectrum in order to give a hint on  the 
selection criteria adopted in the present work.
 The physical parameter  and element abundance  sets which  characterize the models are shown in 
 Tables 3, 4, 6, 8, 9  and 10. The relative abundances of the most significant elements to H
calculated in the present work  are compared with the results obtained by the  other 
methods in the bottom of the tables. 
The results  obtained by the detailed modelling method  are selected by a compromise between calculation precision 
and  observed uncertainties.
Moreover, a further approximation  cannot be avoided  considering that the observations  cover an entire galaxy    
including gas in  different physical  conditions. 
The physical picture is not homogeneous throughout the galaxies. Therefore, the observed data for 
each  object cannot be always satisfactorily fitted with high precision  by the results of a single-cloud model  
which  represents a specific physical situation.

\begin{table*}
\centering
\small{
	\caption{Comparison of observations for I20 survey and J1234 (I19b) galaxy with model results. \Hb=1}
\begin{tabular}{ccccccccccccccccc} \hline  \hline
\     & [OII] &[NeIII]& [OIII]& HeI  & HeII &[OIII]&HeI   &[OI]  &[SIII]& [NII]& [SII]&[SII] &[ArIII]\\
\     & 3727+ &3868   & 4363  &4471  & 4686 &5007+ &5876  &6363  &6213  & 6584 &6717  & 6731 & 7136\\ \hline 
\ {\bf J0007}& 0.24  &0.57    & 0.197 &0.035 &0.013 &10.35 &0.107 &0.005 &0.009 &0.01  &0.029 &0.022 &0.037 \\
\  mis1a &0.23 &0.78    &0.10   &0.047 &0.026 &11.   &0.127 &3e-5  &0.05  & 0.011&0.018 &0.015 &0.09 \\
\  mis1b &0.20 &1.2     &0.14   &0.040 &0.04  &11.88 &0.13  &2e-4  &0.05  & 0.02 &0.017 &0.029 &0.09 \\
\ {\bf J0159} &0.184 &0.58    &0.22   &0.042 &0.014 &8.44  &0.11  &0.005 &0.007 &0.012 &0.013 &0.013 &0.02\\
\ mis2a &0.20  & 0.5    &0.19   & 0.002&0.07  &8.37  &0.005 &4.e-5 &0.04  &0.013 &0.015 &0.014 &0.02\\
\ mis2b &0.21  &0.4     &0.04   &0.03  &0.026 &8.3   &0.09  &2.e-5 &0.02  &0.011 &0.009 &0.009 &0.02\\
\ {\bf J0820} &0.28  &0.5     &0.21   &0.03  & 0.0  &7.6   &0.076 &0.0   & 0.0  &0.0   &0.03  &0.014 &0.032 \\
\ mis3a &0.24  &0.58    &0.17   &0.012 &0.07  &7.69  &0.03  &2e-5  &0.039 &0.01  &0.007 &0.006 &0.03\\ 
\ mis3b &0.25  &0.4     &0.05   &0.03  &0.03  &7.2   &0.086 &4e-5  &0.016 &0.019 &0.012  &0.011 &0.025\\
\ {\bf J0926} &0.23  &0.48    & 0.198 & 0.04 &0.009 &8.97  &0.12  &0.0   &0.009 &0.016 &0.021 &0.025 &0.027\\
\ mis4a &0.21  &0.70    &0.23   &0.001 &0.09  &8.6   &0.003 &4.e-5 &0.052 &0.01  &0.019 &0.016 &0.028\\
\ mis4b &0.18  &0.70    &0.10   &0.05  &0.011 &9.67  &0.13  &4.e-6 &0.03  &0.013 &0.005 &0.003 &0.03 \\
\ {\bf J1032} &0.25  &0.48    & 0.198 &0.04  & 0.012&8.15  &0.12  &0.009 &0.011 &0.02  &0.026 &0.025 &0.03 \\
\ mis5a	&0.3   &0.4     &0.24   &0.03  &0.047 &8.0   &0.01  &5.4e-5&0.034 &0.01  &0.012 &0.01  &0.24\\
\ mis5b	&0.23  &0.4     &0.06   &0.05  &0.011 &8.0   &0.14  &1.4e-5&0.03  &0.012 &0.009 &0.008 &0.03\\
\ {\bf J1205} &0.20  & 0.28   & 0.133 &0.043 &0.038 &5.49  &0.12  &0.015 &0.006 &0.05  &0.011 &0.03  &0.018\\
\ mis6a &0.3   &0.26    & 0.12  &0.069 &0.05  &5.8   &0.18  &4e-5  &0.044 &0.042 &0.011 &0.01  &0.05  \\
\ {\bf J1242} &0.20  &0.3     &0.196  &0.046 &0.022 &6.76  &0.093 &0.0006&0.007 &0.012 &0.024 &0.025 &0.023\\
\ mis7a &0.182 &0.4     &0.22   &0.002 &0.07  &6.8   &0.005 & 3.e-5&0.04  &0.011 &0.012 &0.010 &0.018\\
\ mis7b &0.18  &0.3     &0.043  &0.049 &0.026 &6.8   &0.14  & 1.e-5&0.02  &0.011 &0.008 &0.007 &0.02 \\
\ {\bf J1355} &0.22  &0.54    &0.21   &0.03  &0.027 &7.96  &0.087 &0.009 &0.012 &0.006 &0.023 &0.026 &0.026\\
\ mis8a &0.24  &0.58    &0.17   &0.012 &0.07  &7.69  &0.03  &2e-5  &0.039 &0.01  &0.007 &0.006 &0.03\\ 
\ mis8b &0.20  &0.5     &0.04   &0.05  &0.022 &7.65  &0.14  &2e-5  &0.02  &0.009 &0.008 &0.008 &0.026\\ 
\ {\bf J1234} &0.127 &0.128   &0.077  &0.038 &0.024 &2.7   &0.1   &-     &-     &-     &-     &-     \\ 
\ mis9a &0.12  &0.15    &0.06   &0.036 &0.026 &4.55  &0.1   &-     &-     &-     &-     &-     &-     \\
\ mis9b &0.1   &0.11    &0.04   &0.067 &0.02  &2.9   &0.2   &-     &-     &-     &-     &-     &-     \\ \hline
\end{tabular}}

\end{table*}

\begin{table*}
\centering
\small{
\caption{Models adopted in Table 2}
\begin{tabular}{lcccccccccccccccc} \hline  \hline
\                  & mis1a &mis2a &mis3a    &mis4a  &mis5a &mis6 &mis7a&mis8a    &mis9a \\\hline
\ \Vs (\kms)       & 100   &150   &100      &100    &100   &100    &100    &100   &100    \\
\ \n0 (\cm3)       & 50    &63    &52       &65     &62    & 90    &64     &52    &150   \\
\ $D$ (0.01pc)     & 87    &0.7   &1.4      &0.8    &0.9   & 0.5   &0.8    &1.4   &0.5   \\
\ \Ts (10$^4$K)    & 6.6   &7.6   &5.0      &9.5    &6.2   &4.5    &7.     &5.0   &4.0      \\
\ $U$  -           & 0.05  &0.06  &0.095    &0.03   &0.05  & 0.09  &0.06    &0.095 &0.3      \\
\ N/H calc ($^1$)  & 0.1   &0.2   &0.12     &0.1    &0.1   & 0.3   &0.2     &0.12  &0.1    \\
\ N/H  ($^{1,2}$)  &0.02   &0.017 & -       &0.024  &0.023 &0.048  &0.011   &0.0063&-      \\
\ O/H calc ($^1$)  & 1.6   &2.7   &1.8      &2.0    &2.6   & 1.7   &2.5      &1.8   &1.7    \\
\ O/H  ($^{1,2}$)  & 0.654 &0.366 &0.308    &0.48   &0.39 &0.277  &0.268    &0.359 &0.2    \\
\ Ne/H calc($^1$)  & 0.4   &0.4   &0.4      &0.4    &0.4   &0.4    &0.4      &0.4    &0.4    \\
\ Ne/H ($^{1,2}$)  & 0.11  &0.071 &0.056    &0.074  &0.066 &0.040  &0.054    &0.068   &0.0139\\
\ S/H calc ($^1$)  & 0.2   &0.3   &0.25     &0.2    &0.2   &0.2    &0.3       &0.25   &0.01   \\
\ S/H ($^{1,2}$)   &0.0144 &0.0057&-        &0.0094 &0.007 &0.0047 &0.0052    &0.0081 &-        \\
\ Ar/H calc($^1$)  &0.023  &0.01  &0.023    &0.01   &0.01  &0.023  &0.01      &0.023  &0.01  \\
\ Ar/H ($^{1,2}$)  &0.00316&0.00114&0.00134 &0.0018 &0.00157&0.00087&0.001    &0.00128 &-        \\ 
\ He/H calc        & 0.1   &0.01  &0.03     &0.01    &0.01  & 0.1   &0.018     & 0.03  &0.1 \\ \hline
\end{tabular}}

$^1$ in units of 10$^{-4}$; $^2$ evaluated by I20 for the first 8 galaxies and  by I19b for J1234.

\end{table*}

\begin{table*}
\centering
\small{
\caption{Alternative models adopted in Table 2}
\begin{tabular}{lcccccccccccccccc} \hline  \hline
\                  &mis1b &mis2b &mis3b    &mis4b  &mis5b   &mis6 &mis7b &mis8b   &mis9b\\\hline
\ \Vs (\kms)       &120   &100   &100      &80     & 80     &100    &100  &100    &70    \\
\ \n0 (\cm3)       &300   &90    &70       &50     & 80     & 90    &50   &80     &180    \\
\ $D$ (0.01pc)     &7     &8     &6.67     &96.7   & 16.67  & 0.5   &20   &10     &0.27    \\
\ \Ts (10$^4$K)    &7.9   &5.5   &5.5      &5.1    & 4.8    &4.5    &5.3  &5.3    &3.6     \\
\ $U$  -           &0.13  &0.07  &0.05     &0.08   & 0.08   & 0.09  &0.07 &0.08   &0.4   \\
\ N/H calc ($^1$)  &0.1   &0.13  &0.2      &0.1    & 0.1    & 0.3   &0.1  &0.1    &0.1    \\
\ N/H  ($^{1,2}$)  &0.02  &0.017 & -       &0.024  &0.023   &0.048  &0.011&0.0063 &-     \\
\ O/H calc ($^1$)  &1.3   &5.0   &5.7      &1.4    & 2.6    & 1.7   &2.4  &4.3    &1.2     \\
\ O/H  ($^{1,2}$)  &0.654 &0.366 &0.308    &0.48   &0.39    &0.277  &0.268&0.359  &0.11    \\
\ Ne/H calc($^1$)  &0.4   &0.4   &0.4      &0.4    & 1.0    &0.4    &0.7  &0.8    &0.4     \\
\ Ne/H ($^{3,2}$)  &0.11  &0.071 &0.056    &0.074  &0.3     &0.040  &0.054&0.068  &0.0139\\
\ S/H calc ($^1$)  &0.1   &0.3   &0.25     &0.2    &0.3     &0.2    &0.3   &0.25  &0.01    \\
\ S/H ($^{1,2}$)   &0.0144&0.0057&-        &0.0094 &0.0074  &0.0047 &0.0052&0.0081&-      \\
\ Ar/H calc($^1$)  &0.01  &0.007&0.023     &0.0073 &0.007   &0.023  &0.007 &0.0073 &0.01   \\
\ Ar/H ($^{1,2}$)  &0.00316&0.00114&0.00134&0.0018 &0.00157 &0.00087&0.001 &0.00128&-      \\ 
\ He/H calc        &0.1   &0.1   &0.08     &0.1    & 0.1    & 0.1   &0.1   & 0.08  &0.1\\ \hline
\end{tabular}}

$^1$ in units of 10$^{-4}$; $^2$ evaluated by I20   for the first 8 galaxies and by I19b for J1234.

\end{table*}

\subsubsection{I20 survey}

The observed (reddening corrected) line ratios from the I20 survey are shown in Table 1 (Balmer lines) and 
in Table 2 (other lines). 
The observations by the Cosmic Origins Spectrograph onboard the Hubble
Space Telescope were focused on star-forming galaxies at z = 0.028-0.0065 with low oxygen abundances
and extremely high emission-line  ratios [OIII]5007/[OII]3727 $\sim$ 22-39.
Characteristic of this survey are the simultaneous low [OII]3727+/\Hb~ and  high [OIII]5007+/\Hb.
 Generally,  in the spectra shown by SFG surveys in the local universe the  [OIII]/[OII] line ratios 
 reach values of  $\sim$6 e.g. for the
 Berg et al (2012) survey of galaxies at z $\leq$ 0.023, while 
 for the  Marino et al (2013) Califa HII complex   at  0.005$<$z$<$0.03    
 [OIII]5007+/[OII]3727+ is  $\leq$ 1 (see also Contini 2017 and references therein).
[OII]3727+/\Hb~ in a few cases  shows values similar to or  even lower  than  [OIII]4363/\Hb.  
We consider that the  relatively  large errors as high as 50 percent which are allowed in the fit of the  
weak line ratios to \Hb~ can be 
applied to  [OII]3727+/\Hb~  because in the present spectra  they  are abnormally low (Table 2). 
The observed [OIII]5007+/[OII]3727+ line ratios   range between 45.8 for J0159 and 27.45 for J1205 
and  the HeII4686/\Hb~  between 0.09 and 0.038.
[OIII] 4363/\Hb~ covers a small range (0.196-0.22)  except   for J1205  with [OIII]4363/\Hb=0.13. 
[OIII]4363 can be  blended with \Hg. 
The  [OIII]5007+/[OIII]4363 ratios (hereafter \RO3) range from 34.5 for J1242+4851  to 52.5 
for J0007+0226.
Such  low \RO3 values indicate that the  [OIII] lines, in particular [OIII]4363, are emitted from a relatively 
hot gas when the effect of the shock dominates on photoionization (Aldrovandi \& Contini 1985).
All the line ratios to \Hb~
are low compared to [OIII]5007+/\Hb~ by a factor  $\geq$ 5  in contrast  to the spectra generally presented for starburst 
and SFG at  low redshift. We have added in Tables 2, 3 and 4 the galaxy J1234+3901 (I19b) even if it is 
located at a higher redshift (z=0.13297) because the HeII line has been observed
constraining the model as  for  the other SFG of the  I20 sample.

\begin{figure}
\centering
\includegraphics[width=4.1cm]{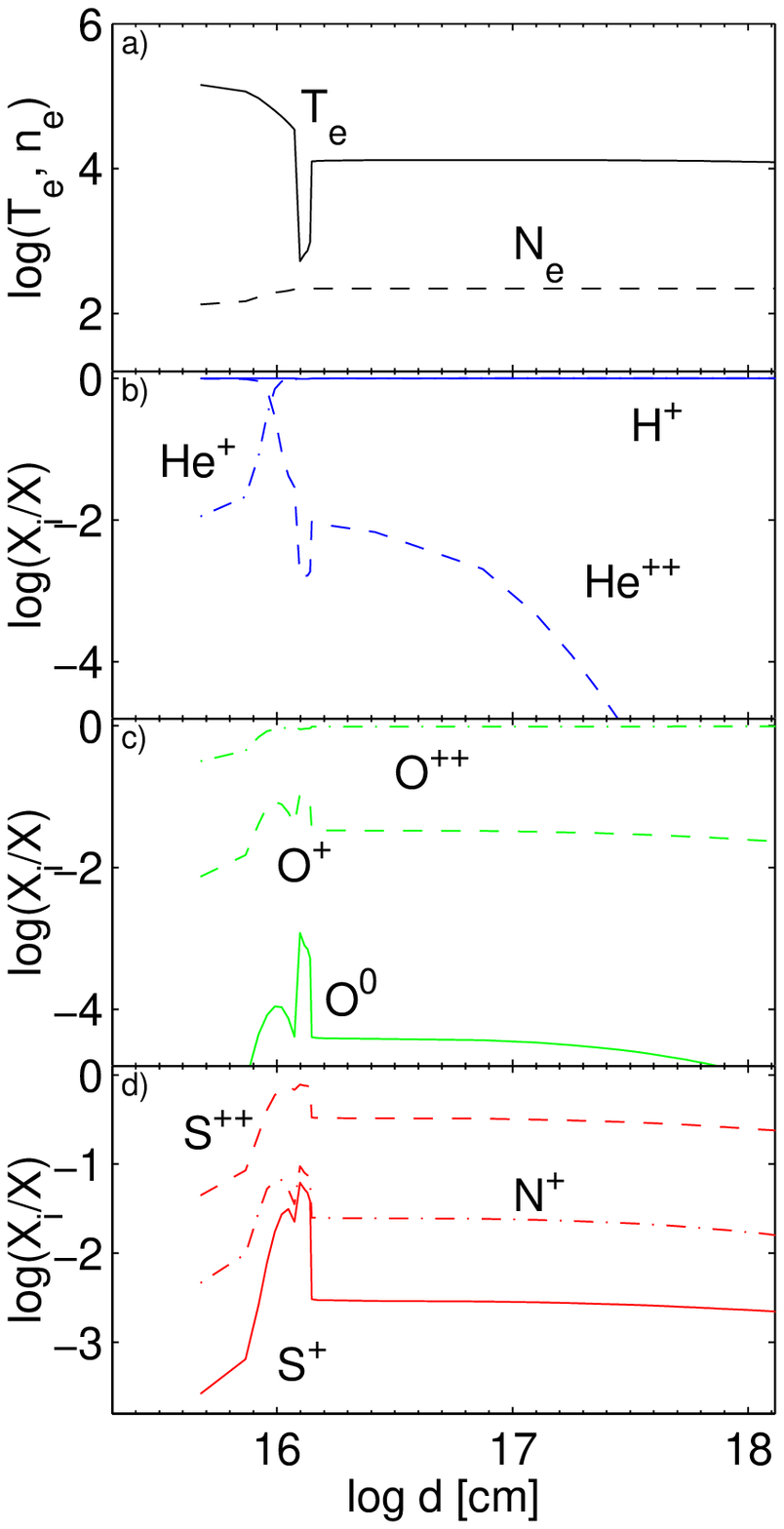}
\includegraphics[width=4.1cm]{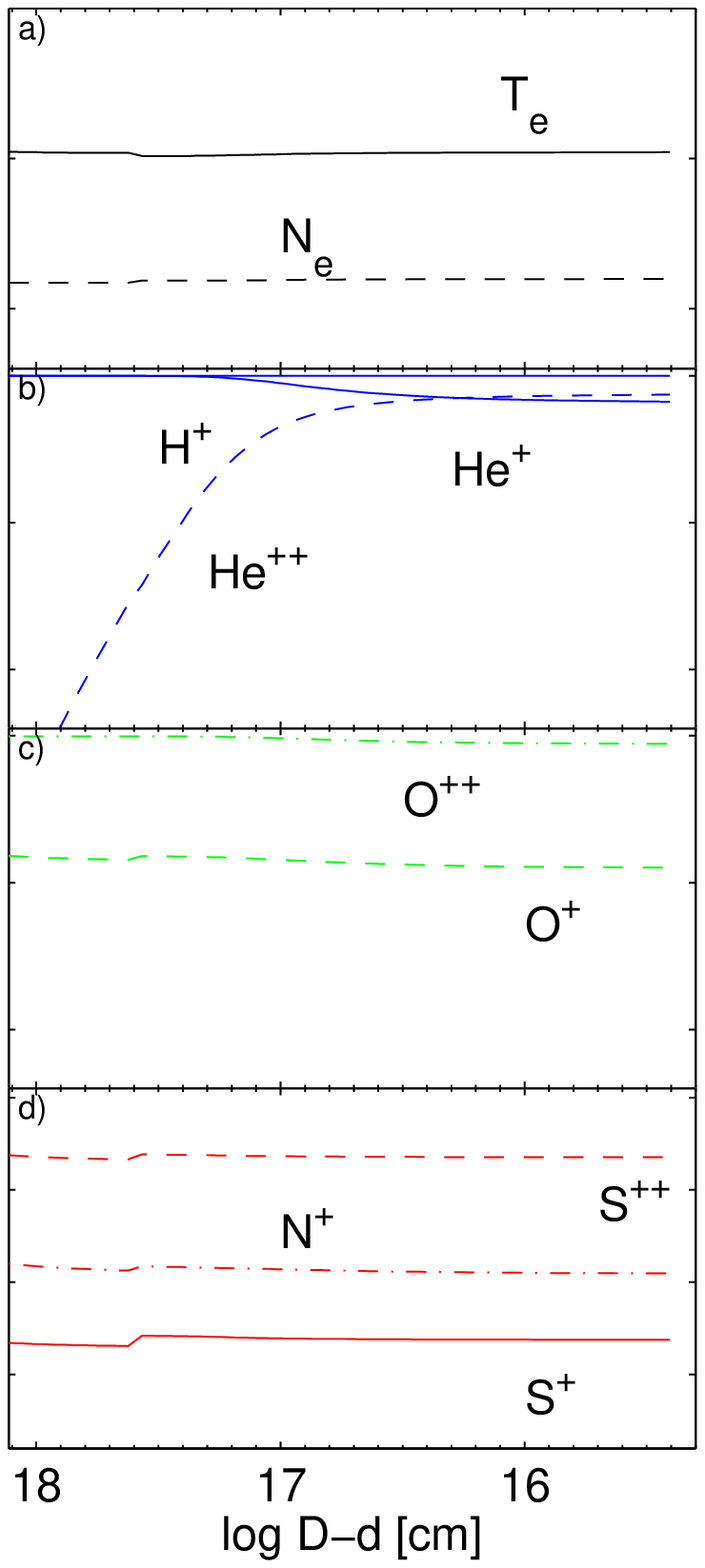}
\includegraphics[width=4.1cm]{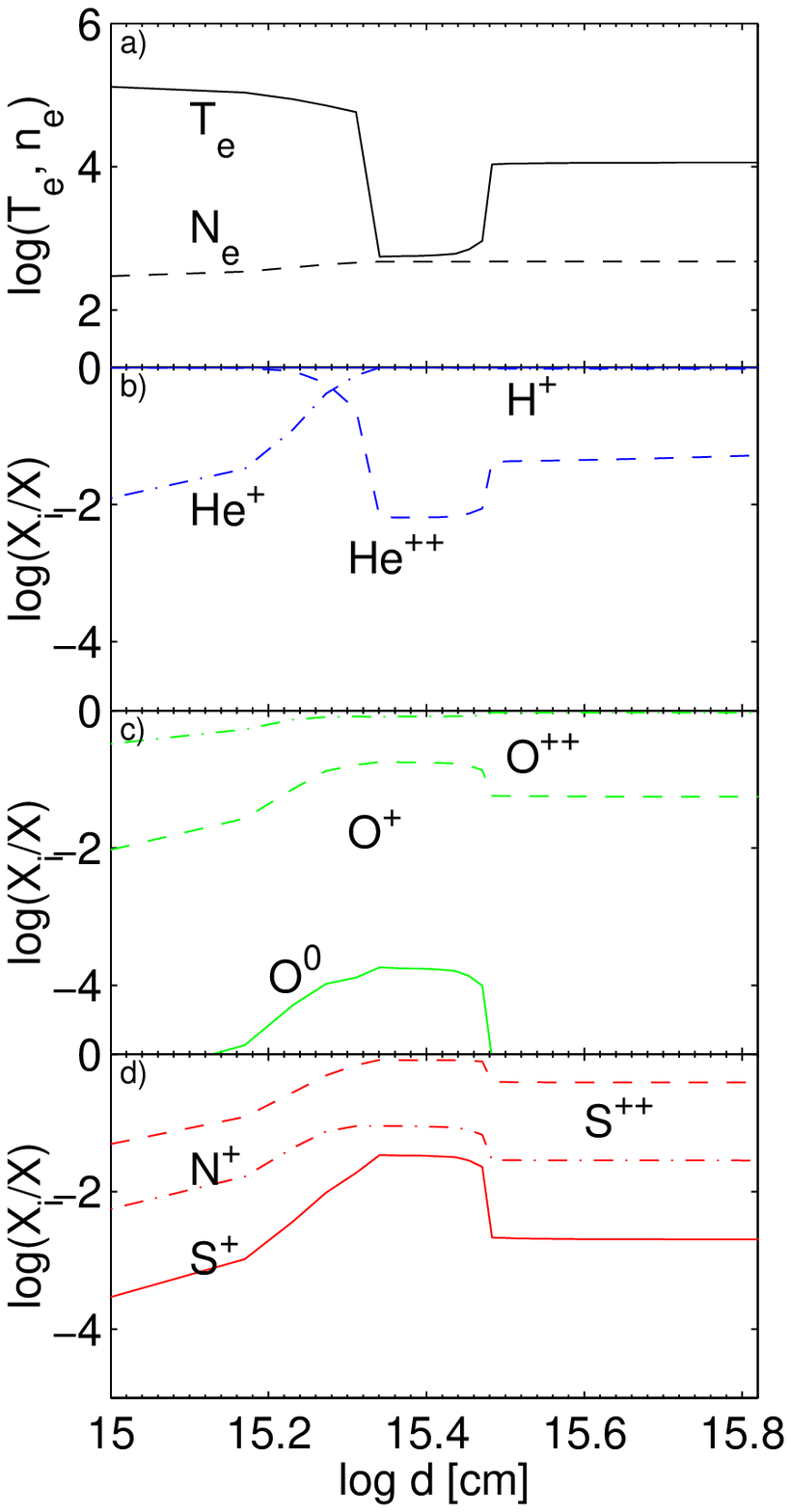}
\includegraphics[width=4.1cm]{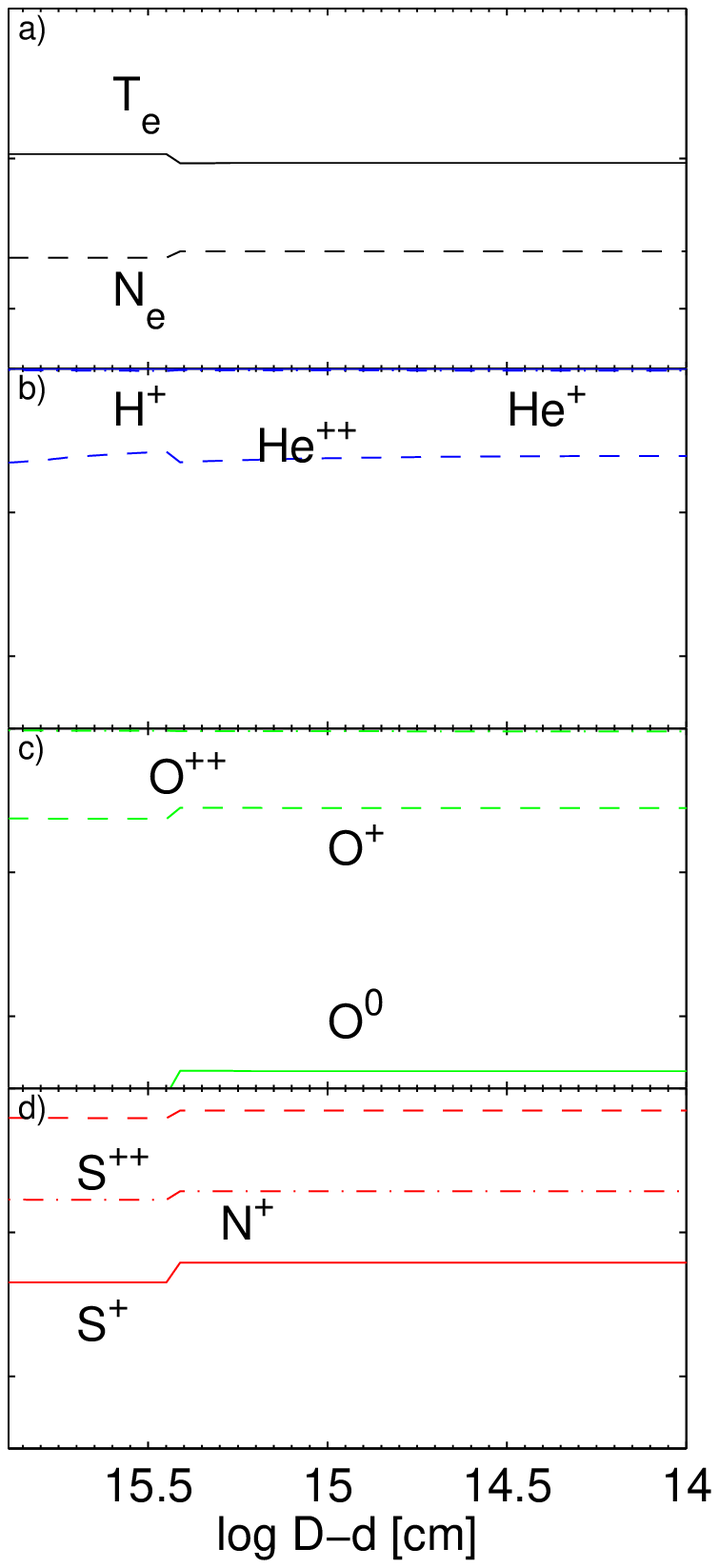}
\caption{The profiles of the electron temperature  the electron density (top panels)
and of the fractional abundances of the H, He, N, O and  S different level ions
throughout the emitting cloud. 
Top diagrams:  J007+0226; bottom diagrams: J1205+4551.
The clouds are divided in two halves. In the left panel
the shock front is on the left and the X-axis scale is logarithmic. In the right panel
the right edge is reached by the flux from the starburst. The X-axis scale is logarithmic
in reverse, in order to have the same detailed view of the cloud edges (the shock-dominated
and the radiation-dominated one).}
\end{figure}

 To better understand the modelling results we show
 in Fig. 3  diagrams the profiles of \Te, \Ne and of the  fractional abundance of the most significant ions 
 throughout the clouds 
 corresponding to J0007 (top, model mis1a) and J1205 (bottom, model mis6a).
 For each galaxy the emitting cloud is divided into two equal and antisymmetrical parts in order to  see the 
 results with  relatively high precision at the two  opposite edges.
 When the cloud propagates outwards from the starburst 
 the shock front is on  one edge (at the left of the left panels in Fig. 3) while the opposite edge (at the right  
 of the right panels) is reached by the photoionization flux from the starburst. 
  Using  composite models   the gas  can reach relatively high temperatures \Te near the shock front
  depending on the shock velocity  ($T\sim$1.5$\times 10^5$ (\Vs/100 \kms)$^2$).
  Radiation from the stars heats the opposite  edge  to no more than 2-3$\times$10$^4$K.  
  The  \Te profile throughout the emitting clouds is  not straightforward, decreasing  towards the cloud centre 
  following the cooling rate.  
 Fig. 3 shows that the  high temperature downstream close to the shock front would   correspond 
 to a not negligible  O$^{3+}$/O fractional abundance. The gas recombines and O$^{++}$/O increases, while 
  He$^{++}$/He decreases.  He$^{++}$/He from the edge of J0007 clouds reached by radiation decreases due to radiation
  transfer. \Te is maintained at $\leq$10$^4$K by secondary radiation.
 The effective starburst temperature \Ts  affects the black-body radiation flux and, combined with the ionization parameter $U$, they yield different 
 [OIII]5007+/[OII]3727+ line ratios. 
HeII line intensity depends strongly on the photoionization flux from the radiation source. Therefore, for J0007
this line comes in particular from the radiation dominated zone (right panel), while for J1205 the only zone of gas
with a temperature high enough to emit a strong  HeII line is downstream, close to the shock front (left panel)
because \Ts is relatively low.

The  J0007  spectrum  is  fitted satisfactorily enough   by model mis1a. 
The disagreements between  calculated and observed [OIII]4363/\Hb~ and HeII4686/\Hb~  ratios reach  a factor  $\leq$ 2, 
while  HeI4471/\Hb~ and HeI5876/\Hb~ are well reproduced.
 The results indicate that for J0007 Ar/H should be reduced by a factor of $\geq$2.
In the same  spectrum the [OI]/\Hb~ and [SIII]/\Hb~ are underestimated and overestimated by  large factors, respectively.
The [SII]/\Hb~ line ratios are both underestimated. They are calculated  with  solar  S/H relative abundances.
Therefore, the sulphur line  fit cannot improve  by  increasing  S/H.
As it was explained by Congiu et al (2017), [OI] refers to a recombined oxygen  and [SII] depend on S$^+$ which
has a ionization potential lower than that of H$^+$.  They suggest that [OI] and [SII] lines  are contaminated by  
the ISM  which is in the same characteristic physical conditions.
The J0007 spectrum shows the maximum observed [OIII]5007+/[OII]3727+ line ratio  relatively to the objects in 
the I20 survey, but the other line ratios are  similar. 
We have tried to reproduce the J0007 spectrum with another model (mis1b) described  in Table 4  with a different
approach i.e. to
reduce the  successful fit of some strong line ratios to \Hb~ and to improve that of the weak ones e.g. [OIII]4363/\Hb.
The calculated  [OIII]4363/\Hb~ line ratio is lower than observed  by a factor of $\sim$1.4, while
the calculated HeII/\Hb~  is higher than  observed  by a factor of $\sim$ 3.
Moreover,  the [SII]6717/6731 line ratio is $<$1 while the observed one is $>$ 1 for model mis1b due to the 
high preshock density.
The S/H ratio adopted  for model mis1b is lower than solar indicating that S is included into dust grains.
The  geometrical thickness $D$  (Table 3)  calculated by model mis1a exceeds  $D$  resulting for  all the other  I20 
survey galaxies 
 by a factor of $\sim$ 10, while $D$ for  model mis1b is similar to that calculated for the other I20 survey objects. 
We select model mis1a to represent J0007  spectrum because [OIII]5007+/\Hb~ and HeII/\Hb~ are  reproduced with  
smaller errors.
The J0007 spectrum analysis however suggests that
 it is unrealistic to reproduce the spectrum of each galaxy by a single model. Therefore in
the following we will present two  sets of models (mis1a-mis9a) and (mis1b-mis9b) for I20  survey galaxies 
and for J1234 in Tables 3 and 4, respectively.
Both sets satisfactorily reproduce the [OIII]5007+/\Hb,  [OII]3727+/\Hb~ and [NII]/\Hb~ ratios. 
Models mis1a-mis5a and 
mis7a-mis8a better fit  [OIII]4363/\Hb, while  mis2b-mis5b and mis7b-mis8b reproduce satisfactorily  HeII/\Hb.
The two sets of models show  extreme  [OIII]4363/\Hb~ and HeII/\Hb~  which indicate that averaged models
may produce  a better fit to the observed  weak line ratios.
J1205 spectrum is well reproduced by one  model because  mis6a is identical to mis6b. Moreover, [OIII]4363/\Hb~ and 
HeII/\Hb~ line ratios  in J1234 are  well reproduced
by both mis9a mis9b, but mis9a should be dropped because [OIII]5007+/\Hb=4.55 is unacceptably high.

Table 3 shows that the shock velocities are $\geq$100 \kms for this galaxy  sample, while Table 4 indicates  
\Vs $\leq$100 \kms for a few objects.
The preshock densities are all $\leq$ 100 \cm3, except for model mis1b with 
\n0=300 \cm3, which although  yielding a better fit of the [OIII]4363/Hb~ line ratio, 
it reproduces the other ones with  larger errors. For mis9b \n0=180 \cm3 leads to a good fit of the J1234 spectrum.
The star effective temperatures for the I20 sample range between 50000K and 95000K, except for 
mis5a and mis6a with \Ts = 48000K and  45000K, respectively.  
The ionization parameter $U$ is in the norm for SF galaxies at these redshift (e.g. Berg et al 2012).
The geometrical thickness of the emitting clouds ranges  between 0.005 pc for J1205
and  0.97pc for J0926. This large $D$
would imply a large region of neutral gas within the cloud  yielding a high [OI]6360/\Hb~
ratio. For all the  models  the calculated
[OI]/\Hb~ is nearly null. Considering that the [SIII]6312 line lies
within the [OI]6300,6363 doublet, that these lines are blended for \Vs $\geq$100\kms,
 that [SIII]/\Hb~ calculated  by the models (Table 3) are all exceeding
the data and that the   observed [OI] lines are contaminated by the ISM component,
such a low [OI]/\Hb~ is acceptable. The   sum of  [OI] and [SIII] lines should be considered.  
With regards  to the  element relative  abundances (Tables 3 and 4), we obtain lower than 
solar O/H by factors of $\sim$1.2-5  ([O/H]$_{\odot}$=6.6$\times$10$^{-4}$), 
N/H by factors between 3 and 10 ([N/H]$_{\odot}$=10$^{-4}$) and Ar/H
by factors 2-6.5 ([Ar/H]$_{\odot}$=4.6$\times$10$^{-6}$). On the other hand S/H is about solar 
([S/H]$_{\odot}$=2$\times$10$^{-5}$). This result strengthens our  suggestion that the observed [SII] lines 
have  a strong  ISM contribution.

An unusual result for the I20 sample galaxies  consists in  the He/H  
([He/H]$_{\odot}$=0.1) adopted in the mis2a-mis5a and mis7a-mis8a models  which are lower than solar by 
different factors.  They were used to  better reproduce the HeII/\Hb~ line ratios. However, they reduce  
 the  HeI5876/\Hb~ calculated line ratios in some objects. 
The HeI5876/\Hb~ observed values  range between $\leq$0.1 and $\leq$0.15 in Table 2 objects and in
most of the  observed  spectra  from  different types of galaxies at different redshift. 
 In fact \Hb~ and HeI 5876 are both recombination lines and they behave similarly throughout large regions
 of the clouds at temperatures $\leq$10$^4$K (Fig. 3) even if matter-bounded.
The models presented in Table 4  better reproduce  the HeII/\Hb~ data   adopting
He/H near solar for  most of them.
The [OIII]4363/\Hb~ line ratios are generally  underpredicted by these models. 
Even considering that \Hg 4340 and [OIII]4363
lines could be partly blended at \Vs$\sim$ 100 \kms, this does not resolve the problem because  
\Hg/\Hb~ are nearly constant (Table 1)  and show suitable values  and
[OIII]4363/\Hb~ ratios  show similar values  for  most of the I20  spectra.

Concluding, we have found   by fitting the [OIII] 5007/\Hb, [OIII]4363/\Hb, [OII]/\Hb, [NII]/\Hb~
and [SII]/\Hb~ line ratios  and  selecting  the physical parameters and the relative abundances, 
that the HeII/\Hb~ line ratios constrain the models by the geometrical thickness of the cloud
 and by the He/H abundance ratio.
Table 2  shows that both  [OIII]4363/\Hb~ and HeII4686/\Hb~  line ratios depend on $D$, therefore to better 
reproduce the observed spectra pluri-cloud models  seem  more realistic. 
 The  large range of the calculated different geometrical thickness of the clouds  is  in agreement with cloud 
fragmentation due to turbulence created by the shocks at the shockfront.  
 Recall that the  observations cover regions with different physical and chemical conditions. 
Therefore a pluri-cloud
 model should be used to reproduce each spectrum.  Here, we try to reduce a pluri-cloud ensemble to a minimum
 of two clouds which both show a  good fit of  calculated to  observed strong line ratios (to \Hb)
  and of a few significant weak ones (e.g. [OIII]4363/\Hb~ or HeII4686/\Hb).
  The worst  fit is still acceptable due to the observation error and to the uncertainty of the
  data used in the calculations. Therefore, an average of these two models represents the best option.

\subsubsection{I18b survey} 

\begin{table*}
\centering
\small{
\caption{Comparison of observations for I18b survey, J0811 (I18a)  and HS0837 (P04) galaxies with model results. \Hb=1}
\begin{tabular}{ccccccccccccccccc} \hline  \hline
\       &MgII & [OII] &[NeIII  & [OIII]& HeI  &HeII &[OIII]&HeI   &[OI]  &[SIII]   & [NII]& [SII]&[SII] &[ArIII]\\
\       & 2796+ & 3727+ &3868    & 4363  &4471  &4686 &5007+ &5876  &6360  &6312  & 6584 &6717  & 6731 & 7136  \\ \hline 
\ {\bf J0901} & -     & 0.815 &0.46    & 0.076 &0.039 &-    &8.75  &0.096 &0.032 &-     & 0.093&0.055 &0.054 &-      \\
\ miss1 & 0.21  &0.81   &0.46    &0.11   &0.032 &-    & 8.8  &0.091 &3e-4  &0.046 & 0.088&0.052 &0.046 & -     \\   
\ {\bf J1011} & -     &0.297  &0.497   &0.146  &0.028 &-    &10.68 &0.117 &0.022 &0.043 &0.07  & -    & -    & -     \\
\ miss2 & 0.08  &0.28   &0.6     &0.16   &0.024 &-    & 10.78&0.07  &4e-5  &0.024 &0.073 &  -   & -    &-      \\
\ {\bf J1243} & -     &0.537  &0.49    &0.155  &0.031 &-    & 9.65 &-     & -    &-     &0.058 & -    &-     &-      \\
\ miss3 & 0.15  &0.57   &0.53    &0.147  &0.035 &-    & 9.66 &0.1   &1.e-4 &0.04  &0.06  &0.026 &0.023 &-     \\
\ {\bf J1248} & 0.17  &0.49   &0.46    &0.177  &0.042 &-    & 7.77 &0.10  &0.022 &-     &0.057 &0.026 &0.031 & 0.037\\
\ miss4 &0.11   &0.49   &0.12    &0.14   &0.04  &-    &7.76  &0.11  &3e-5  &0.038 &0.046 &0.02  &0.02  &0.046 \\
\ {\bf J1256} &0.26   &0.443  &0.55    &0.164  &0.04  &-    &9.73  &0.11  &-     &-     &0.079 &0.038 &0.056 &-     \\
\ miss5 &0.26   &0.44   &0.54    &0.17   &0.03  &-    &9.97  &0.07  &-     &-     &0.05  &0.038 &0.033 & -    \\ 
\ {\bf J0811} & -     &0.167  &0.13    &0.063  &0.036 &0.023 & 2.27 &0.095 & -    &0.004 &0.0072& 0.014&0.0123&0.01 \\
\ miss6  & -    &0.141  &0.12    &0.04   &0.07  &0.008 &2.66  &0.23 &3.e-6  & 0.047& 0.007&0.0046&0.04 &0.02   \\
\ {\bf HS0837} &-     &  0.41 & 0.42   &0.166  &0.037&0.019 &7.674 &0.13  &0.024  &0.014 & 0.024&0.041 &0.035&0.034\\
\ mpa         &-&0.43   & 0.40   &0.15  & 0.007&0.06  &7.76  &0.02  &1.3e-4 &0.07  & 0.03 &0.04  &0.043 &0.040\\ 
\ mpb         &-&0.43   & 0.35   &0.042 & 0.039&0.016 &8.0   &0.11  &1.0e-4 &0.054 &0.029 &0.037 &0.037 &0.039\\ \hline
\end{tabular}}

\end{table*}

%

\begin{table*}
\centering
\caption{Models adopted in Table 5}
\begin{tabular}{lcccccccccccccccc} \hline  \hline
\                  & miss1&miss2 &miss3 &miss4    &miss5& miss6 & mpa   &mpb\\\hline
\ \Vs (\kms)       & 100  &100   &100   &100      &100  &  70   & 130    &100\\
\ \n0 (\cm3)       & 64   & 63   &64    &63       &95   &   190 & 100    &100\\
\ $D$ (0.01pc)     & 1.96 &1.33  &1.43  &1.33     &0.6  &  0.2  & 0.4    &6.67\\ 
\ \Ts (10$^4$K)    & 6.2  &6.3   &6.2   &5.2      &7.0  &  3.4  & 6.5    &5.5\\
\ $U$  -           & 0.016&0.055 &0.025 &0.044    &0.03 & 0.3   & 0.035  &0.04\\
\ N/H calc ($^1$)  & 0.17 &0.55  &0.2   &0.2      &0.3  &  0.1  & 0.1    &0.1\\
\ N/H  ($^{1,2}$)  & 0.087&0.14  &0.056 &0.035    &0.072&0.0028 &0.067  &0.067 \\
\ O/H calc ($^1$)  & 2.7  &2.7   &2.6   &2.6      &2.8  &  1.0  & 2.5    & 2.8\\
\ O/H  ($^{1,2}$)  & 1.44 & 0.98 &0.77  &0.44     &0.74 & 0.095 & 0.44   &0.44   \\   
\ Ne/H calc($^1$)  &0.3   & 0.3  &0.3   & 0.3     &0.5  &  0.1  & 0.3    &0.3\\
\ Ne/H ($^{1,2}$)  &0.25  &0.14  &0.12  &0.07     &0.012&0.015  & 0.067  &0.067 \\   
\ S/H calc ($^1$)  &0.16  &0.16  &0.16  &0.2      &0.3  &   0.3 &0.24   &0.28\\
\ S/H ($^{1,2}$)   & -    &-     & -    &  -      &  -  & 0.002 &0.017 &0.017 \\
\ Mg/H calc ($^1$)  &0.16  &0.16  &0.16  &0.3      &0.3 & 0.3   &0.3    &0.3\\
\ Mg/H ($^{1,2}$)  &   -  & -    & -    &0.014    &0.04 & -     &-      & - \\
\ Ar/H calc ($^1$) &0.007 &0.007 &0.007 &0.007    &0.007& 0.007 &0.005  & 0.005\\
\ Ar/H ($^{1,2}$)  &  -   &  -   & -    & 0.0016  & -   & 2.66e-4&0.0023&0.0023 \\
\ He/H calc ($^1$) &0.07  &0.08  &0.08  &0.08     &0.08  & 0.1  &0.02   &0.08\\  \hline 
\end{tabular}

 $^1$ in units of 10$^{-4}$; $^2$ relative abundances  are evaluated by I18b for the first 5 galaxies
presented in Table 5,  
by I18a for J0811 and by P04 for HS0837.

\end{table*}

\begin{table*}
\centering
\tiny{
\caption{Comparison of observations for G20 survey galaxies with model results. \Hb=1}
\begin{tabular}{cccccccccccccccccccc} \hline  \hline
\  &MgII&[OII] &[NeIII]& [OIII]& HeI& HeII &[ArIV]&[OIII]&HeI &[OI]&[SIII]&[NII]&[SII]&[SII]&[ArIII]&[OII]&[SIII]\\
\  &2789  &3727+ &3868   & 4363  &4471& 4686 & 4713 &5007+ &5876&6360+&6213  &6584+&6717 & 6731& 7136&7320&9304 \\ \hline 
\ {\bf J0901}&0.225&0.9  &0.55    &0.11   &0.04& 0.009&0.01  &9.27   &0.11&0.045&0.013&0.15 &0.077&0.069&0.059&0.034&0.24 \\
\ mG1  &0.24&0.92 &0.47    &0.07   &0.034&0.02  &0.02  &9.45   &0.10&5e-4&0.05  &0.12 &0.073&0.064&0.050&0.24 &0.2 \\ 
\ {\bf J0925}&0.215&1.1  &0.49    &0.072  &0.04&0.011 &0.01  &7.9    &0.11&0.04&0.013 & 0.14&0.09 &0.078&0.06 &0.038&0.28\\
\ mG2&0.20 &1.19&0.45    &0.06   &0.04 &0.019&0.01  &8.28      &0.11&1e-3&0.06  &0.16 &0.10 &0.09 &0.05&0.032&1.0\\
\ {\bf J1154}&0.19&0.49&0.52   &0.13   &0.039&0.02 &0.05  &7.62   &0.11&0.03&0.011 &0.067&0.046&0.036&0.035&-&0.043 \\
\ mG3a&0.2  &0.54&0.5     &0.11   &0.03 &0.19 &0.04  &7.68   &0.08&2e-4&0.067&0.07  &0.05 &0.04 &0.05&0.023&0.9\\
\ mG3b&0.25&0.53&0.6     &0.072  &0.037&0.017&0.02  &7.58   &0.1 &3e-4&0.06 &0.08  &0.06 &0.05 &0.03&0.016&0.7\\
\ {\bf J1442}&0.38  &0.94&0.73 &0.106  &0.035&0.01 &0.018 &8.53   &0.1 &0.04&0.013&0.10  &0.087&0.065&0.05&0.03 &0.38 \\
\ mG4a  &0.38  &0.95&0.9  & 0.106  &0.022&0.07 &0.01  &8.32   &0.06&5e-4&0.05 &0.11  &0.074&0.065&0.04&0.032&0.9\\
\ mG4b & 0.36 &1.0 &0.6  &0.063  &0.023 &0.03 &0.007 &8.59   &0.07&7e-4&0.056&0.13 &0.095&0.083 &0.05 &0.028&0.9\\\hline
\end{tabular}}

\end{table*}

\begin{table*}
\centering
\small{
\caption{Models adopted in Table 7}
\begin{tabular}{lcccccccccccccccc} \hline  \hline
\                  & mG1   &mG2    & mG3a &mG3b   &mG4a &mG4b  \\ \hline
\ \Vs (\kms)       & 100   &100    & 80   &80     &100  &100\\
\ \n0 (\cm3)       & 64    &64     & 68   &60     &64   &64\\
\ $D$ (0.01pc)     & 10    &8.33   & 1.53 &46.6   &2.07 &6.66\\
\ \Ts (10$^4$K)    & 6.2   &6.0    & 6.7  &6.4    & 6.4 &6.4\\
\ $U$  -           & 0.016 &0.013  &0.014 &0.019  &0.012&0.012\\
\ N/H calc ($^1$)  & 0.18  &0.18   & 0.14 &0.12   &0.17 &0.17\\
\ N/H  ($^{1,2}$)  & 0.088 &0.08   &0.04  &0.04   &0.06 &0.06\\
\ O/H calc ($^1$)  & 2.7   &2.5    &1.5   &1.5    & 2.6 &2.6\\
\ O/H  ($^{1,2}$)  & 1.11  &1.3    &0.56  &0.56   &0.97 &0.97\\
\ Ne/H calc($^1$)  & 0.3   &0.3    & 0.3  &0.3    & 0.6 &0.6\\
\ Ne/H ($^{1,2}$)  & 0.2   &0.26   &0.09  &0.09   &0.19 &0.19 \\
\ S/H calc ($^1$)  & 0.16  &0.16   & 0.16 &0.16   &0.16 &0.16\\
\ S/H ($^{1,2}$)   &0.024  &0.024  &0.01  &0.01   &0.02  &0.02   \\
\ Ar/H calc($^1$)  &0.007  &0.007  & 0.007&0.007  &0.007&0.007\\
\ Ar/H ($^{1,2}$)  &0.004  &0.004  &0.002 &0.002  &0.0032&0.0032\\
\ Mg/H calc ($^1$) & 0.16 &0.16   & 0.1  &0.1     & 0.22&0.22\\
\ He/H calc        &0.07   &0.08  & 0.07  &0.07   & 0.05&0.05\\ \hline
\end{tabular}}

$^1$ in units of 10$^{-4}$; $^2$ evaluated by G20.

\end{table*}

I18b presented observations obtained by the Cosmic Origin Spectrograph onboard the Hubble Space 
Telescope of galaxies within  the z range  0.2993-0.4317 and high [OIII]5007+/[OII] 3727+.
The comparison of  the observed line ratios with the results of calculation by
detailed modelling  is shown in Table 5. The models are   described in Table 6.
[OII]/\Hb~ ratios  for the I18b survey spectra are higher by  factors  $>$ 2 than for the I20 survey galaxies. 
The spectra contain the MgII 2796,2803 lines.
Table 5 shows that  models miss1-miss4 are similar for galaxies J0901, J1011, J1243 and J1248
but for J1256 miss5 shows  a different set of parameters. In particular, \Ts  and \n0 are higher than for the 
other survey galaxies by a factor $\leq$ 1.5.   The  element relative abundances to H 
for all the galaxies are lower than solar, except for S/H that even exceeds the solar value in miss5.
In particular, O/H are $\sim$0.4 solar and Ar/H are 0.15 solar. 
HeII4686 lines  were not observed. We constrain  the models by  an HeII/\Hb~ upper limit $<$0.1. The best fit  to 
the observed HeI/\Hb~ ratios is  obtained  by He/H=0.08, slightly lower than solar (0.1).
Comparing our results with those derived by I18b, the relative abundances of all the elements 
 are higher by  factors  $\sim$ 2 - 6 for O/H, 2-4 for N/H, 1.2-5 for Ne/H and for Ar/H by a 
 factor of $\leq$28 for J1354.  S/H and He/H  were not indicated by I18b.

\subsubsection{J1234+3901 (I19b)}

We have added to the I20 survey the spectrum of the most metal-poor galaxy J1234+3901 at z=0.133.
Tables 2-4  include the modelling of  J1234+3901 galaxy spectrum (I19b).
Optical spectroscopy of this galaxy in the local Universe was  obtained by the LBT/MODS telescope.
The J1234+3901 galaxy spectrum and the  models are described in the bottom three lines of Table 2 and in the
last  column of Tables  3 and 4.  Model mis9a, which
reproduces at most  the HeI and HeII line ratios to \Hb, overpredicts  [OIII]5007+/\Hb~  by  68 percent. 
The [OIII]5007 line is the  strongest one, therefore it requires a more accurate model. 
Model mis9b satisfactorily reproduces the [OIII]5007+/\Hb~ line ratio. It  underpredicts  [OIII]4363/\Hb~  by a
 factor $<$2 which, however, is within the accepted error,  being  [OIII] 4363  a weak line.
 Model mis9b shows  relatively high \n0=180 \cm3 and   low \Vs=70 \kms. The effective temperature of
 the stars is the lowest for all the sample galaxies (36000K) and  the high ionization parameter
 $U$=0.4 compensates the modelling of the strongest line ratios. N/H and O/H are particularly low, 0.1 solar 
 and 0.18 solar, respectively, while He/H=0.1.
 Model mis9b is selected because it reproduces all the weak line ratios within a factor of 2 (Table 4).
In model mis9b  $D$=0.0027 pc is very thin and indeed  it leads to more acceptable results.
However,  changing $D$, all the line ratios change as well and, in particular  the good fit of the
[OIII]5007+/\Hb~ and [OII]3727+/\Hb~ calculated to observed line  ratios  may  be lost. 
We suggest that to improve the modelling of the  survey galaxy spectra, the final line ratios should be calculated
from the average of single-cloud  spectra. 

\subsubsection{J0811+4730 (I18a)}

 I18a present  a rich observation  spectrum of galaxy J0811+4730  at z=0.044  from the
 Data Release I3 (DRI3) of Sloan Digital Sky Survey (SDSS) LBT/MOD  with high
 signal-to-noise ratios. By the direct \Te strong line method they obtain 12+log(O/H)=6.98.
 I18a explain the low metallicity by infall  of poor metallic gas from the galactic halos mixing
 with the metal rich gas in the central region (Ekta \& Chengalur 2010).
 We have revisited the spectrum by the detailed modelling of the line ratios. The results which appear in 
 Tables 1, 5 and 6  (model miss6)  well reproduce   the  [OII]/\Hb, 
 [OIII]5007+/\Hb,  [NeIII]/\Hb~ and [NII]/\Hb~ ratios, roughly fit  [OIII4363/\Hb~ and  HeI/\Hb~ line
 ratios, but they underpredict HeII/\Hb~ by a factor  $\leq$ 3. The S lines are even less fitted.  
Model miss6 shows (Table 6) that although the O/H and N/H calculated results  are  0.15 and 0.1 solar, 
respectively, they  are  still higher than those evaluated  by I18a (0.014 and 0.0028 solar, respectively).
 Also for this spectrum we  suggest that the [OI], [SII] and HeI lines  are most probably contaminated by the ISM.

\subsubsection{HS0837+4717 (P04)}

We report  Pustilnik et al (2004) observations  which
show the results of high S/N long-slit spectroscopy with the Multiple Mirror and the 
SAO 6-m telescope, optical imaging with the Wise 1-m telescope and the HI observations with the Nancay Radio
Telescopes of the very metal deficient (12+log(O/H)=7.64) luminous blue compact galaxy  HS0837+4717
at z=0.041950. This galaxy should belong to the galaxy group at z $<$0.2 presented in Table 2, but 
it is dislocated in Table 5  as well as J0811  for sake of space.
By the  direct method analysis  of the line ratios  P\'{e}rez-Montero et al. (2011) claim that nitrogen is
overabundant.
We have obtained an N/H abundance ratio of 10$^{-5}$ by the detailed modelling of the observed spectrum,
$\sim$1.5 times higher than that  evaluated by P04. 
Table 6 columns 8 and 9 show the models which better reproduce the data.
We have found that model mpa better fits the [OIII]4363/\Hb~ line ratio while model mpb reproduces
 HeII 4686/\Hb~ and both the HeI/\Hb~ lines. We think that mpa and mpb  must be averaged  because they represent 
 extreme cases in the starburst.
We have found that - as for the other samples - the O/H ratios are by a factor $\sim$ 2.5 lower than solar in
agreement with the results  obtained by fitting I20 and I19a local SF galaxies.  N/H= $\sim$10$^{-5}$ 
relative abundance is  similar  to those calculated   for other SF galaxies. Moreover, our results show that  
in agreement with Pustilnik et al the starburst is young  as revealed from the relatively high \Ts= 62000K.

\subsubsection{G20 survey}

Using the (Very Large Telescope) VLT/Shooter spectral observations Guseva et al (2020) presented the
spectra of five SF galaxies at z$\sim$0.3-0.4 
which contain a rich number of lines from different elements in different ionization levels. 
In Tables 7 we  compare the observed line ratios with model results calculated by  detailed modelling.
The models are described in Table 8.
Two galaxies  J0901 and J1011 in the  G20 survey  are included in the I18b survey. We have repeated
the J0901  modelling because G20 observations contain more  lines  e.g. [OII]7320+
and [SIII]9304.
J0901 and J0925 spectra  are reproduced by single models, mG1 and mG2, respectively.
For J1154 and J1442 we show the results of two models   calculated by different $D$. The other input parameters
are similar.  [SIII]9304/\Hb~ is  well fitted only for J0901.
The results confirm that $D$ is the key parameter which can  better reproduce the HeII/\Hb~
line ratios even   if it affects [OIII]4363/\Hb.  
Comparing the present results for metallicities with those evaluated by  G20  Table 8 shows that N/H calculated in 
this  work
are by a factor  $\geq$1.5-3 higher than those evaluated by  G20, O/H by a factor between 2 and 3 higher, Ne/H by a factor
1.5-3, S/H by $\sim$ 8 and Ar/H by a factor of $\sim$ 2 higher. We have found that He/H are between  0.5 and 0.8 solar.

\begin{table*} 
\centering
\small{
	\caption{Comparison of observations  (I19a survey galaxies) with model results. \Hb=1} 
\begin{tabular}{ccccccccccccccccc} \hline  \hline
\        & [OII]  & [OIII] & [OIII] & [NII] &[OII]& \Vs  &\n0  &$D$   &\Ts     &$U$ & N/H     & O/H       \\
\        & 3727+  & 4363   & 5007+  & 6584  &7320 & \kms &\cm3 &0.01pc&10$^4$K &-   & 10$^{-4}$&10$^{-4}$\\\hline
\ {\bf J0314}  & 0.494  & 0.084  & 3.304  & 0.113 &0.021&-     &-    &-     &-       &-    &-        &0.2$^1$  \\
\ misss1  & 0.47   &0.076   & 4.4    & 0.09  &0.023& 100  &100  &1.8   & 5      &0.036   &0.25 &1.8      \\
\ {\bf J1433}  & 0.914  &0.078   &3.352   & 0.066 &0.04 &-     &-    &-     &-       &-       &-    &0.2$^1$    \\ 
\ misss2  & 0.97   &0.08    & 3.65   & 0.05  &0.046&   100&100  & 1.6  &4.7     &0.019   &0.1  &2.0        \\ \hline

\end{tabular}}

$^1$  obtained by the direct method  (I19a)

\end{table*}

\subsubsection{I19a survey}

The lines presented in this survey are only the most significant oxygen ones [OIII]4363, [OIII]5007+ and seldom 
[OII]3727. However, the spectra contain  [OII]7322  which  constrains  the models.
In Table 9  both the comparison of calculated to observed line ratios and the models are shown. 
We selected  from the rich I19a observed survey only two galaxies
J0314 at z=0.02741 and J1433 at z=0.02027 which show  a set of  line ratios adapted to  constrain the models.
In  columns 2-5 of Table 9 the line ratios to \Hb~ are shown. The following columns contain  the model sets.
I19a have  selected in their survey only galaxies with 12+log(O/H) $<$ 7.4.  They  compare the results obtained by 
different direct  methods. 
By the detailed modelling of the spectra we find O/H= 1.8 -2.0$\times$10$^{-4}$
and N/H= 0.25 - 0.1$\times$10$^{-4}$ for J0314 and J1433, respectively. 
I19a calculated O/H=0.2$\times$10$^{-4}$ for the two galaxies J0314 and J1433,
which  corresponds to  O/H lower by a factors of 9  and 10  than calculated by  detailed modelling.
The comparison  about the results calculated by the strong line methods and detailed modelling is discussed in the follow.

\subsubsection{B16 survey}

 Berg et al (2016) present UV spectrophotometry of 12 nearby, low metallicity, high ionization HII regions in 
dwarf galaxies at z=0.003-0.04
obtained by the Cosmic Origins Spectrograph on the Hubble Space telescope. They selected seven galaxies 
to analyse the O$^{+2}$ and C$^{+2}$ ions. 
We try to reproduce the  data from the Berg et al (2016) sample  by the detailed modelling of both the UV and optical
 spectra in order to  calculate
the C/O  and C/N relative abundances for local star-forming galaxies. Modelling results are  presented in Table 10. 
B1o-B7o refer to observations and B1m-B7m to model calculations.  
 Our results confirm that  the C/H and N/H 
are lower than solar, yet they are by a factor $>$ 10 higher that those calculated by the strong line method.

\begin{table*} 
\centering
\small{
\caption{Comparison of observations  (Berg et al 2016) with model results. \Hb=1} 
\begin{tabular}{lllllllllllllcccc} \hline  \hline
\     J          &082555&       &104457&      &120122&      &124159 &      &122622&     &122436&    &124827&\\ 	
\                &B1o   & B1m   & B2o  &B2m   &B3o   &B3m   & B4o   &B4m   &B5o   &B5m  &B6o  &B6m  &B7o  &B7m \\ \hline
\ CIV1548+1550   &0.97  &0.82   &3.59  &3.9   &-     &2     &-      &-     &-     &-    &-    &-    &1.22 &0.97\\
\ HeII1640       &0.38  &0.41   &0.7   &0.64  &-     &0.48  &0.55   &0.55  &-     &-    &-    &-    &0.96 &0.75\\
\ OIII]1660      &1.33  &0.9    &2.34  &0.86  &1.6   &1.56  &1.92   &0.73  &0.01  &0.54 &1.8  &1.3  &0.77 &1.5\\
\ SiIII]1883+1892&2.41  &2.3    &1.41  &-     &-     &-     &2.4    &1.36  &-     &-    &1.29 &-    &-    &4.2\\
\ CIII]1906      &2.63  &3.     &2.83  &2.5   &3.89  &3.94  &2.94   &2.8   &0.022 &0.019&3.38 &3.42 &2.12 &2.04 \\
\ CIII]1909      &3.44  &2.     &1.12  &1.6   &1.8   &2.6   &1.8    &1.8   &0.010 &0.013&2.96 &2.24 &1.05 &1.33 \\
\ [OII]3727+3729 &-     &0.12   &-     &-     &-     &-     &-      &-     &-     &-    &0.97 &0.9  &0.78 &0.79\\
\ [NeIII]3869+   &0.276 &0.6    &-     &-     &-     &-     &-      &-     &0.48  &0.5? &-    &-    &-    &1.35\\
\ \Hg 4340       &0.476 &0.46   &0.506 &0.46  &0.5   &0.46  &0.49   &0.46  &0.49  &0.46 &0.46 &0.45 &0.47 &0.45 \\
\ [OIII]4363     &0.116 &0.15   &0.146 &0.15  &0.11  &0.14  &0.105  &0.136 &0.116 &0.116&0.111&0.21 &0.129&0.25\\
\ [OIII]5007+4959&4.83  &5.2    &6.0   &6.1   &4.8   &4.7   &6.4    &6.17  &7.6   &8.0  &7.4  &7.23 &7.9  &7.88\\
\ [NII]6548+6584 &0.019 &0.01   &0.01  &0.013 &0.023 &0.02  &0.05   &0.02  &0.053 &0.04 &0.059&0.064&0.04 &0.048\\
\ \Ha            &2.76  &2.96   &2.75  &3.0   &2.78  &2.96  &       &      &2.79  &2.97 &2.79 &3.34 &2.78 &3.26\\
\ [SII]6717      &0.026 &0.005  &0.022 &0.017 &0.052 &0.02  &0.088  &0.015 &0.106 &0.017&0.087&0.04 &0.084&0.058\\
\ [SII]6731      &0.022 &0.004  &0.018 &0.014 &0.037 &0.017 &0.076  &0.014 &0.074 &0.019&0.066&0.03 &0.059&0.048\\
\ [OII]7330      &0.012 &0.007  &0.01  &0.02  &0.015 &0.028 &0.11   &0.02  &0.051 &0.026&0.028&0.059&0.019&0.05\\ 
\ \Vs (\kms)     & -    & 50    &  -   & 80   &  -   &70    &  -    & 80   &  -   & 60  &  -  &80   &   - & 70 \\
\ \n0 (\cm3)     & -    & 80    &  -   & 70   &  -   &80    &  -    & 90   &  -   & 180 &  -  &80   &   - & 90 \\
\ $D$ (0.01pc)   & -    & 1.08  &  -   & 1.1  &  -   &0.33  &  -    & 0.67 &  -   & 0.43&  -  &0.6  &   - & 0.53 \\
\ $U$            & -    & 0.14  &  -   & 0.08 &  -   &0.12  &  -    & 0.08 &  -   &0.05 &  -  &0.06 &   - & 0.04\\
\ \Ts (10$^4$K)  & -    & 4.1   &  -   & 4.6  &  -   &4.2   &  -    & 4.6  &  -   &4.6  & -   &4.2  &   - &5.0\\
\ (C/H)$^1$      & -    & 1.3   &  -   & 0.08 &  -   &1.8   &  -    & 1.1  &  -   &0.03 &  -  &1.7  &   - &0.36\\
\ (C/H)$^2$      & -    & 0.118 &  -   & 0.055&  -   &0.1   &  -    & 0.08 &  -   &0.135&  -  &0.162&   - &0.151\\
\ (N/H)$^1$      & -    & 0.2   &  -   & 0.07 &  -   &0.12  &  -    & 0.1  &  -   &0.26 &  -  &0.16 &   - &0.13\\
\ (N/H)$^2$      & -    &0.012  &  -   &0.01  &  -   &0.013 &  -    &0.021 &  -   &0.013&  -  &0.023&   - &0.023\\
\ (O/H)$^1$      & -    & 1.8   &  -   & 1.3  &  -   &1.8   &  -    & 1.4  &  -   &4.6  &  -  &1.3  &   - &1.35\\
\ (O/H)$^2$      & -    & 0.23  &  -   & 0.282&  -   &0.282 &  -    &0.537 &  -   &0.794&  -  &0.692&   - &0.646\\
\ (S/H)$^1$      &  -   & 0.3   &  -   & 0.3  &  -   &0.3   &  -    & 0.3  &  -   &0.3  &  -  &0.3  &   - &0.3\\
\ (S/H)$^2$      &  -   &0.009  &  -   &0.010 &  -   &0.10  &  -    &0.015 &  -   &0.015&  -  &-    &   - & - \\ \hline
\end{tabular}}

$^1$ in 10$^{-4}$ units calculated in this paper; $^2$ in 10$^{-4}$ units calculated by Berg et al. (2016)

\end{table*}

\section{Discussion}

In  previous sections we have tried to reproduce the observed line ratios in the spectra of   local
galaxies from different surveys.
We have selected  from each of them the spectra which could lead to the  most reliable results
i.e. those  which present the highest number of significant lines. 
Nevertheless, when  the observed [OIII]4363 and HeII4686 lines   were  reported in  one spectrum,  
we  had some problem in reproducing both of them with a single-cloud model.  
We have found that the [OIII]4363/\Hb~ and HeII4686/\Hb~ line ratios strongly depend on the geometrical 
thickness of the clouds.  $D$ ranges between $\sim$0.002pc and $\sim$1pc. 
 With regards to the other parameters,  the detailed modelling  method  yields
\Vs $\leq$ 100 \kms (in  a few cases \Vs $\leq$150 \kms), \n0 $\sim$ 50-100 \cm3, \Ts=34000-95000K,
$U$=0.012-0.4, N/H=0.1-0.55$\times$10$^{-4}$ and O/H=1.2-2.8 $\times$10$^{-4}$. Three objects show O/H=
4.3, 5.1 and 5.7 $\times$10$^{-4}$ which are  closer to solar.  
He/H ranges between 0.01 and 0.1, but  most of the spectra were reproduced adopting He/H= 0.08 in agreement with 
the predicted value (see e.g. Morton 1968).

\begin{figure*}
\centering
\includegraphics[width=6.4cm]{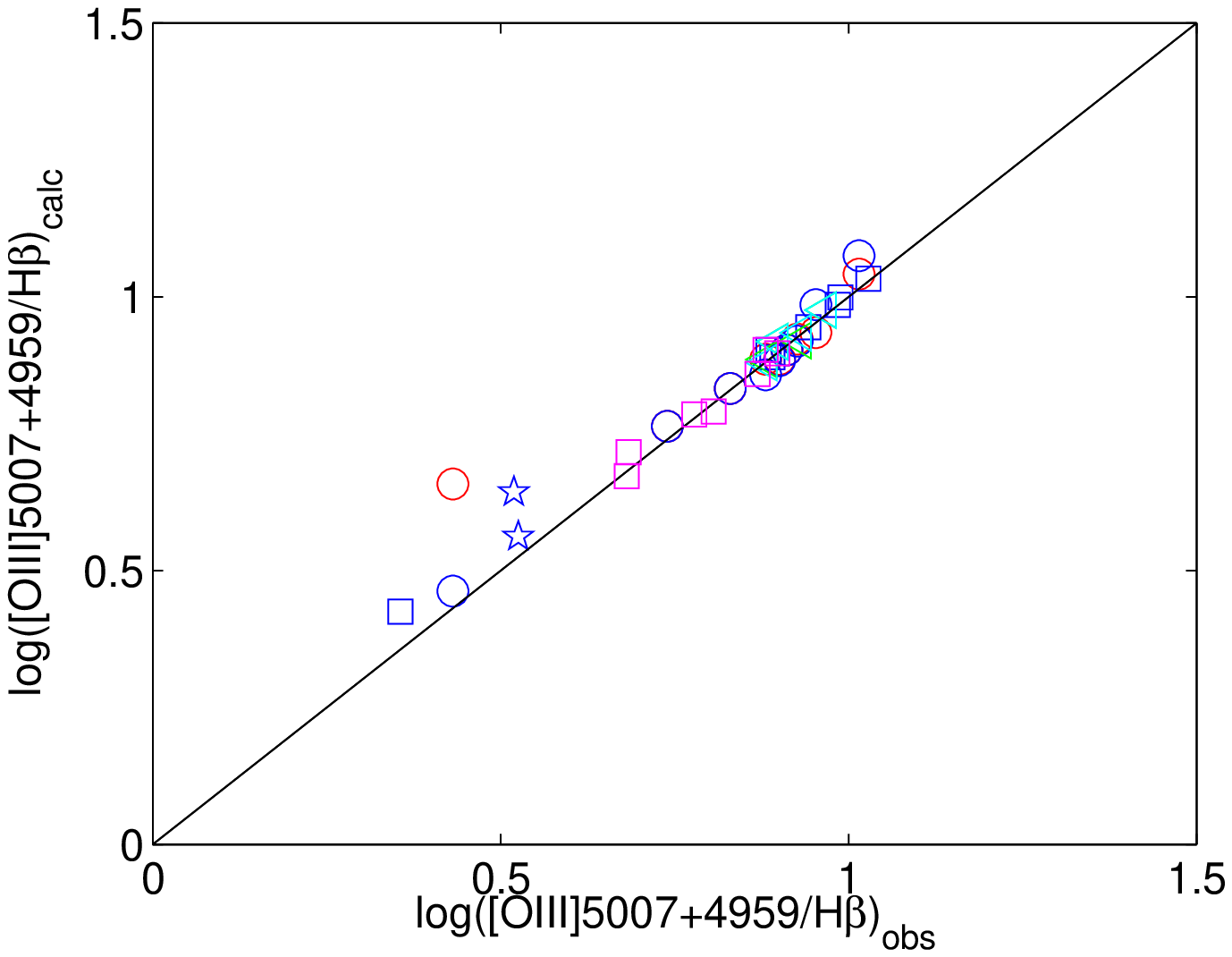}
\includegraphics[width=6.4cm]{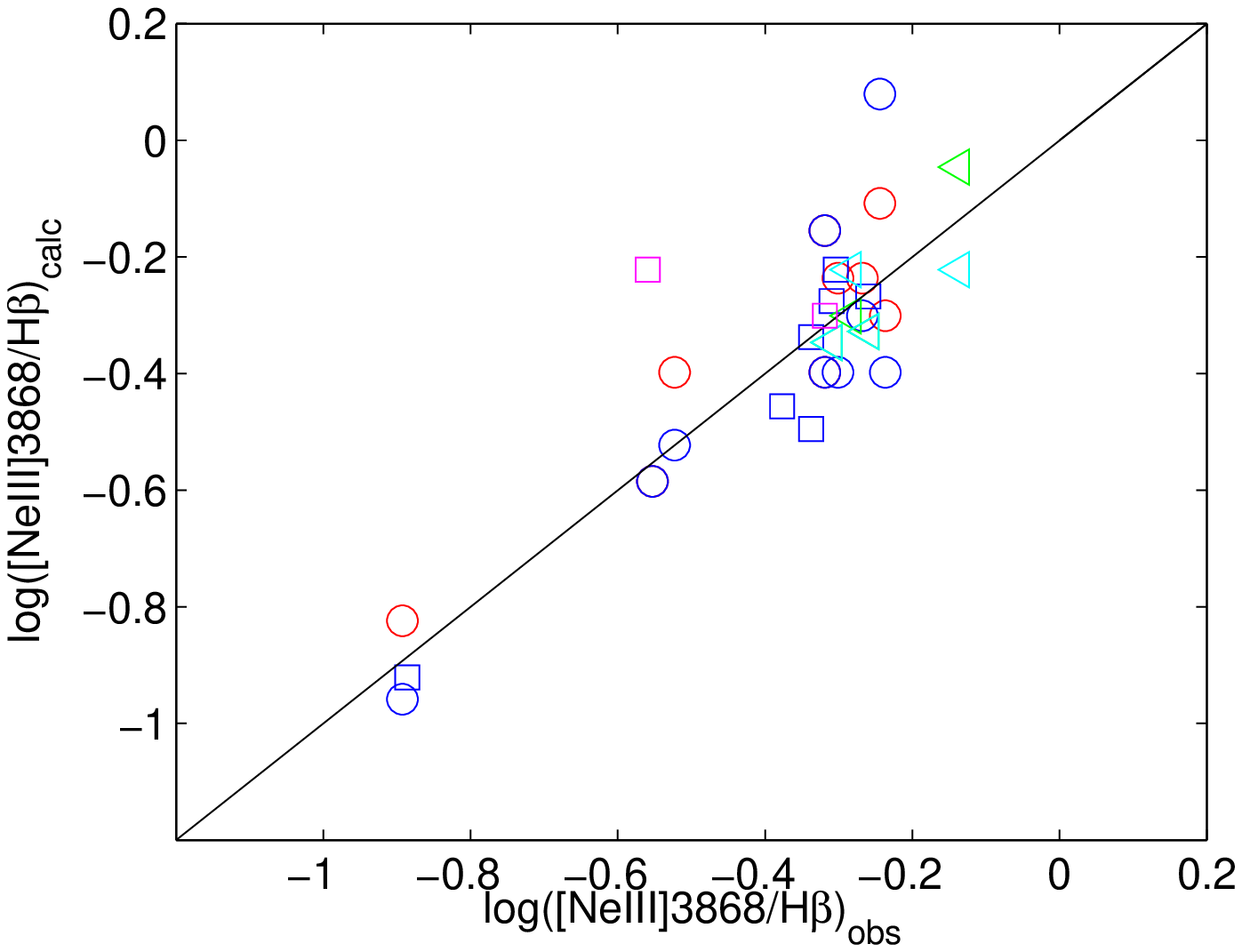}
\includegraphics[width=6.4cm]{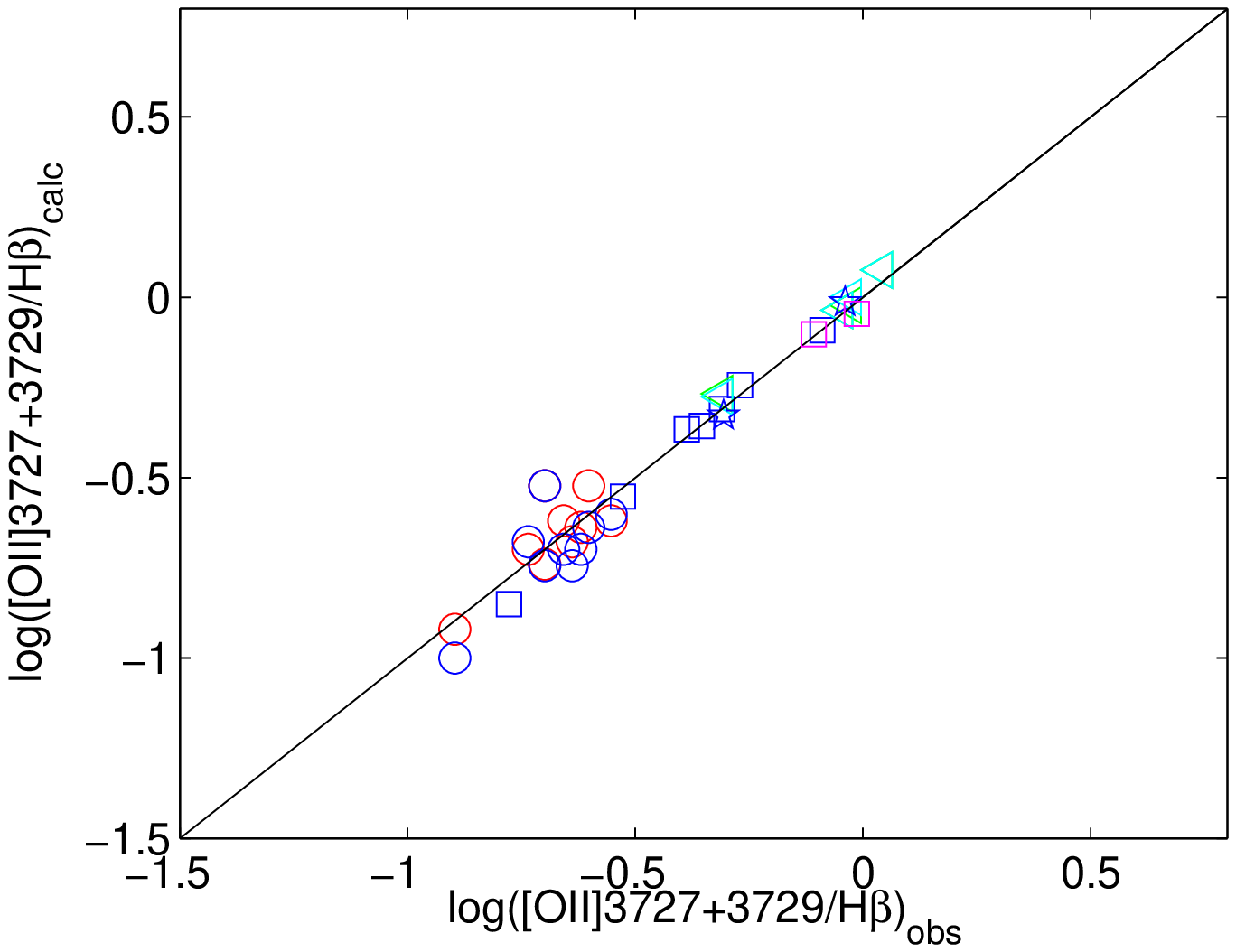}
\includegraphics[width=6.4cm]{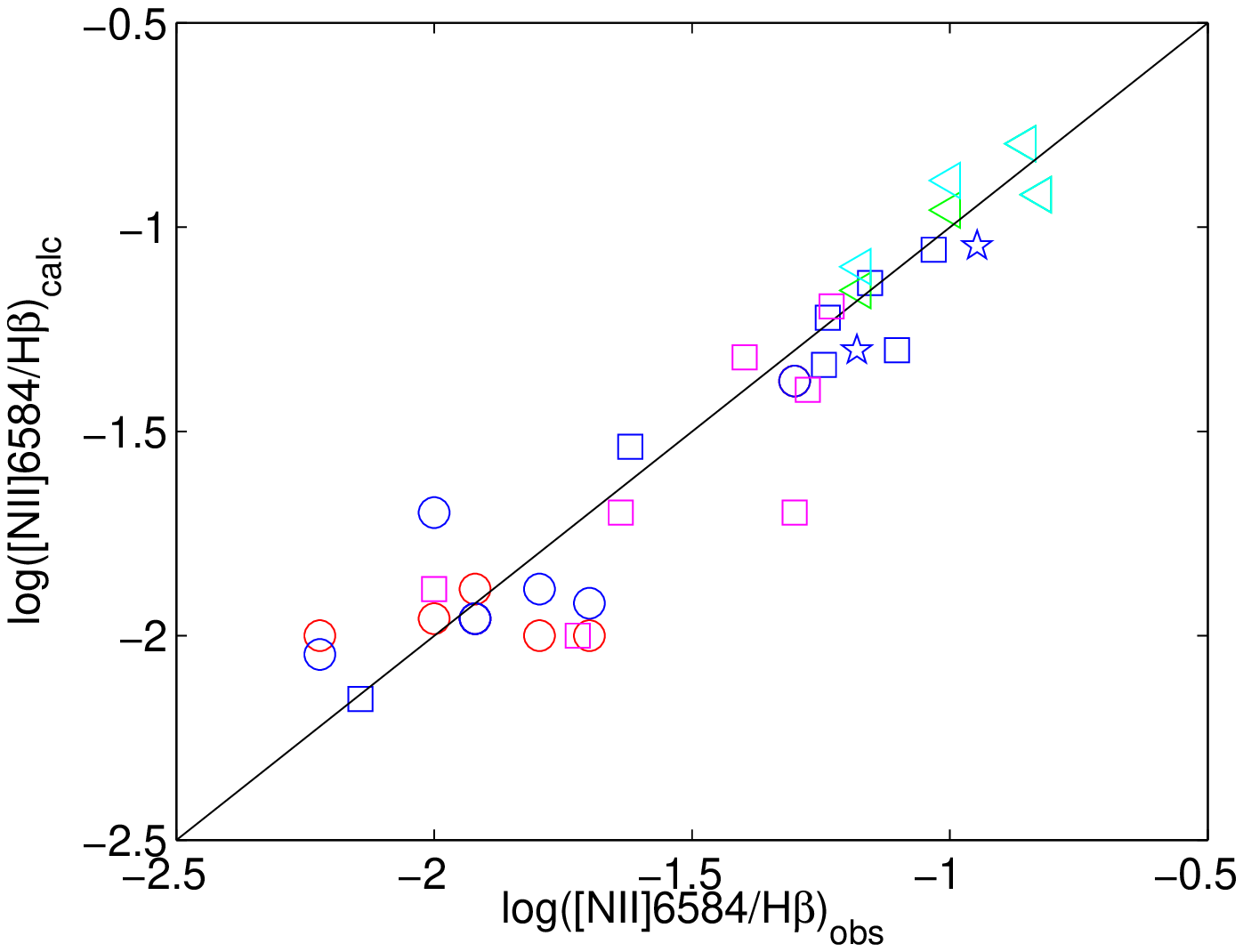}
\includegraphics[width=6.4cm]{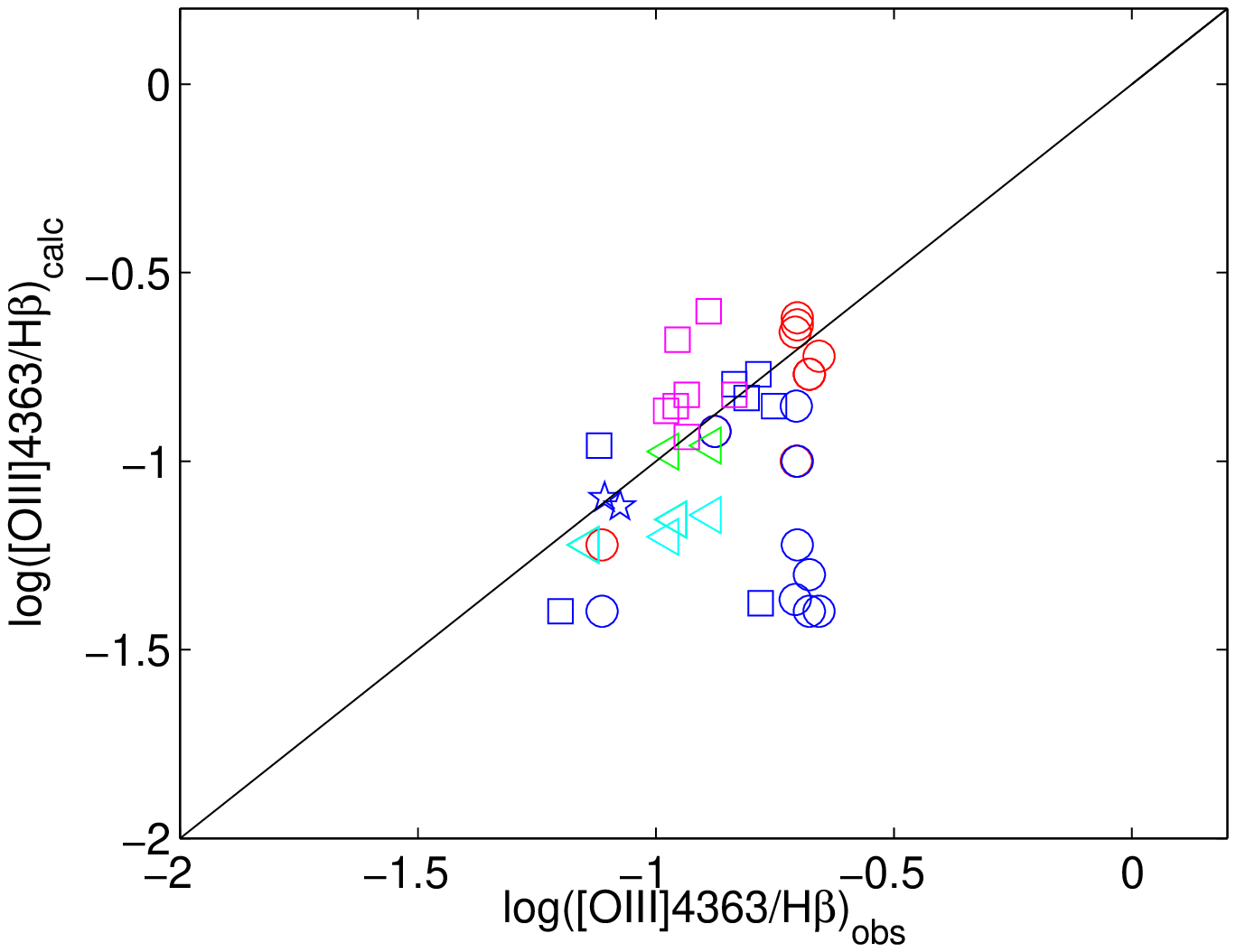}
\includegraphics[width=6.4cm]{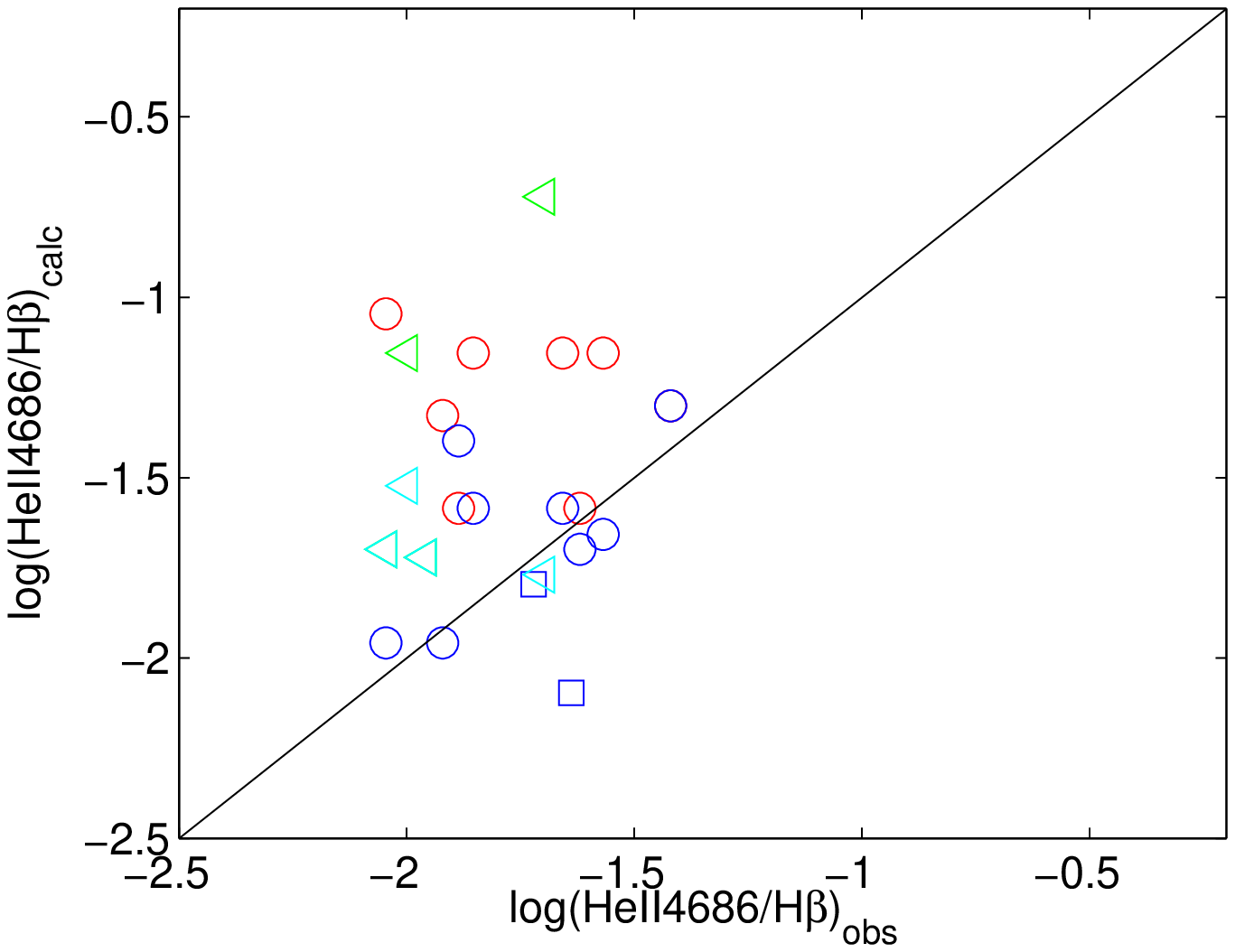}
\includegraphics[width=6.4cm]{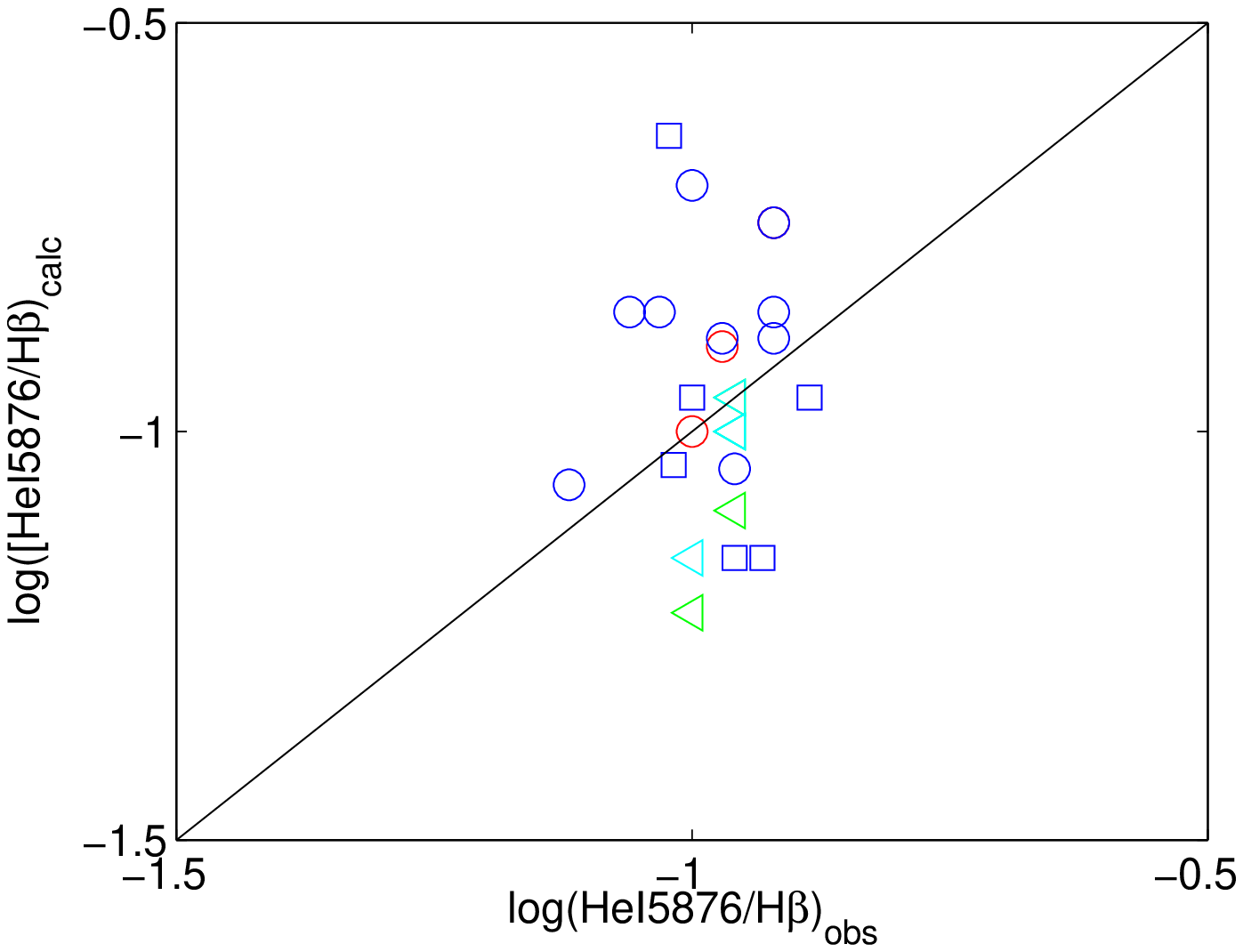}
\includegraphics[width=6.4cm]{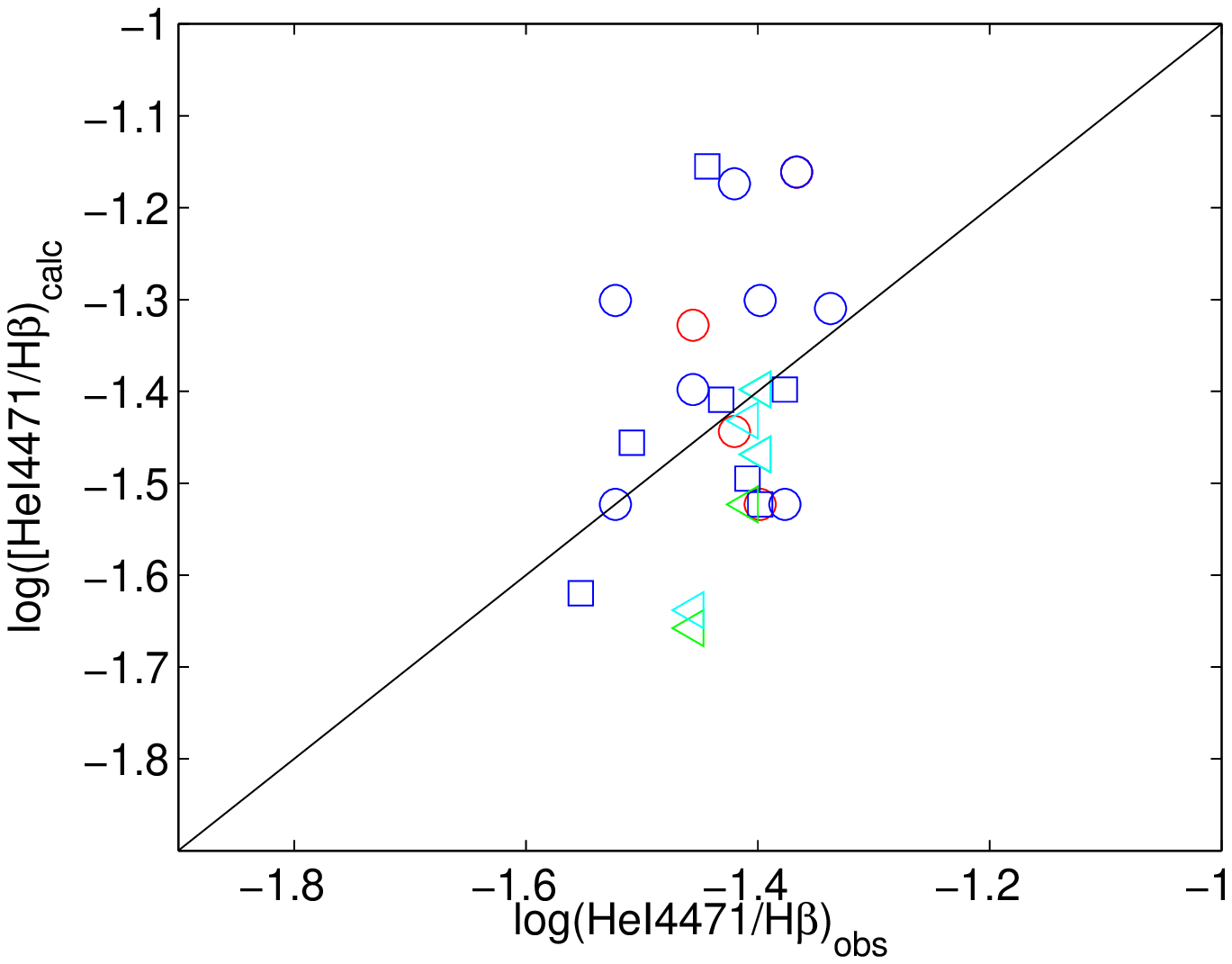}
\includegraphics[width=6.4cm]{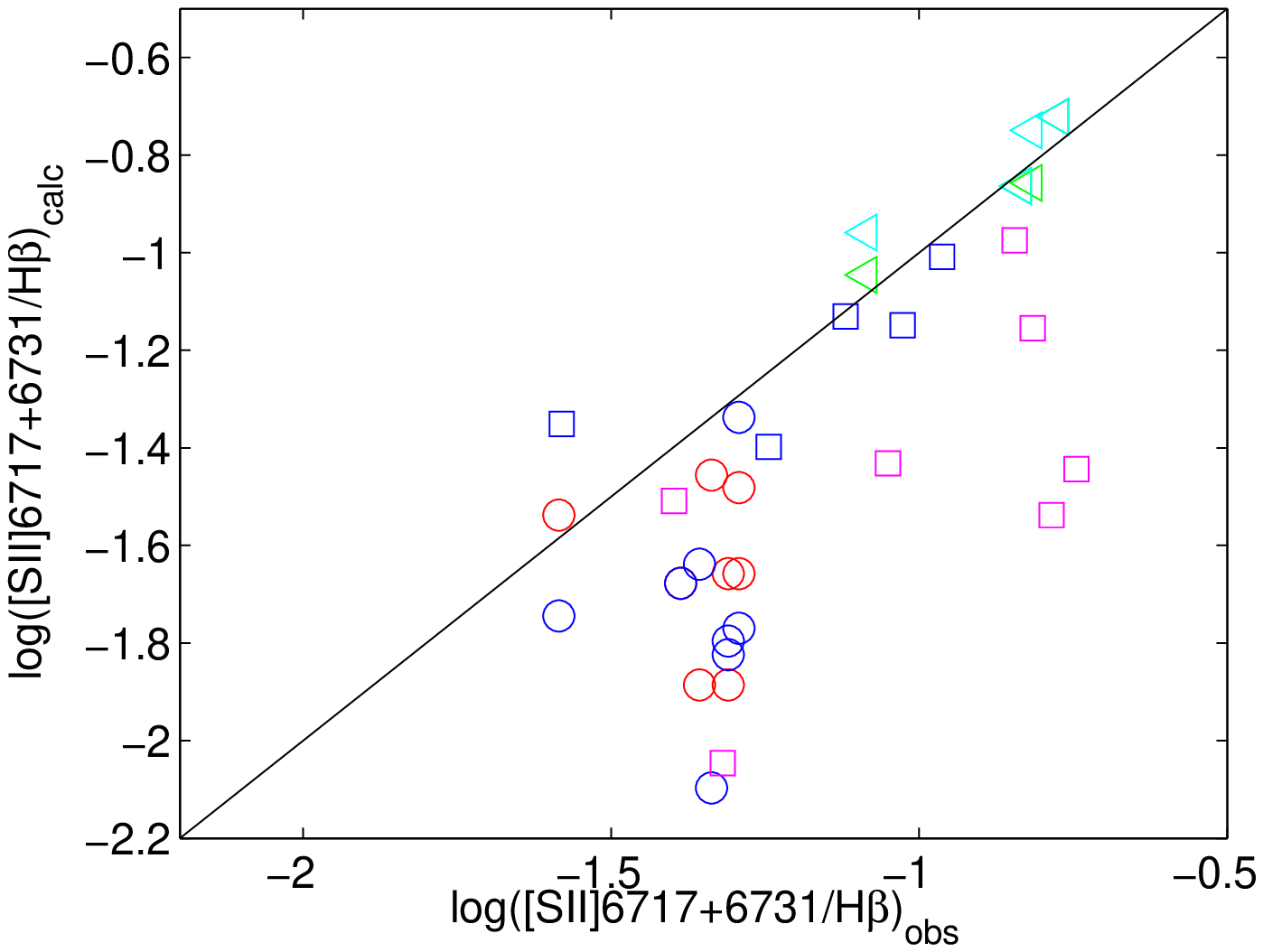}
\includegraphics[width=6.4cm]{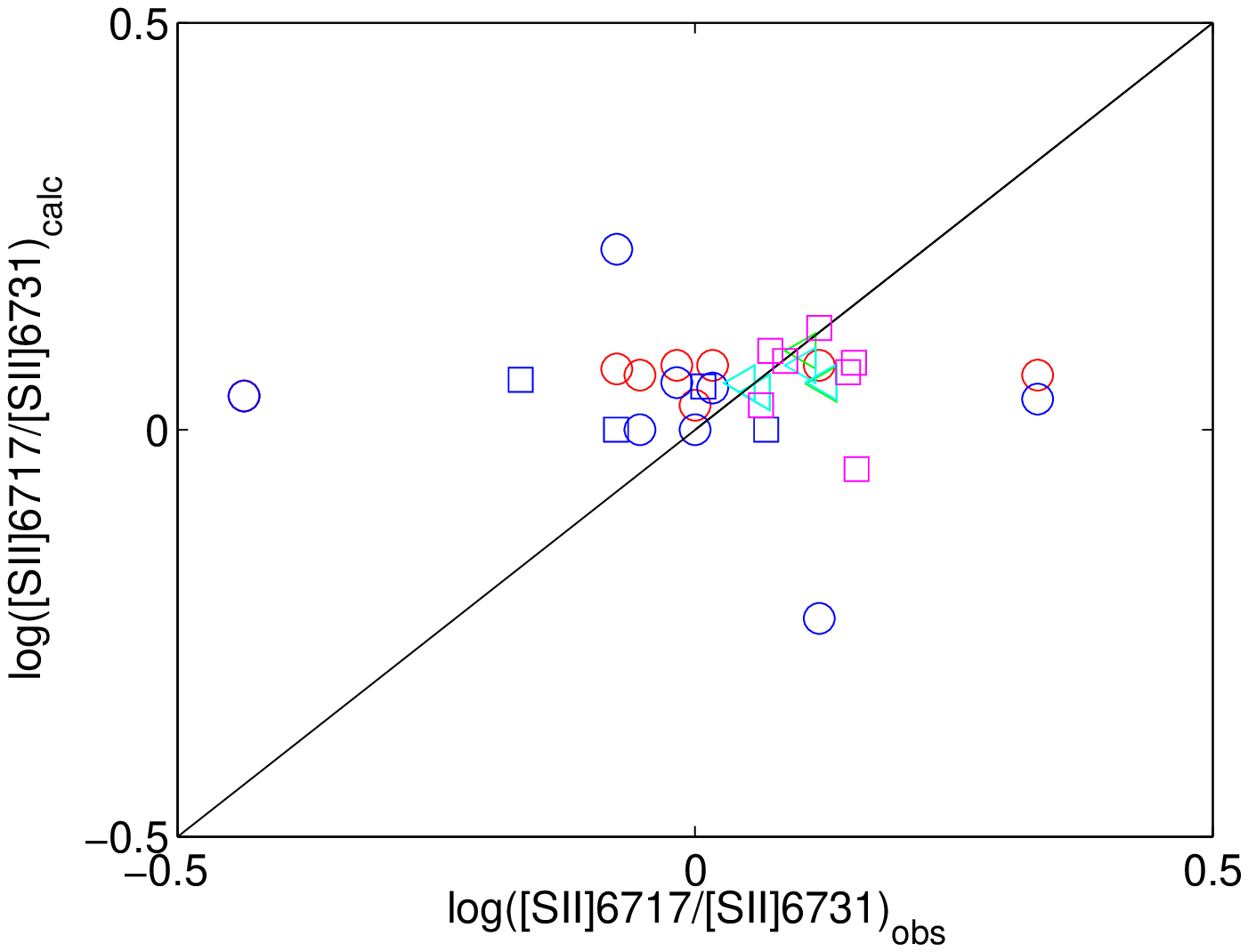}
\caption{Comparison of model results with the  data for
I20 survey and J1234 galaxies (red circles: models mis1a-mis9a and  blue circles: models mis1b-mis9b), for
I18b survey, J0811 and HS0837 (blue squares), for G20 survey (green triangles: models G1, G2, G3a, G4a
	and cyan triangles: models G3b and G4b),  for I19a galaxies (blue stars)  and B16 (magenta squares).
}
\end{figure*}

\subsection{Line ratios}

In Fig. 4   model results  are compared with the observed line ratios.
In the  two top left and right  diagrams  we show the results for the most  significant lines,
which, however, in SFG they are not always the strongest ones.  [OII]3727+/\Hb~  can be very  low.
The observed  [NeIII]/\Hb~ are less precisely reproduced because
[NeIII]3868  may be  blended with [NeIII]3967 and with the [[SII]4070,4077 lines  which are not negligible  
at the physical conditions  calculated by  models which adopt \Vs$\sim$100 \kms.
About the [NII] line, the error in the observed values is relatively high for low [NII]/\Hb~ ($\sim$0.01),
reaching 60 percent in the J1355 galaxy. Therefore, the fit in Fig. 4 is  less straightforward.
The [OIII]4363/\Hb~ ratios are underpredicted by the models reported in Table 4 and the HeII4686/\Hb~ are overpredicted
by models presented in Table 3. Fig. 4 (middle diagrams) suggest that pluri-cloud models averaged on the
Table 3 and Table 4 results could improve the fit.
The summed [SII]6717+6731 line ratios to \Hb~   (Fig. 4, bottom left) are underpredicted by  the models even adopting  
S/H close to solar.  Higher S/H relative abundances are not realistic  because
sulphur   is more  often  trapped into dust grains and depleted from the gaseous phase.
We suggest that a large contribution to [SII] comes from  ISM regions where dust grains were  
sputtered by previous strong shock events.
The [SII]6717/[SII]6731  (Fig. 4, bottom right)  depend  on the gas electron density and temperature. 
 The calculated  line ratios are nearly constant  $\sim$1, while the observations
range between $\sim$ 0.3 and 2.2. Even  considering the ISM gas inhomogeneous conditions, 
the extreme low and high observed [SII]6717/6731  indicate \n0 $>$ 10$^5$ \cm3 and $<$10 \cm3,
respectively (Osterbrock 1974), which are not adapted to reproduce the other observed line ratios.

\subsection{He lines and He/H relative abundance}

In our sample 15 out of 28 spectra  include the  HeII4686, HeI4471 and HeI5876 lines.  
In the  modelling process we started by  [OIII]5007+/\Hb~ and [OII]3727+/\Hb. We have found that
we could reproduce more precisely  [OII]3727+/\Hb~ 
reducing the geometrical thickness of the clouds, however, the fit  of HeII4686/\Hb~ deteriorated
even adopting He/H as low as 0.01 ((He/H)$_{\odot}$=0.1).
The HeI5876 line  is present  in the spectra of  nearly all types of galaxies with HeI5876/\Hb~  ratios in the 
$\sim$0.1-0.15 range. 
Fig. 4 diagrams in the row  next to the last  suggest  that the nearly constant observed HeI5876/\Hb~  ratios 
may  show a high
contamination by the ISM    and  that they are less affected by radiation  from the stars and
 by  radiation  from gas collisionally heated  by the shocks.
The same is valid for HeI4471/\Hb~ which is also less affected  by the He/H relative abundance 
than HeI5876/\Hb.
Guseva et al (2020)  refer to HeII4686 lines as to  hard ionizing radiation indicators. They claim that
HeII 4686 cannot be emitted from gas photoionized
by  normal stellar populations with the generally observed temperatures.  G20  invoke X-ray from binary stars 
and shocks as alternatives.    Izotov et al (1997) also suggest that the observed HeII 4686/\Hb~  cannot be easily 
reproduced  adopting  pure photoionization models or \Te methods. 
Shaerer et al (2019) addressing the HeII4686  issue also claim  that these lines can be due to X-ray from
binary stars. 
They note  that in low metallicity SFG both  the empirical data  and the  theoretical models suggest that high mass
X-ray binaries are the main source of nebular HeII emission.
In agreement with  Guseva et al  we have adopted that shocks  collisionally  heat the emitting gas to  
high temperatures. 
We have explained  the HeII/\Hb~ and the other line ratios by photoionization from the stars coupled to shocks.
 The sub-solar He/H ratios  are  calculated consistently with the other results because,  
heating the gas by the shock,
we obtain HeII4686/Hb~ higher than observed (see e.g. Table 2).

We suggest that the wind from the starburst region  collides with  the gaseous clouds throughout the
 galaxy.  Not only the cloud dynamics but also the element  composition of the  gas  within the clouds
is affected by the wind from the  star atmosphere.
In particular, the various He/H abundance ratios found even in  nebulae  far away from the starburst, trace 
the WD atmosphere  element composition which is conveyed by  strong winds.  
Lauffer et al (2018) explain that WD in the DA spectral class have H-rich atmospheres, DO types show  strong
HeII lines and \Tef $\sim$ 45000-200000K, while DB types have strong HeI lines and \Tef $\sim$ 11000K-30000K.
He/H ratios  and other heavy element abundances are   connected with the  star atmosphere temperatures.
Barstow et al (1994)  found   \Tef $\sim$ 50000K-90000K, in agreement with our results.
WD in the present star-forming regions  seem of DA type because   the HeI lines come most probably
from the ISM gas. However they do not exclude contamination by DO and DB types. 
For most of  the models (mis2a-mis8a, Table 3)  we could  not find  any acceptable  fit to the observed
HeII/\Hb~ line ratios by solar He/H. The fit improved reducing the He/H relative abundance by a factor of 
$\sim$5-10, in agreement with
Morton (1968) who  reported that 'Miss Underhill believes that He/H for O and B stars lies somehow between
0.05 and 0.01'.  
This implies a  decrease of the  HeI 5876/\Hb~ line ratio (calculated from the clouds) by a factor  $\leq$ 2. 
However, the fit  also improved  by  using models mis2b-mis8b (Table 4) with a less depleted He/H. 
 For a   few galaxies we found He/H = 0.08 in agreement with  Morton et al (1968)  results which  show a minimum 
of He/H=0.077 in early A- and B-type stars.
 With regards to the components of close binary systems, Lyubimkov (1995) showed that 
the low original  He/H is maintained
throughout the first half of their main sequence evolution, 
there is no He/H monotonic increase as in the hot isolated stars. 
The HeII4686 lines  are  generally weak  in galaxies which do not  host an AGN. When an AGN is present
 the spectra  show  HeII/\Hb~  higher than those  observed  for the I20 survey objects due to the power-law flux
 from the active nucleus.

\begin{figure*}
\centering
\includegraphics[width=8.4cm]{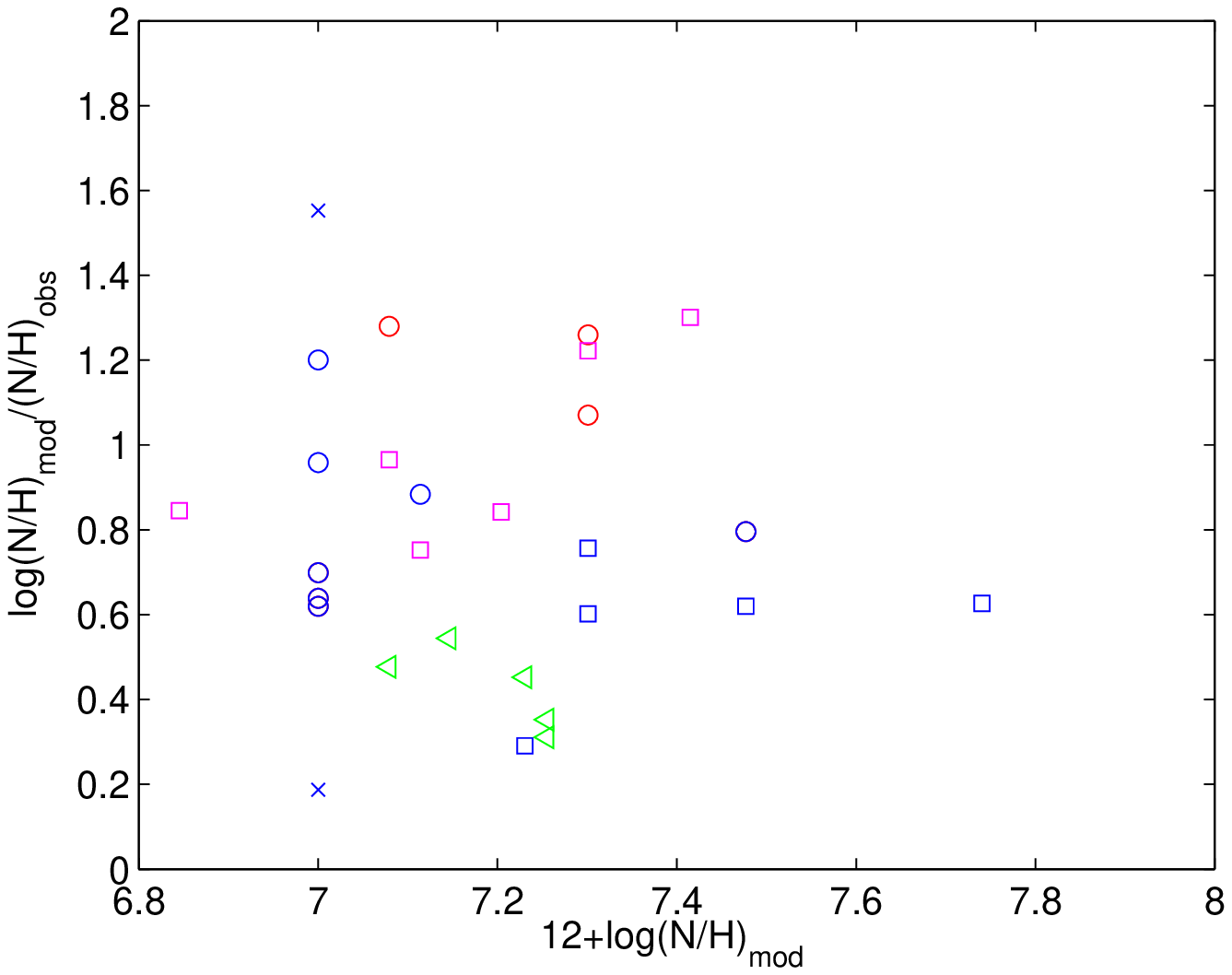}
\includegraphics[width=8.4cm]{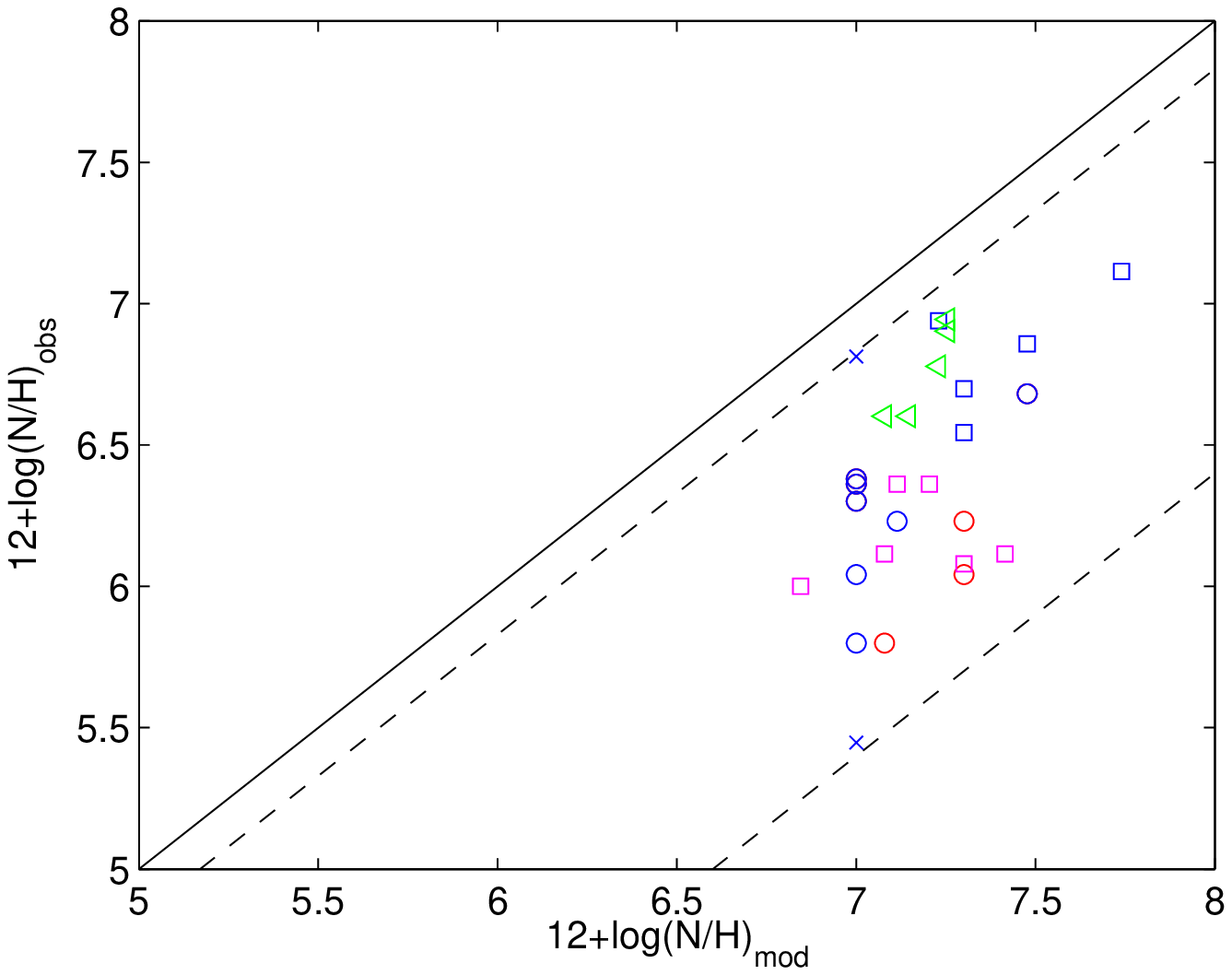}
\includegraphics[width=8.4cm]{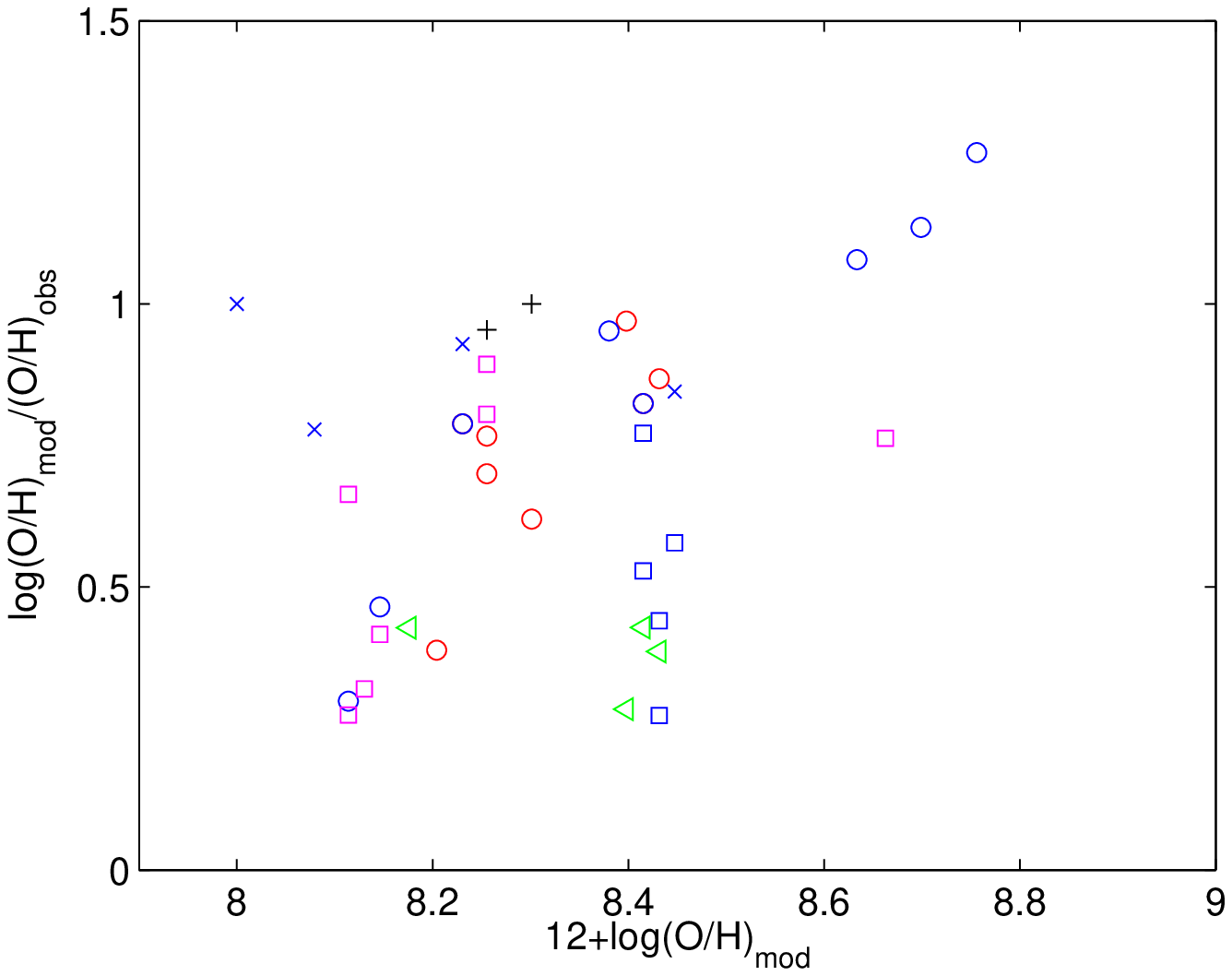}
\includegraphics[width=8.4cm]{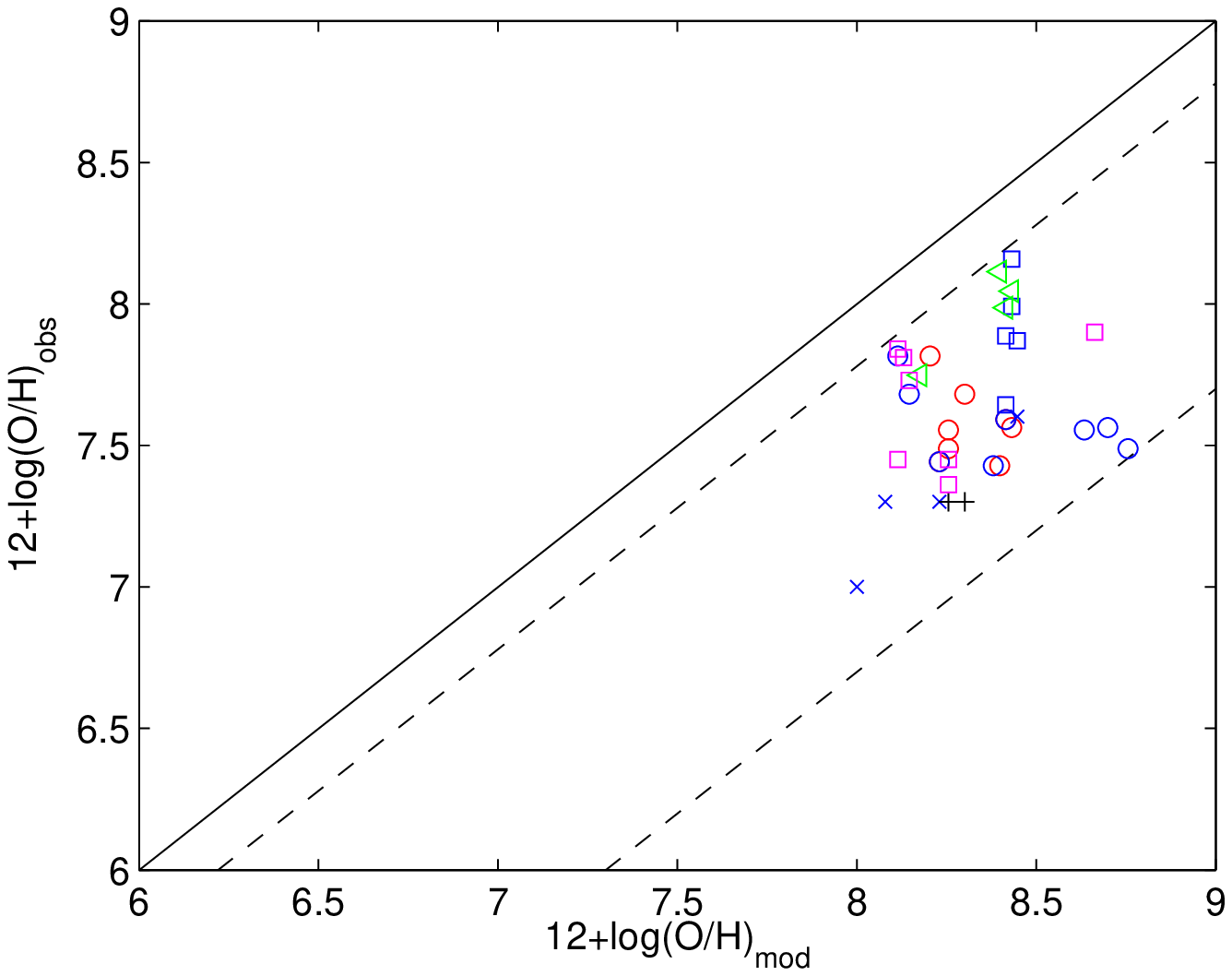}
\includegraphics[width=8.4cm]{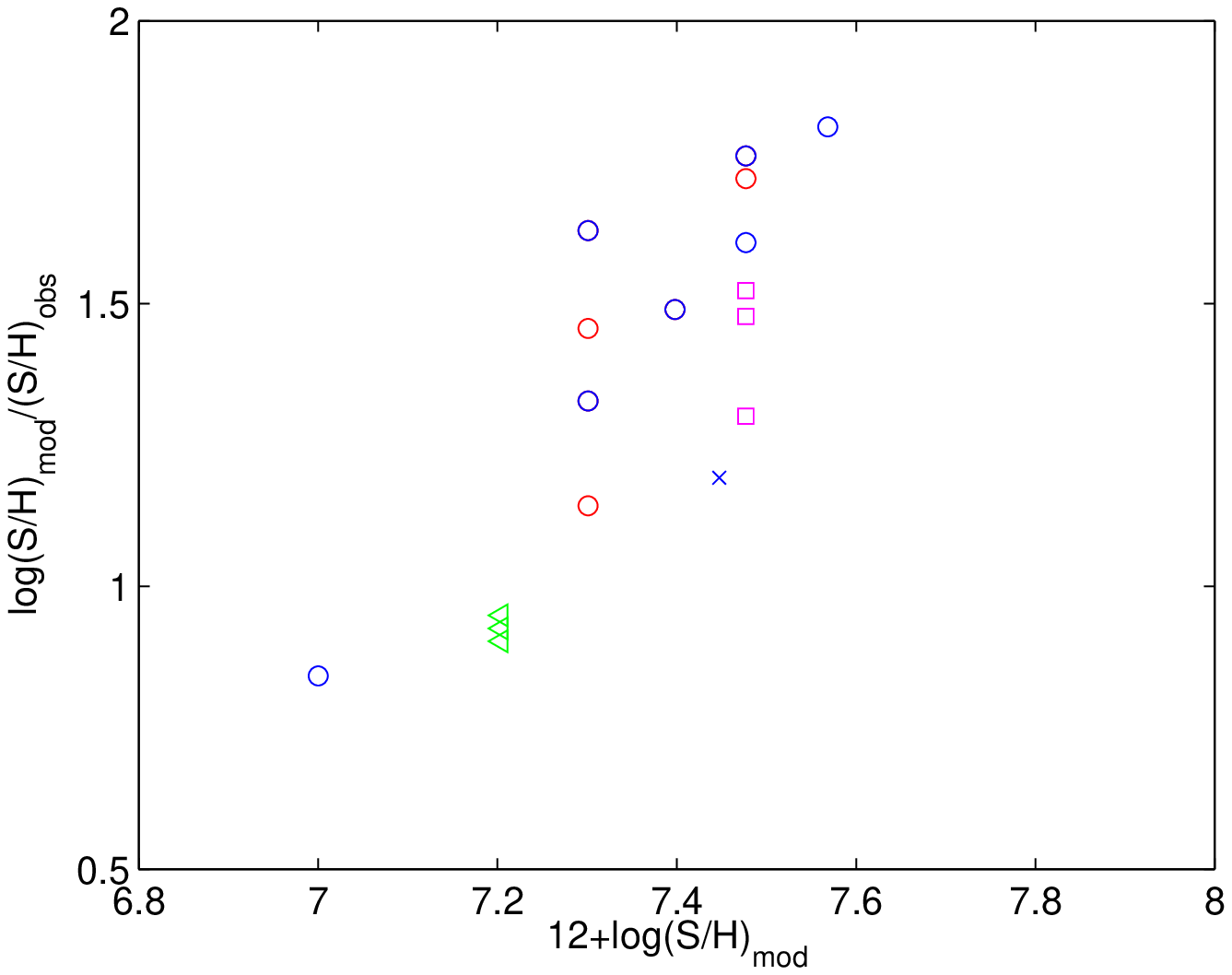}
\includegraphics[width=8.4cm]{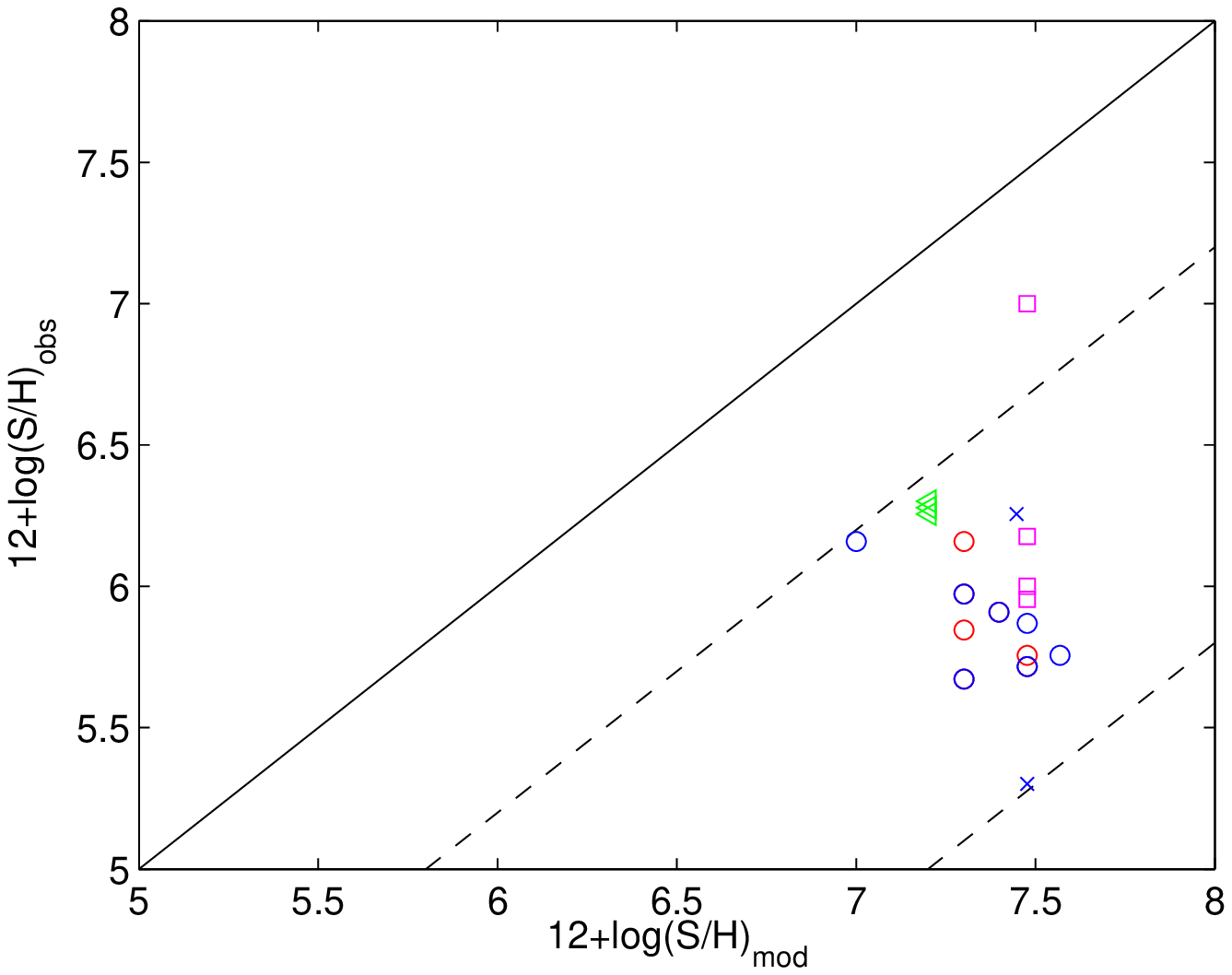}
\caption{Comparison of relative abundances calculated in this work with those 
evaluated by the observers  using the strong line method: I20 
 presented in Table 3 (red circles) and in
Table 4 (blue circles), I18b  presented in Table 6 (blue squares), G20  presented in
 in Table 8  (green and cyan triangles),
 I19a presented in Table 9 (black plus) and for galaxies
J1234, J0811 and HP0837  presented in Tables 4 and 6 (blue crosses). Top:  N/H, middle: O/H and bottom:
S/H.
}
\end{figure*}

\subsection{Relative abundances of  the heavy elements}
\subsubsection{N/H, O/H and  S/H}

The comparison of the N, O and S metallicities calculated by  the detailed modelling and by the direct strong line 
methods are shown in Fig. 5 (right diagrams). 
The relative abundances evaluated directly by the observers are all lower
than those calculated by detailed modelling.
Fig. 5 right diagrams  indicate that the relative abundances  closest to our model results are given by G20.
  They  also show  that O/H calculated by detailed modelling are  by  factors between 1.6 and 20  higher than 
 those calculated by other methods, N/H by factors $\sim$1.5 and 40 and S/H by factors $\sim$ 6 and 63.

 The log((N/H)$_{mod}$/(N/H)$_{obs}$) versus 12+log(N/H)$_{mod}$, log((O/H)$_{mod}$/(O/H)$_{obs}$) 
 versus 12+log(O/H)$_{mod}$  and log((S/H)$_{mod}$/(S/H)$_{obs}$) versus 12+log(S/H)$_{mod}$
 are  reported in Fig. 5 left diagrams.  
 They represent the ratios of N/H, O/H and S/H  calculated by detailed 
modelling to those calculatd using the strong line 
methods by the observers  as function of the N/H, O/H and S/H  metallicities calculated by detailed method  
for the sample objects.
 The trend in Fig. 5 for N/H  (left top diagram) is opposite to those of O/H and S/H.
 N$^+$ and H$^+$  ions are linked by charge exchange reactions. Accordingly,
 [NII]/\Hb~  depends  strongly on N/H. At high N/H, modelling and strong line method results converge
 because N appears  only in the [NII]6584,6548 lines and  observation uncertainties are lower.
 If we omit in the Fig. 5 middle left diagram the three galaxies with a near solar
O/H calculated by models mis2b, mis3b and mis8b the trend becomes ambiguous. These models 
were already criticised  because underpredicting [OIII]4363/\Hb~ (Table 2).  
For  elements like oxygen which corresponds to  a relatively high number of significant lines 
([OIII]5007+, [OIII]4363, [OII]3727+,
[OI]6300+, etc) in a single optical spectrum, the  O/H relative  abundances by the strong line methods may
 lose some important components and underpredict O/H  for high O/H.
In fact, by detailed modelling the  electron temperature and density profiles are 
calculated throughout  the entire clouds, whereas by the strong line methods some regions may be  excluded. 
With regards  to S/H, also for S$^+$ and H$^+$ ions  charge exchange reactions are taken into consideration. 
However, the poor fit of the [SII]/\Hb~ ratios   calculated by detailed modelling was explained by a strong
contribution from the ISM with S/H at their maximum value. Therefore, the trend is opposite to
that for N/H.

\subsubsection{C/O and N/O relative abundances}

 A general trend of increasing C/O abundance ratio with O/H  and a  
constant C/N were noticed by Berg et al (2016). 
We  obtain the  C/O and N/O ratios by modelling (see Table 10) consistently both the
 UV and the optical emission lines observed by Berg et al (2016).  
 The C/O ratios calculated in the present paper are  reported on top of the log(C/O) versus 12+log(O/H) 
 diagram presented
 by Berg et al (2016, their fig. 8) in Fig. 6.  In  particular, they show that log(C/O) rapidly increases at 
 12+log(O/H)=8.
 Henry et al (2000)  claim that C and N production in the Universe
 are decoupled and originate from separate sites. 
Carbon is mainly produced by massive stars (M$>$ 8\msol),
while primary nitrogen production at low O/H is dominated by intermediate-mass
stars between 4 and 8 \msol. 
At 12+log(O/H)$>$8.3 secondary nitrogen becomes prominent.
The N/O ratio positively correlates with stellar mass (Perez-Montero et al 2013).
 Hayden-Pawson et al (2021)   report  that N/O maintains
a constant value  log(N/O)$\sim$-1.5 at low metallicities
while at higher metallicities the N/O ratio increases rapidly with O/H.
Henry et al suggest that the flat N/O is  due to low star formation rate in contrast with Izotov \& Thuan (1999) who 
claim that these systems are very young because of their low metallicity.
Izotov \& Thuan (1998) results for blue compact galaxies show that none of the heavy element-to-oxygen abundance 
ratios depend on O/H. 
 An investigation  of  many  host galaxy types indicates  that 
secondary  nitrogen  becomes evident throughout the
redshift range (Contini 2017, fig. 7)  at z$\leq$0.1. 
N/O ratios in local HII regions appear at the bottom of  the N/O versus z diagram 
well separated from the AGNs.

\begin{figure}
\centering
\includegraphics[width=8.8cm]{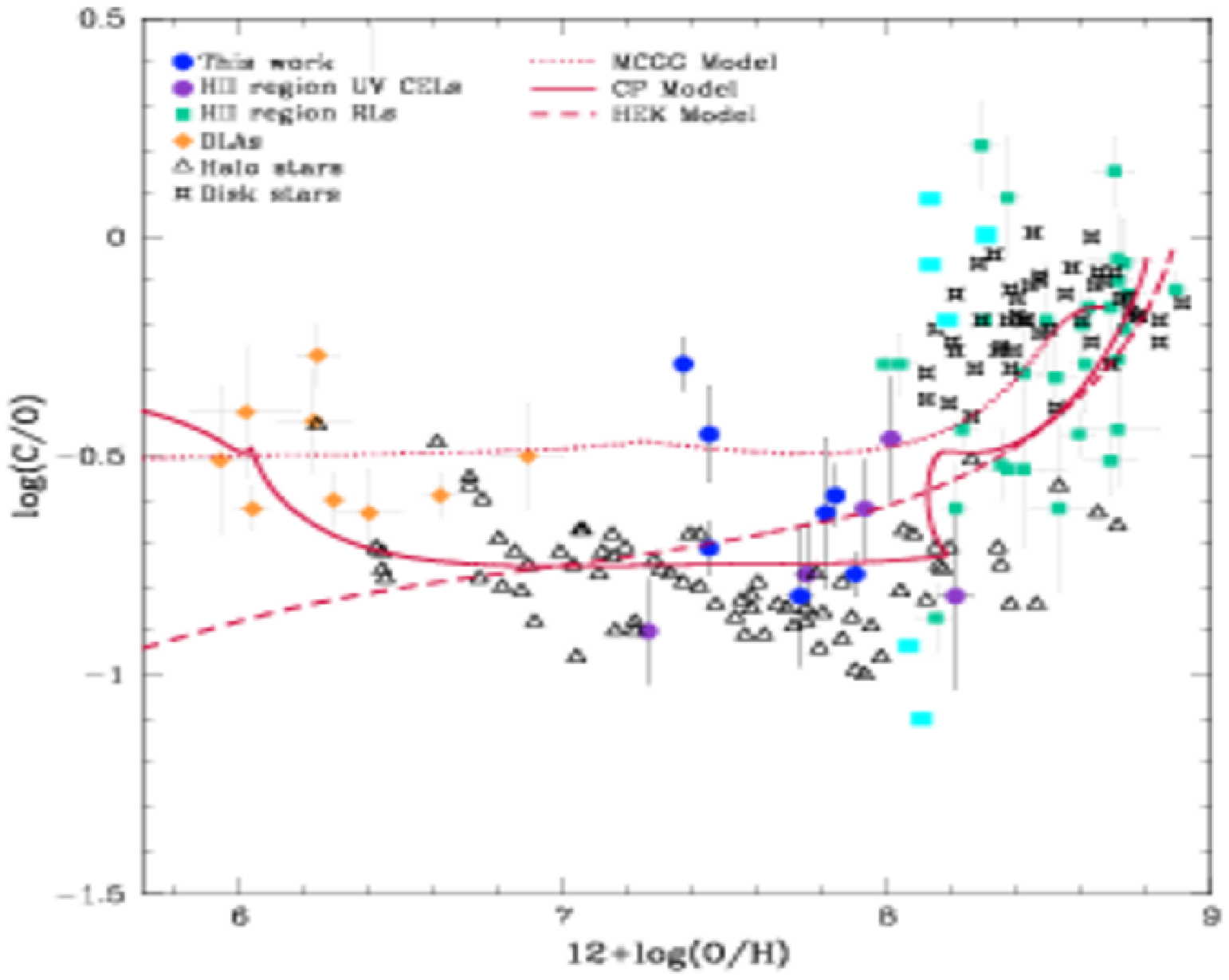}
\caption{Comparison of C/O relative abundances versus metallicity calculated in this paper (Table 1) (cyan rectangles)
with  the results  reported  from Berg et al (2016, their fig. 8)}
\end{figure}

\begin{figure*}
\centering
\includegraphics[width=7.2cm]{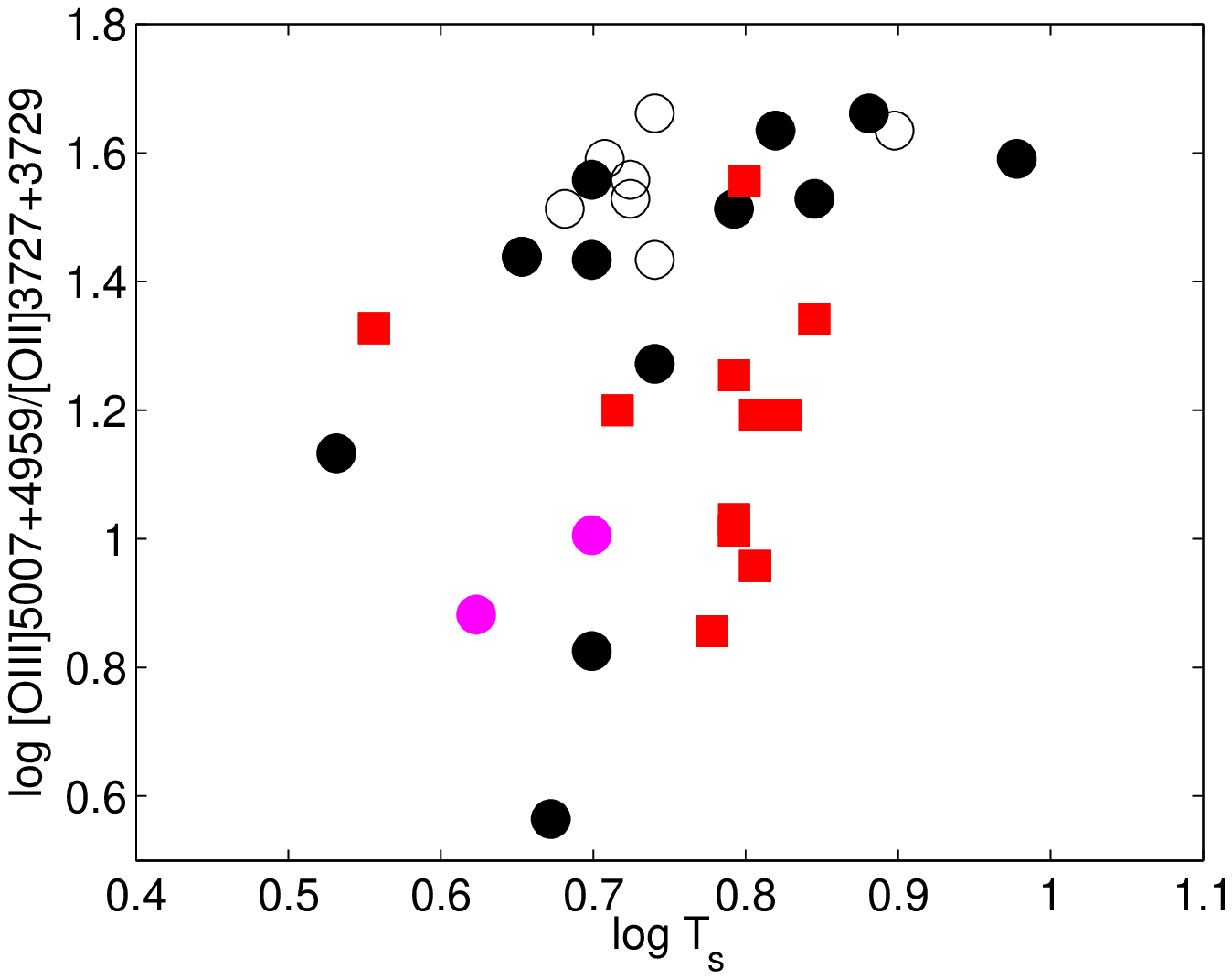}
\includegraphics[width=7.2cm]{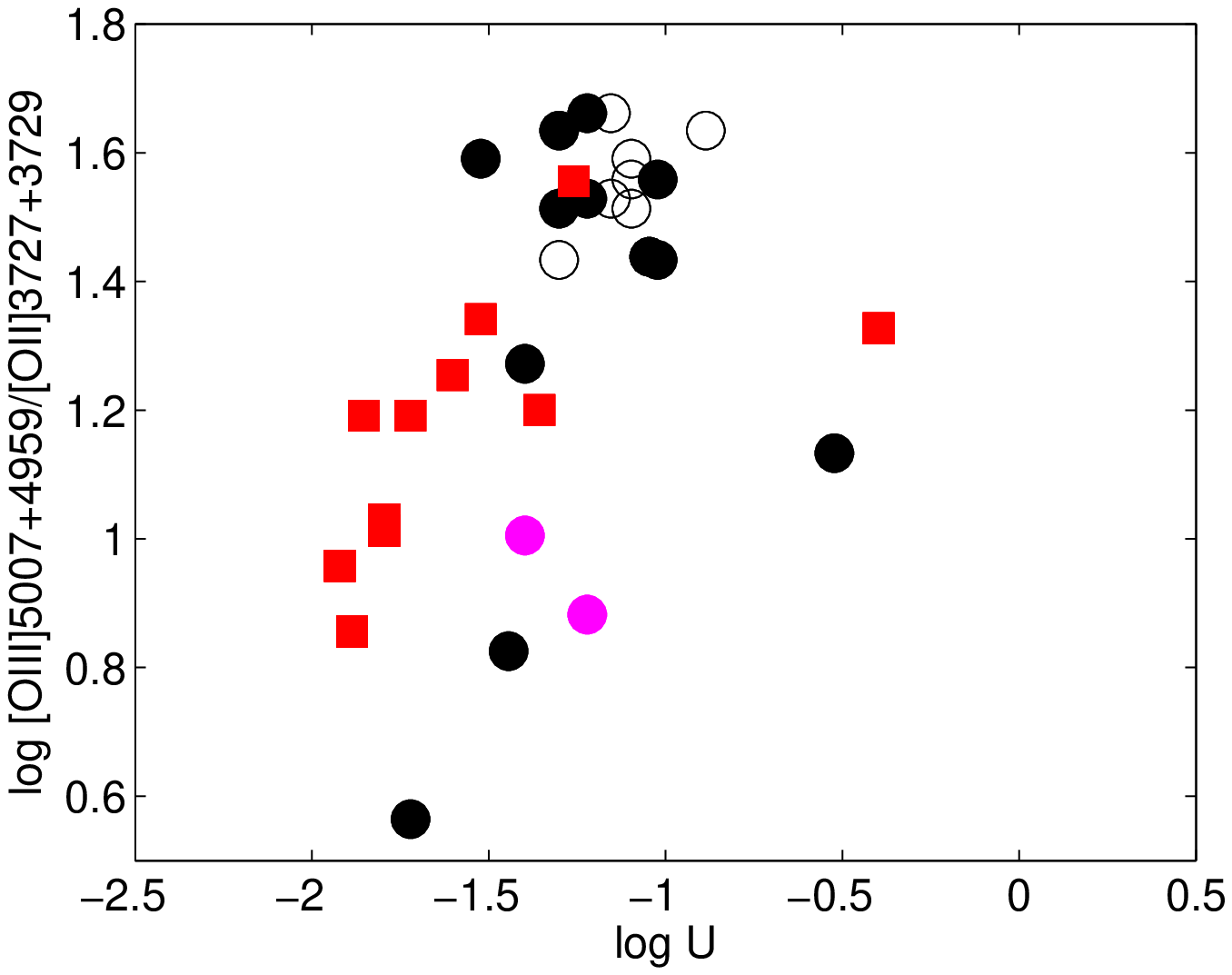}
\includegraphics[width=7.2cm]{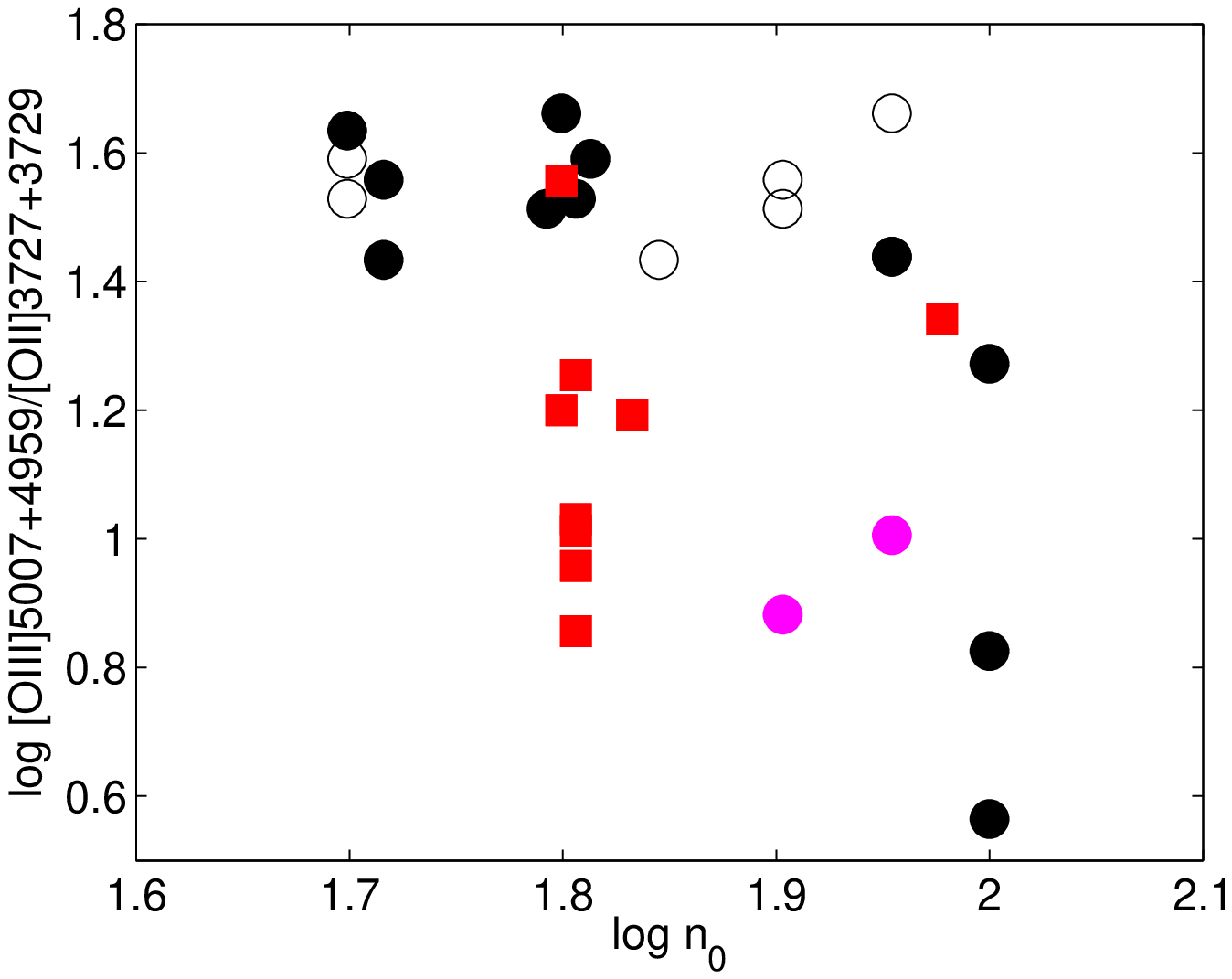}
\includegraphics[width=7.2cm]{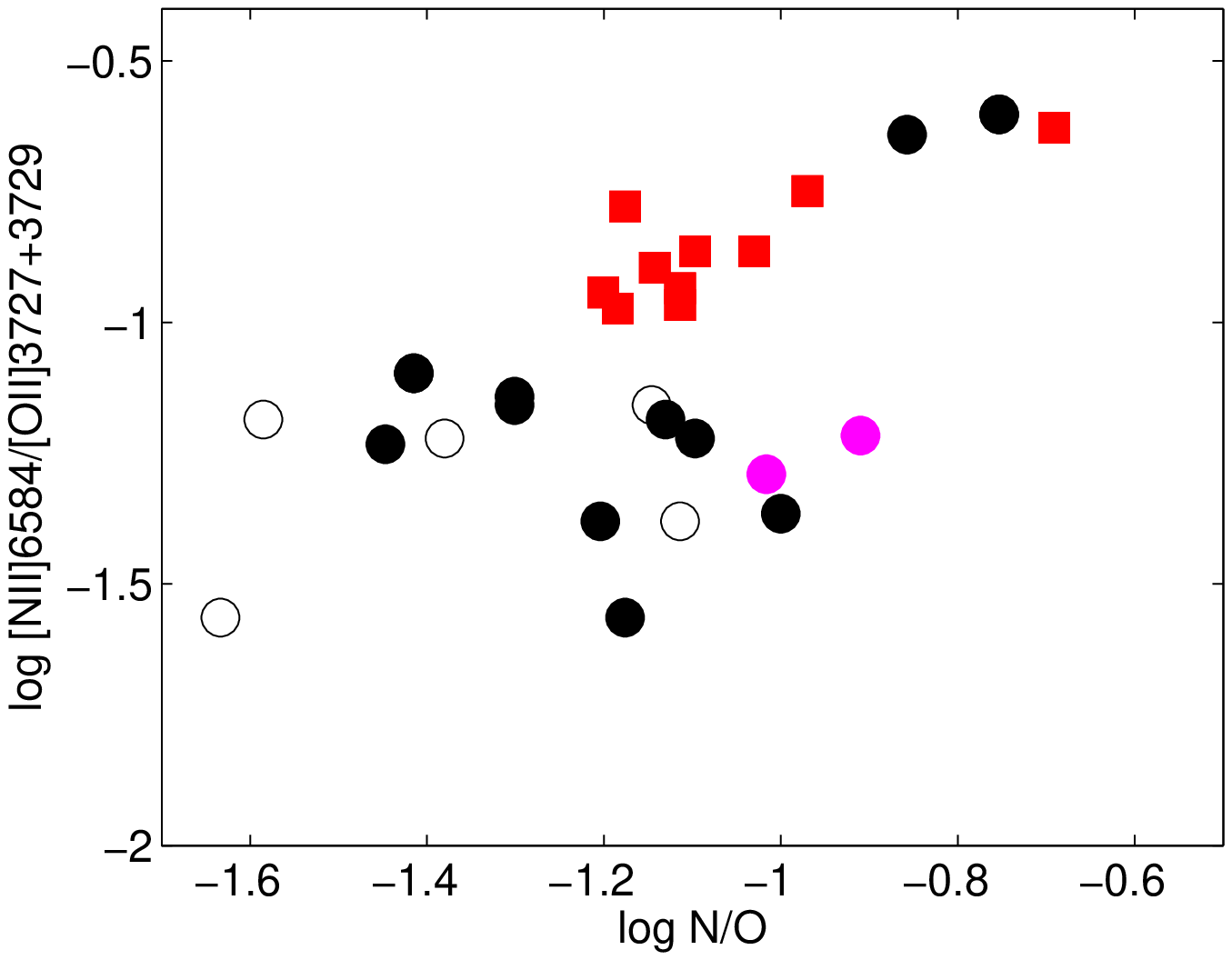}
\includegraphics[width=7.2cm]{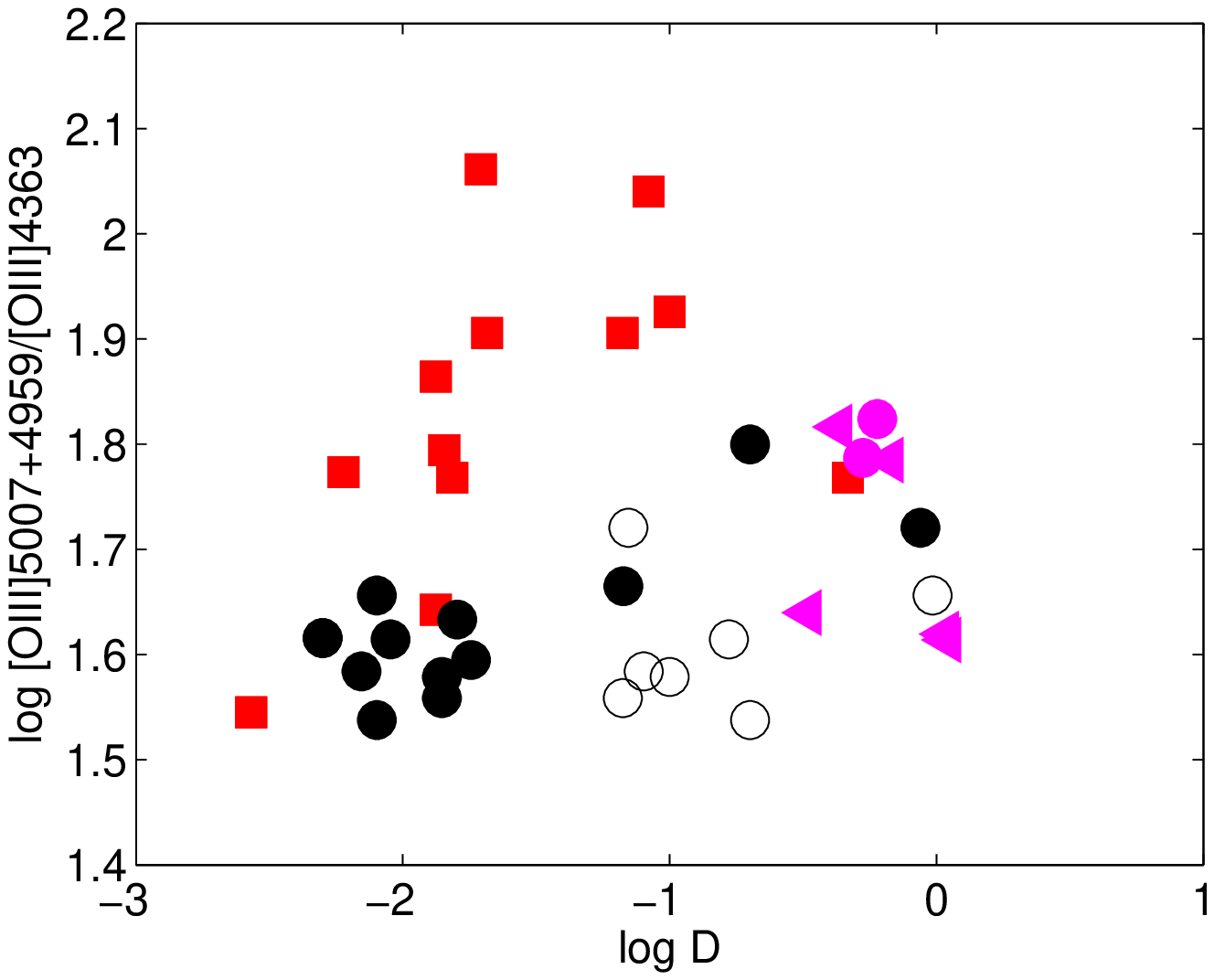}
\includegraphics[width=7.2cm]{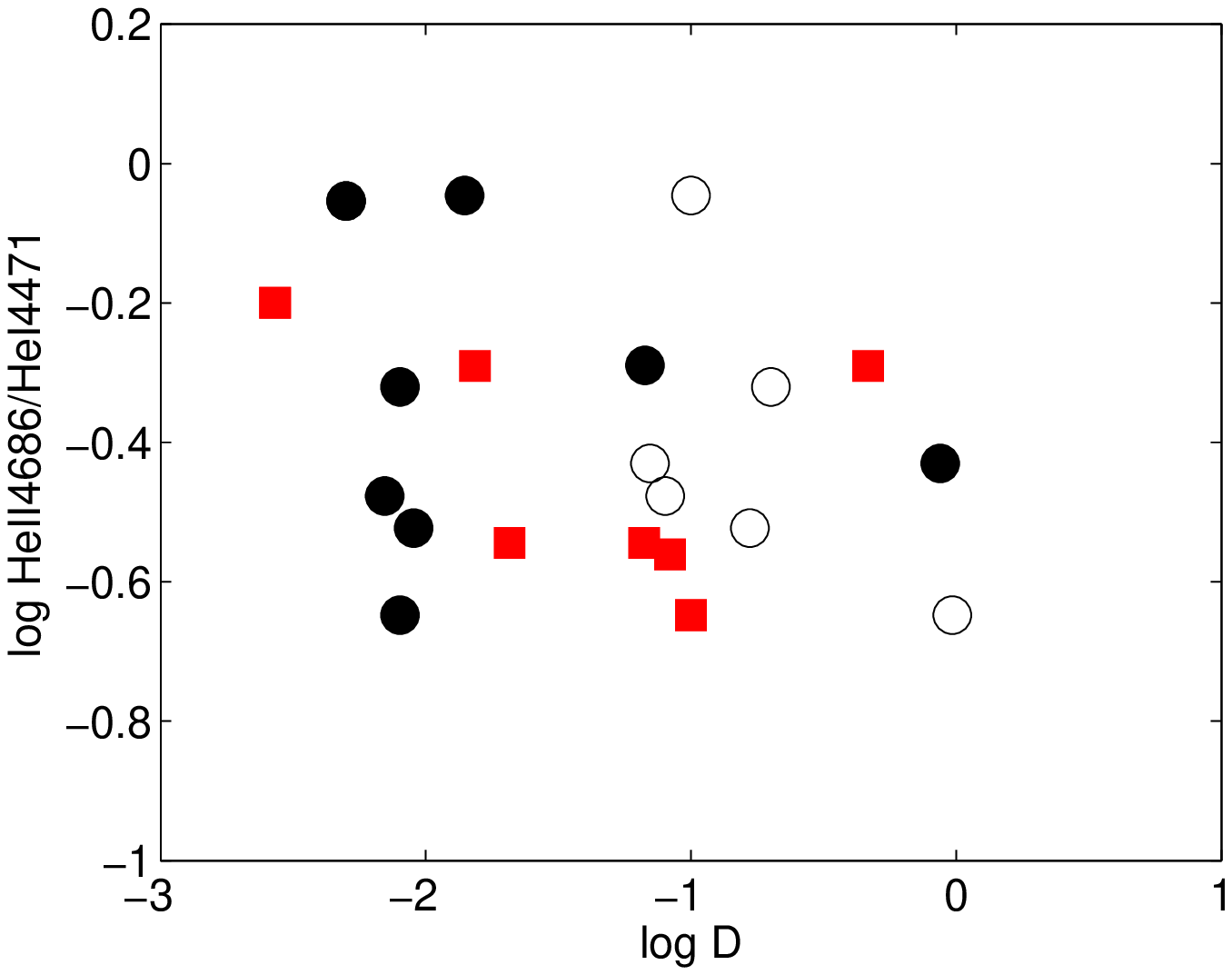}
\caption{Some significant line ratio trends as function of  
the different parameters. Black filled circles: galaxies in the
0.020-0.065 z range; 
black open circles and magenta:  galaxies in the 0.020-0.065 z range referring to Table 4
parameters; red filled squares: galaxies in the 0.13-0.43 z  range. 
}
\end{figure*}

\section{Concluding remarks}

We have  collected some significant spectroscopic data observed from star-forming galaxies in the local Universe.
We have selected the objects showing a relatively rich spectrum in number of lines from different elements. 
In particular we have focused on  
the  weak lines such as [OIII]4363, HeII4686, HeI5876, etc. trying to reproduce 
 all the line ratios within  each of the  observed spectra by the detailed modelling. 
 The results are summarised in the following.

$\bullet$ We confirm that  a  relatively low  metallicity characterizes all the elements, except S.  
A solar S/H is explained by the contribution to the [SII] lines  from the ISM. 
 However, the calculated relative abundances for the other elements (O, N, Ar, etc.) are not as low  as evaluated 
 by the observers using  the strong line  and \Te methods.  

$\bullet$ 
A  main contribution to the  neutral lines (e.g. HeI, [OI])  from the ISM is revealed  by comparing calculated to
observed line ratios.

$\bullet$ We have found by  modelling   the observed line ratios that  in  a not negligible number of galaxies 
 He/H $\leq$0.08  is lower than solar ((He/H)$_{\odot}$= 0.1). 
This result  suggests that the wind from the star-forming region not only collisionally affects the dynamics of 
the emitting cloud throughout the galaxy, but also  the  gas composition of the emitting clouds.

$\bullet$ For most of the spectra a single model is not  enough  to fit contemporarily the
HeII4686/\Hb~ and the [OIII]4363/\Hb~ line ratios, whereas the [OIII]5007+/\Hb, [OII]3727+/\Hb, [NII]6584/\Hb~ and 
[NeIII]3868/\Hb~
line ratios are all  satisfactorily reproduced. The geometrical thickness of the clouds plays  a critical role. 
Pluri-cloud models are  adopted in agreement with cloud fragmentation due to turbulence created by the shocks.  

$\bullet$  Although by a reduced number of objects, we  could assemble some diagrams which 
can be  useful to guess the physical condition and element relative abundance  ranges
in  SF galaxies.
In Fig. 7  the trends of the observed line ratios are displayed as function of the  calculated parameters. 
The trends   are disturbed because a not  negligible  number of  parameters are  interacting with each other.
Nevertheless, Fig. 7 diagrams  can  give  some  hints to the interpretation of   SFG spectra.
In the bottom diagrams  the trends of the main line ratios are shown for  models calculated
 in the different  geometrical thickness ranges in order to point out  the role of $D$ in  
the  [OIII]4363/\Hb~ and HeII4686/\Hb~ modelling.

\section*{Data availability}
The data underlying my work are available in the manuscript

\appendix
\section{Calculation method}

The code {\sc suma}  simulates the physical conditions of an emitting gaseous 
cloud under the coupled effect of photoionization from a radiation
source and shocks assuming a plane-parallel geometry.
Two  cases  are considered relative to the cloud propagation : the photoionizing 
radiation reaches  the gas on the  cloud edge  corresponding to the shock front (infalling) 
or on the   edge  opposite to the shock front (ejection). 

To calculate the line flux and the continuum emitted from a gas the physical
conditions and the fractional abundances of the ions must be known.
In a shock dominated  regime
the calculations start at the shock front where the gas is compressed 
and thermalized adiabatically, reaching the maximum
temperature (T $\propto$ V$_s^2$, where \Vs, is the shock velocity)
in the immediate post-shock region. Compression is 
calculated by  the Rankine-Hugoniot equations (Cox 1972) for the 
conservation of mass, momentum and energy throughout the shock front and downstream. 
Compression strongly affects the cooling rate and consequently,
the distribution of the physical conditions downstream, as 
well as that of the element fractional abundances.   
The downstream region is automatically cut in many plane-parallel slabs 
(up to 300) with different geometrical widths   in order 
to account for the temperature gradient throughout the gas.  Thus, the change of the physical 
conditions downstream from one slab to the next is minimal. 
In each slab the fractional abundances of all the ions is calculated resolving
the  ionization equations which
account for the ionization mechanisms (photoionization by the primary and diffuse radiation and
collisional ionization) and recombination mechanisms (radiative,
dielectronic recombinations) as well as charge transfer effects.
The ionization equations are coupled to the energy equation (Cox 1972),
when collisional processes dominate, and to the thermal balance equation if
radiative processes dominate. This latter balances the heating
of the gas due to the primary  and diffuse radiations reaching
the slab  and the cooling due to recombinations and collisional
excitation of the ions followed by line emission and
thermal bremsstrahlung. The coupled equations
are solved for each slab, providing the physical conditions necessary
to calculate  the slab optical depth  and  the  line and
continuum emissions. The slab contributions are integrated
throughout the nebula.
The calculations stop when the electron temperature is as low as 200 K, if the nebula is 
radiation-bounded or  at a given value of the nebula geometrical 
thickness, if it is matter-bounded. 
The fractional abundances of the ions are calculated resolving the ionization equations
for each element (H, He, C, N, O, Ne, Mg, Si, S, Ar, Cl, Fe) in each ionization level.
Then, the calculated line ratios, integrated throughout the cloud thickness, are compared with the
observed ones. The calculation process is repeated
changing  the input parameters until the observed data are reproduced by the model results,  at maximum
within 10-20 percent.

 On this basis we calculate a grid of models. 
When one or two line ratios  underpredict or overpredict the data by a factor of
$\geq$ 10, we  change  drastically the input parameters e.g. the geometrical thickness of the emitting cloud.
It is clear that one only set of parameters is not enough to represent the whole galaxy emission. The observations
give an average, but an average cannot be fitted by a single spectrum
because an average spectrum has not a physical meaning in terms of the line ratios.

The main input parameters of the code  are  those referring  to the shock
 as well as those characterizing 
the source ionizing radiation spectrum, and the chemical abundances 
of He, C, N, O, Ne, Mg, Si, S, Ar, Cl and Fe, relative to H. 
In our model the line and continuum emitting  regions throughout the galaxy cover  an ensemble of fragmented clouds.
The geometrical thickness of the clouds is  an input parameter of the code ($D$) which is  calculated
consistently with the physical conditions and element abundances of the emitting gas.
The effect of dust present in the gas, characterized by the dust-to-gas
 ratio, $d/g$,  and the initial grain radius, \agr, are also  consistently
taken into account. 
 The  set of selected  parameters  finally determined for a particular spectrum  represents
the {model}.

\subsection{Parameters depending on the shock}

In the turbulent regime created throughout a starburst (SB) shocks are ubiquitous. 
The preshock density n$_0$, the shock velocity \Vs, the magnetic field \B0
(for  all  galaxy models \B0=10$^{-4}$Gauss is adopted) represent the shock.
They  are  used in the calculations  of the Rankine-Hugoniot equations
  at the shock front and downstream.
They  are combined in the compression equation  which is resolved
throughout each slab of the gas
in order to obtain the density profile downstream.
Generally, \Vs is constrained by the FWHM of the line profiles,
\n0 by the ratio of the characteristic lines.
The relative  abundances of the elements  are constrained by the line ratios.
In the case where shock and 
photoionization act on opposite sides of a plan-parallel nebula, 
the geometrical width of the nebula, $D$, is a critical input parameter.
The diffuse radiation bridges the two sides, and  the smaller $D$
 the more entangled are the photoionized and 
the shocked regions on the opposite sides of the nebula. In this
case, a few iterations are necessary to  obtain the 
physical conditions downstream.

\subsection{Photoionizing radiation flux}

 We adopt for the primary radiation  a  black-body (bb) corresponding to an
 effective temperature \Ts and  a ionization parameter $U$.
 A pure bb radiation referring to \Ts is a poor approximation for a SB, 
even adopting a dominant spectral type (see Rigby \& Rieke 2004). 
Following Rigby \& Rieke,  "the starburst enriches and heats its ISM as well as the intergalactic medium.
The ionizing spectrum is set by the SB age, IMF and star formation history."  Adopting a single effective
temperature the entire SB field is represented by a single star type. However, 
the observed line spectra for high redshift galaxies at present  cover a  narrow  optical-near-IR range of frequencies,
the lines are few and from few ionization levels, therefore  the bb radiation flux calculated by a dominant
temperature is acceptable, also in view that
 the line ratios (that are  related  to \Ts) in a shock dominated regime also depend  on the
electron temperature, density, ionization parameter, metallicity, on the morphology of the ionized clouds, 
and in particular, they depend on the hydrodynamical field.
Therefore we  will determine  \Ts phenomenologically by  selecting the effective temperature \Ts which  leads 
to  the best fit of all the observed line ratios for each spectrum and we will use it to calculate the  continuum.
The primary radiation source
 does not depend on the  physical conditions throughout the galaxy  but it affects the surrounding gas.   
 This  region  is not considered
as a unique cloud, but as a  sequence of slabs with different thickness calculated automatically
following the temperature gradient. 

The  radiation from a  photoionizing source is characterized by its
spectrum, which is calculated at 440 energies, from a few eV to KeV,
depending on the object studied. Due to  radiative transfer, the
radiation spectrum changes throughout the downstream slabs, each of them
contributing to the optical depth. The calculations assume a steady
state  downstream. In addition to the radiation from the primary
source, the effect of the diffuse secondary radiation created by the gas emission
(line and continuum) is also taken into account (see, for instance,
Williams 1967), using about 240 energies to calculate the spectrum.
The secondary diffuse radiation is emitted from the slabs of
gas heated  by the radiation flux reaching the gas and by the shock.
Primary and secondary radiation are calculated by radiation transfer.

 For an AGN, the primary radiation is the power-law radiation
flux  from the active centre $F$  in number of photons cm$^{-2}$ s$^{-1}$ eV$^{-1}$ at the Lyman limit
and  spectral indices  $\alpha_{UV}$=-1.5 and $\alpha_X$=-0.7. 
 $F$  is combined with the ionization parameter $U$ by
$U$= ($F$/(n c ($\alpha$ -1)) (($E_H)^{-\alpha +1}$ - ($E_C)^{-\alpha +1}$)
(Contini \& Aldrovandi, 1983), where
$E_H$ is H ionization potential  and $E_C$ is the high energy cutoff,
$n$ the density, $\alpha$ the spectral index, and c the speed of light.

If the stars are the photoionization source
the number of ionizing photons cm$^{-2}$ s$^{-1}$ produced by the hot 
source is $N$= $\int_{\nu_0}$ $B_{\nu}$/h$\nu$ d$\nu$, 
where $\nu_0$ = 3.29$\times$10$^{15}$ s$^{-1}$ and B$_{\nu}$  is the Planck function. 
The flux from the star is combined with $U$ and n by $N$ (r/R)$^2$=$U$nc, where r is 
the radius of the hot source (the stars),
 R is the radius of the nebula (in terms of the distance from the stars), n is the density of the nebula and c is the 
speed of light. Therefore, \Ts  and $U$ compensate each other, but only in a qualitative way, because \Ts  determines 
the frequency distribution of the primary flux, while $U$ represents the number of photons per number of electrons 
reaching the nebula. The choice of \Ts and $U$   is obtained  by the fit of the line ratios.

\subsection{Electron temperatures through the nebula}

The temperature in each slab depends on energy gains (G) and losses (L) of the gas.
Close to the shock front downstream, collisional mechanisms prevail
and the temperature is calculated from the energy equation in terms of
the enthalpy change (Cox 1972).
In the slabs  where the temperature is $\leq$ 2 10$^4$ K, photoionization
and heating by both the primary and the secondary radiation dominate
and the temperature is calculated by thermal balance (G=L).
Gains are calculated by the rate at which energy is given to the electrons
by the radiation field (Osterbrock 1974).  The energy of suprathermal
electrons created by photoionization is rapidly distributed among the thermal
electrons through collisions, heating the gas.
Several processes contribute to the gas cooling. The cooling rate (Williams 1967) is given by:
L = L$_{ff}$+L$_{fb}$+L$_{lines}$+L$_{dust}$,
where L$_{ff}$ corresponds to bremsstrahlung, particularly strong
at high temperatures and high frequencies.
Self-absorption is included in the calculations.
 L$_{fb}$ corresponds to free-bound losses due to recombination and
is high at T$\leq$ 10$^5$ K.
L$_{lines}$ is due to line emission  with the bulge between $\leq$ 10$^4$ K and 10$^5$ K.
 L$_{dust}$  represents the energy lost by the gas in  the collisional heating of dust
grains. It is high the higher d/g and \agr.

Immediately behind the shock front the gas is thermalized to
a temperature of T=1.5 10$^5$(\Vs/(100 \kms))$^2$ K.
At high temperatures ($\geq$ 10$^6$ K) recombination coefficients are very
low.  The cooling rate is then low.
At T between 10$^4$ K and 10$^5$ K the UV lines and the coronal lines
in the IR are strong and lead to rapid cooling and compression
of the gas. If the cooling rate is so high to
drastically reduce the  temperature  eluding intermediate ionization-level lines,
the calculated spectrum will be wrong.
Therefore, the slab thickness must be reduced and all the physical
quantities recalculated. The choice of the slab thickness is determined
by the gradient of the temperature.
This process is iterated until the
thickness of the slab is such as to lead to an acceptable gradient of
the temperature (T(i-1) - T(i))/T(i-1) $\leq$ 0.1, 
where T(i) is the temperature  of slab i.
As the temperature drops, a large region of gas with 
temperature  $\sim$10$^4$ K,  which is sustained by  the secondary radiation,
 is present in the radiation-bounded case,
i.e. when the gas recombines completely before reaching the 
 edge of the nebula opposite to the shock front. 
 Due to a lower temperature gradient,  calculations in this zone  
may be performed in slabs with a larger geometrical thickness. 

\subsection{Element abundances}

We start  model calculations by adopting solar abundances. We use solar abundances which are intermediate
between those presented by Anderson \& Grevesse (1989)  and Asplund et al (2009), similar to those
of Allen (1976) for all the elements, in particular for oxygen, trying to
 fit the observations. If  the  [OIII]5007+/\Hb, [OIII]4363/\Hb~ and  [OII]/\Hb~  line
 ratios all satisfactorily reproduce the data, we adopt for O/H  the solar value. If the
 the oxygen lines ratios to \Hb~ are all higher (or lower) than
 observed  by a similar factor, we try to reproduce all the oxygen line ratios to \Hb~ by  reducing (or increasing)  O/H.
 If the  various oxygen line ratios to \Hb~  differ from the data by different factors  we restart  the modelling process
 by changing all the input parameter set.
 When the oxygen to \Hb~ line ratios reproduce successfully the
 data, we change the other  element relative abundances, in particular N/H and S/H, if  necessary.
 The oxygen abundance, in particular, strongly affects the cooling rate downstream in the recombination zone
 throughout the emitting clouds. 
 Therefore, the calculations process must be reinitiated until
 the best fit of  all the line ratios to \Hb~ from all the   elements  is obtained.

In particular, the absolute line fluxes referring to the ionization level i of element K are calculated by the 
term $n_K$(i) which represents the density of the ion i. We consider that $n_K$(i)=X(i)[K/H]$n_H$, where X(i) is 
the fractional abundance of the ion i calculated by the ionization equations, [K/H] is the relative 
abundance of the element K to H and $n_H$ is the density of H (by number \cm3). In models including shock, 
$n_H$ is calculated by the compression equation in each slab downstream. 
So the abundances of the elements are given relative to H as input parameters.

\subsection{Dust grain heating}

Dust grains are coupled to the gas across the shock front by the magnetic field (Viegas \& Contini 1994). 
They are heated by the radiation source  and collisionally by the gas to a maximum temperature which is a 
function of the shock velocity, of the chemical composition and of the radius of the grains, 
up to the evaporation temperature ($T_{\rm dust}$ $\geq$ 1500 K). 
The grain radius distribution downstream is determined 
by sputtering, which depends on the shock velocity and on the density. 
Throughout shock fronts and downstream, 
the grains might be destroyed by sputtering.
The grains are heated  by the primary and secondary radiation, and by gas collisional 
processes.  Details of the calculations of the dust 
temperature are given by Viegas \& Contini (1994). 
When the dust-to-gas ratio $d/g$ is high, 
the mutual heating of dust and gas may accelerate the 
cooling rate of the gas by L$_{dust}$, changing 
the line and continuum spectra emitted from  the gas.
The intensity of dust reprocessed radiation in the IR  depends on $d/g$ and on the radius \agr.
In this work we use $d/g$=10$^{-14}$ by number for all the models which corresponds to
4.1 10$^{-4}$ by mass for silicates.

\end{document}